\documentclass[3p,number,sort&compress,review]{elsarticle}

\usepackage{graphicx}
\usepackage{array}
\usepackage{amsmath,amssymb}
\usepackage{float}
\usepackage{morefloats}
\usepackage{placeins}
\newtheorem{example}{Example}

\newtheorem{assumption}{Assumption}
\DeclareMathOperator{\sign}{sign}
\usepackage{graphicx}
\usepackage{amsmath,amssymb}
\usepackage[suffix=, outdir=./]{epstopdf}

\usepackage{float}
\DeclareMathOperator{\arctg}{arctg}
\usepackage{color}

\usepackage{setspace}
\usepackage{graphicx}

\begin{document}

\begin{frontmatter}

\title{Tutorial on Dynamic Analysis of the Costas Loop.\footnote{Final version of this article is published in Annual Reviews in Control doi:10.1016/j.arcontrol.2016.08.003 http://www.sciencedirect.com/science/article/pii/S1367578816300530}\footnote{This work was supported by Russian Science Foundation (project 14-21-00041) and Saint-Petersburg State University}}

\author[best]{Best~R.E.}
\ead{rolandbest@aol.com}

\author[spbu,jyv]{Kuznetsov~N.V.}
\ead{nkuznetsov239@gmail.com}

\author[spbu,ras]{Leonov~G.A.}
\ead{g.leonov@spbu.ru}

\author[spbu]{Yuldashev~M.V.}
\ead{maratyv@gmail.com}

\author[spbu]{Yuldashev~R.V.}
\ead{renatyv@gmail.com}


\address[best]{Best Engineering, Oberwil, Switzerland}
\address[spbu]{Saint-Petersburg State University, Russia}
\address[jyv]{University of Jyv\"{a}skyl\"{a}, Finland}
\address[ras]{Institute of Problems of Mechanical Engineering RAS, Russia}

\begin{abstract}
  Costas loop is a classical phase-locked loop (PLL) based circuit
for carrier recovery and signal demodulation. The PLL is an automatic control system
that adjusts the phase of a local signal to match the phase of the input reference signal.
This tutorial is devoted to the dynamic analysis of the Costas loop.
In particular the acquisition process is analyzed. Acquisition is most conveniently
described by a number of frequency and time parameters such as lock-in range,
lock-in time, pull-in range, pull-in time, and hold-in range.
While for the classical PLL equations all these parameters have been
derived (many of them are approximations, some even crude approximations), this
has not yet been carried out for the Costas loop. It is the aim of this analysis
to close this gap.
The paper starts with an overview on mathematical and physical models
(exact and simplified) of the different variants of the Costas loop.
Then equations for the above mentioned key parameters are derived.
Finally, the lock-in range of the Costas loop for the case where a lead-lag filter is
used for the loop filter is analyzed.
\end{abstract}

\end{frontmatter}

\renewcommand{\thesection}{\arabic{section}}
\setcounter{section}{0\,} \thispagestyle{empty}

\section{Introduction}\label{ss1.1}
Costas loop is a classical phase-locked loop (PLL) based circuit
for carrier recovery and signal demodulation
\cite{Costas-1962-patent,Waters-1982-patent}.
The PLL is an automatic control system which is designed to generate an electrical signal (voltage),
the frequency of which is automatically tuned to the frequency of the input (reference) signal.
Various PLL based circuits are widely used  in modern telecommunications,
computer architectures, electromechanical systems (see, e.g. \cite{Best-2007,kobayashi1990reduction,lazzari2015enabling}).
Nowadays among the applications of Costas loop there are Global Positioning
Systems (see, e.g., \cite{KaplanH-2006-GPS}), wireless communication (see, e.
g., \cite{Rohde-2000-book}) and others (\cite{Stephens-2001,Couch-2007,Proakis-2007,Godse-2010-book,sidorkina2016costas}).

\begin{figure}[H]
\centering
\includegraphics[scale=0.8]{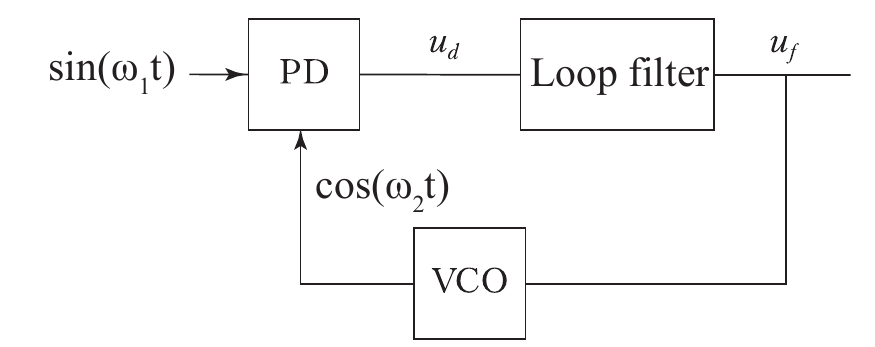}
\caption{Block diagram of a PLL}\label{f1-1}
\end{figure}

Dynamic behavior of the PLL and the Costas loop has been described extensively in the literature \cite{Gardner-1979-book,Best-2007,Rantzer-2001,bianchi2005phase,bizzarri2012periodic,Rohde-2000-book,simon1977optimum,cahn1977improving,LeonovKYY-2015-SIGPRO,BestKLYY-2014-IJAC,KuznetsovLYY-2014-ICUMT,KuznetsovKLNYY-2014-ICUMT-QPSK,KuznetsovLNSYY-2012-IEEE-Costas,KaplanH-2006-GPS}, and a number of key parameters has been defined that describe its lock-in and lock-out characteristics.
 When the PLL is initially out of lock, two different types of acquisition processes can occur, either the so-called lock-in process or the so-called pull-in process.
 The first of those is a fast process, i.e. the acquisition takes place within at most one beat note of the difference between reference frequency  $\omega_1$ and initial VCO (Voltage Controlled Oscillator) frequency  $\omega_2$, cf. Figure~\ref{f1-1} for signal denotations\footnote{Non-sinusoidal signals in PLL-based circuits are considered in \cite{LeonovKYY-2012-TCASII,KuznetsovLYY-2011-IEEE}}.\
 The frequency difference for which such a fast acquisition process takes place corresponds to the lock-in range  $\Delta\omega_L$, and the duration of the locking process is called lock time $T_L$.
 When the difference between reference and VCO frequency is larger than the lock-in range but less than the pull-in range $\Delta\omega_P$, a slow acquisition process occurs.
 The time required to get acquisition is called pull-in time $T_P$.
 In case of the PLL all these acquisition parameters can be approximated by characteristic parameters of the PLL, i.e. from natural frequency  $\omega_n$ and damping factor  $\zeta$.

To the authors knowledge such acquisition parameters have not been analytically derived for the different types of Costas loops.
It seems that most authors only described the static properties of the Costas loop such as the derivation of the phase error in the locked state and the like.
Based on methods developed earlier for the PLL, the authors could now derive similar expressions for all relevant acquisition parameters of the Costas loop.
This enables the designer to determine the lock-in and pull-in ranges, and to estimate
the duration of the corresponding processes.

Because the systems considered are highly non linear, exact computation of such parameters is very difficult or even impossible. Therefore it is necessary to introduce a number of simplifications. This implies that the obtained results are only approximations, in some cases rather crude approximations.

 As will be shown in the following sections there are different types of Costas loops.
 The very first of these loops has been described by J. Costas in 1956 \cite{Costas-1956} and was primarily used to demodulate amplitude-modulated signals with suppressed carrier (DSB-AM). The same circuit was used later for the demodulation of BPSK signals (binary phase shift keying) \cite{Proakis-2007}. With the advent of QPSK (quadrature phase shift keying) this Costas loop was extended to demodulate QPSK signals as well. These two types of Costas loop operated with real signals. In case of BPSK, the input signal $u_1(t)$ is a sine carrier that was phase modulated by a binary signal, i.e.
 \begin{equation}
 \label{input signal}
 u_1(t)=m_1(t)\sin(\omega_1t),
 \end{equation}
where  $\omega_1$ is the (radian) carrier frequency, and $m_1(t)$ can have two different values, either $+1$ or $-1$, or two arbitrary equal and opposite values $+c$ and $-c$, where $c$ can be any value. In case of QPSK, two quadrature carriers are modulated by two modulating signals, i.e.

 \begin{equation}\label{qpsk signal}
 u_1(t)=m_1(t)\cos(\omega_1t) + m_2(t)\sin(\omega_1t),
 \end{equation}
where $m_1$ and $m_2$ can both have two equal and opposite values $+c$ and $-c$. It is obvious that in both cases the input signal is a real quantity. In the following these two types of Costas loop will be referred to as ``conventional Costas loops''.

\begin{figure}[H]
\centering
\includegraphics[scale=0.5]{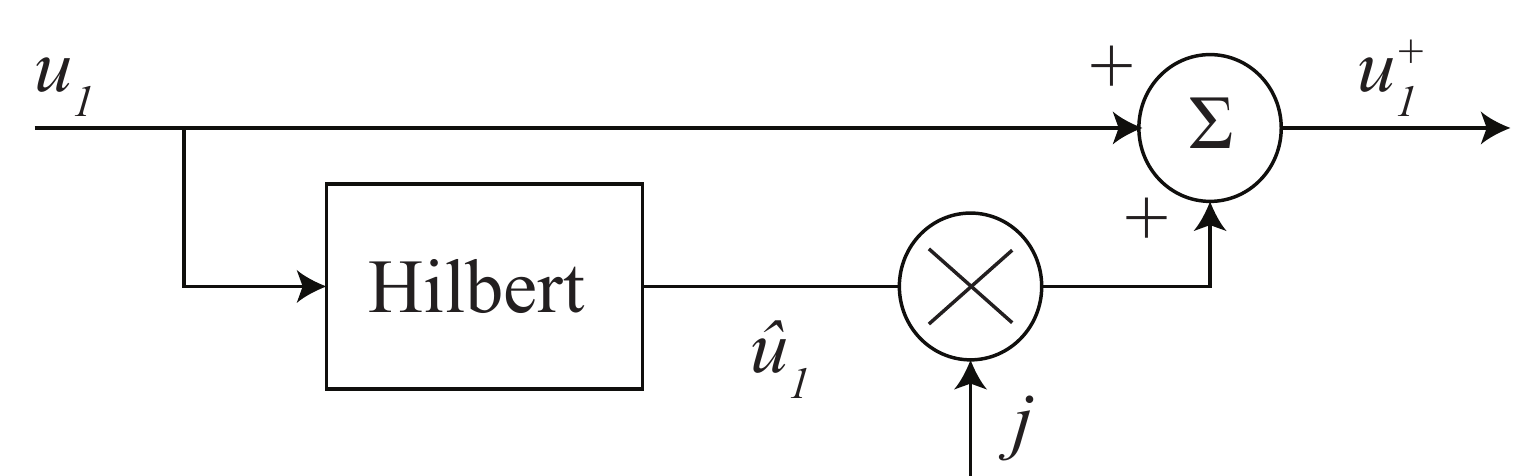}
\caption{Generation of the pre-envelope signal using Hilbert transformer
}\label{f1-2}
\end{figure}

Much later, Costas loops have been developed that operate not on real signals, but on pre-envelope signals \cite{Tretter}. These types of Costas loops will be referred to as ``modified Costas loop'' in the following sections. The block diagram shown in
Figure~\ref{f1-2} explains how the pre-envelope signal is obtained. The real input signal $u_1(t)$ is applied to the input of a Hilbert transformer [2], [5]. The output of the Hilbert transformer $\hat u_1(t)$  is considered to be the imaginary part of the pre-envelope signal, i.e the pre-envelope signal is obtained from
$$
u_1^+(t)=u_1(t)+j\hat u_1(t).
$$
The Costas loops operating with pre-envelope signals will be referred to as ``modified Costas loops'', cf. sections \ref{s4} and \ref{s5}.
Because there are different types of Costas loops the acquisition parameters must be derived separately for each of these types. This will be performed in the following sections. In order to see how good or bad the obtained approximations, we will develop Simulink models for different types of Costas loops and compare the results of the simulation with those predicted by  theory.

\subsection{Classical mathematical models of the Costas loops}
\subsubsection{BPSK Costas loop}
\begin{figure}[thpb]
\centering
 \includegraphics[width=0.8\textwidth]{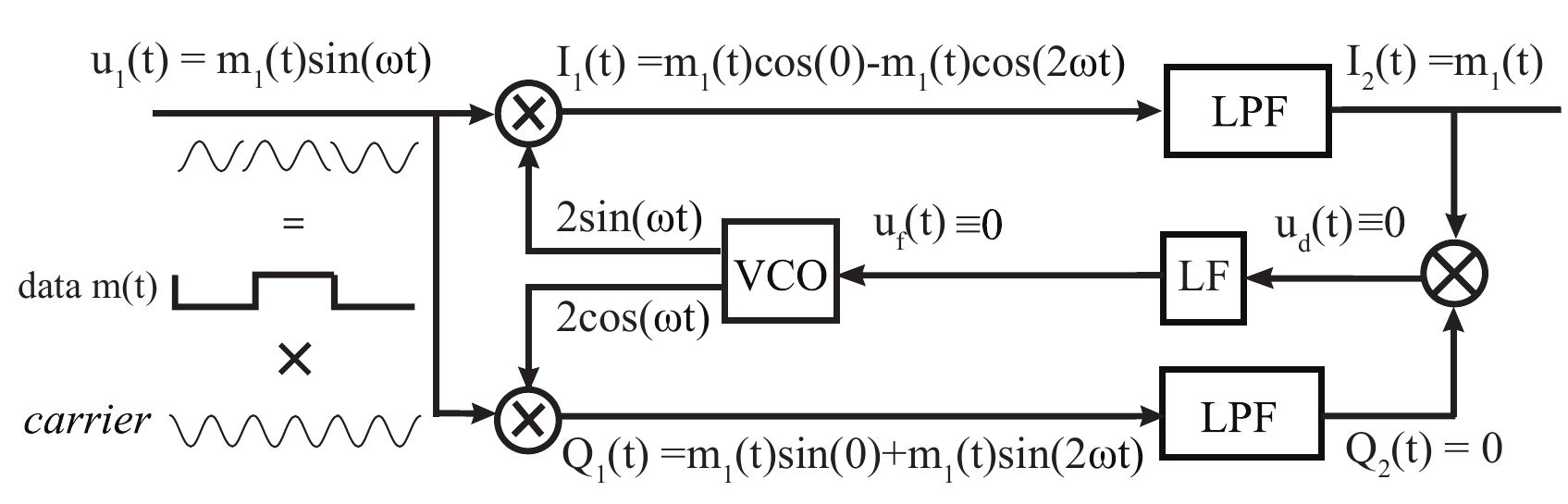}
\caption{Costas loop is locked
(the case of equal phases of input carrier and free running VCO output):
there is no phase difference.
}
\label{costas_locked}
\vspace{-0.4cm}
\end{figure}

The operation of the Costas loop is considered first in the locked state with zero phase difference (see Figure~\ref{costas_locked}),
hence the frequency of the carrier is identical with the frequency
of the VCO.

By \eqref{input signal} the input signal $u_1(t)$ is  the product of a transferred binary data
and the harmonic carrier $\sin(\omega t)$ with a high frequency $\omega$.
Since the Costas loop is considered to be locked,
the VCO orthogonal output signals are synchronized with the carrier
(i.e. there is no phase difference between these signals).
The input signal is multiplied (multiplier block ($\otimes$))
by the  corresponding VCO signal on the upper branch and by the VCO signal, shifted by $90^{\circ}$, on the lower branch.
Therefore on the multipliers' outputs one has
 $I_1(t) =
  m_1(t) - m_1(t)\cos(2\omega t),
  Q_1(t) =
  m_1(t)\sin(2\omega t).$

Consider the low-pass filters (LPF) operation.
\begin{assumption}\label{as-twice-frequency}
  Signals components, whose frequency is about twice the carrier frequency,
  do not affect the synchronization of the loop
  (since they are suppressed by the low-pass filters).
\end{assumption}
\begin{assumption}\label{as-lpf-initstate}
  Initial states of the low-pass filters do not affect the synchronization of the loop
  (since for the properly designed filters, the impact of filter's initial state on its output
  decays exponentially with time).
\end{assumption}
\begin{assumption}\label{as-lpf-data}
  The data signal $m_1(t)$ does not affect the synchronization of the loop.
\end{assumption}

Assumptions~\ref{as-twice-frequency},\ref{as-lpf-initstate}, and \ref{as-lpf-data}
together lead to the concept of so-called \emph{ideal low-pass filter}, which
completely eliminates all frequencies above the cutoff frequency (Assumption~\ref{as-twice-frequency})
while passing those below unchanged (Assumptions~\ref{as-lpf-initstate},\ref{as-lpf-data}).
In the classic engineering theory of the Costas loop it is assumed that
the low-pass filters LPF are ideal low-pass filters\footnote{Note that Assmptions~1--3 may not be valid and require rigorous justification \cite{KuznetsovKLNYY-2015-ISCAS,BestKKLYY-2015-ACC}}.

Since in Figure~\ref{costas_locked} the loop is in lock,
i.e. the transient process is over and the synchronization is achieved,
by Assumptions~\ref{as-twice-frequency},\ref{as-lpf-initstate}, and \ref{as-lpf-data}
for the outputs $I_2(t)$ and $Q_2(t)$ of the low-pass filters LPF one has
$I_2(t) =  m_1(t), \ Q_2(t) = 0.$
Thus, the upper branch works as a demodulator and the lower branch works as a phase-locked loop.

Since after a transient process there is no phase difference,
a control signal at the input of VCO,
which is used for VCO frequency adjustment to the frequency of input carrier signal,
has to be zero: $u_d(t) = 0$.
In the general case when the carrier frequency $\omega$ and
a free-running frequency $\omega_{free}$ of the VCO are different,
after a transient processes
the control signal at the input of VCO has to be non-zero constant:
$u_d(t) = const$,
and a constant phase difference $\theta_{e}$ may remain.
\smallskip
\smallskip
\smallskip

\begin{figure}[thpb]
\centering
 \includegraphics[width=0.8\textwidth]{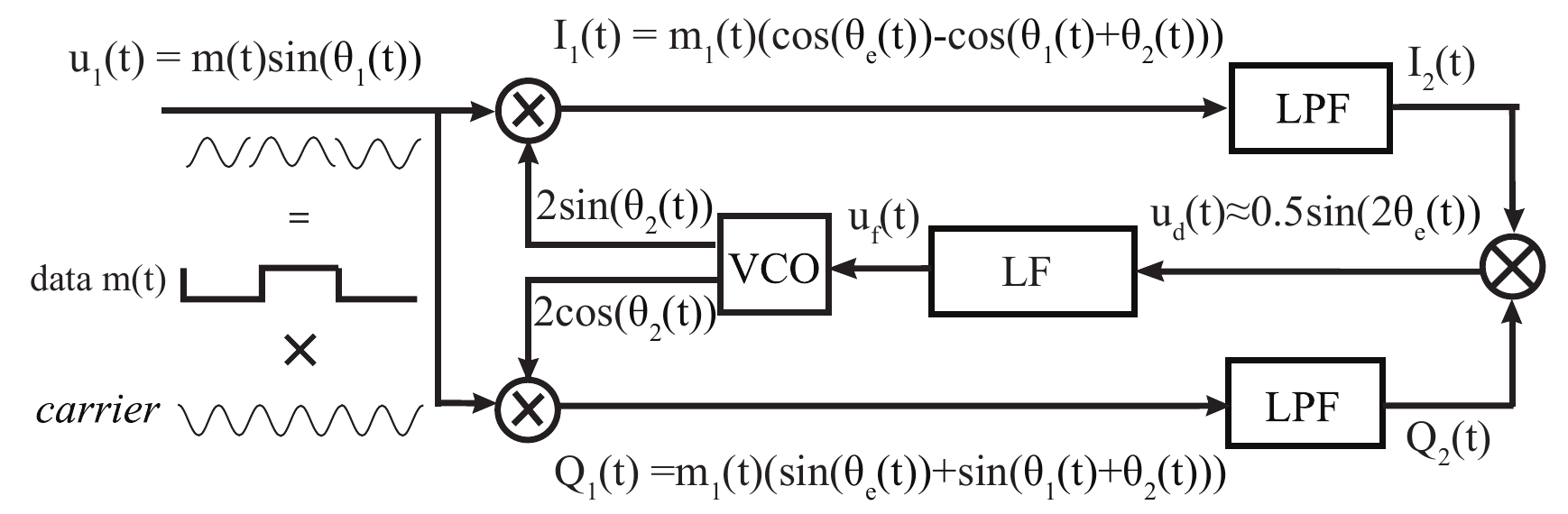}
 \caption{Costas loop is out of lock: there is time-varying phase difference.
 }
\label{costas_before_sync}
\end{figure}

Consider the Costas loop before synchronization
(see Figure~\ref{costas_before_sync}).
Here the phase difference
$
  \theta_{e}(t) = \theta_1(t)-\theta_2(t)
$
varies over time, because the loop has not yet acquired lock
(frequencies or phases of the carrier and VCO are different).
In this case, using Assumption~\ref{as-twice-frequency},
the signals $I_1(t)$ and $Q_1(t)$ can be approximated as
\begin{equation}\label{phi12-approx}
\begin{aligned}
  & I_1(t)
  \approx m_1(t)\cos(\theta_{e}(t)),
  & Q_1(t)
  \approx m_1(t)\sin(\theta_{e}(t)).
\end{aligned}
\end{equation}
Approximations \eqref{phi12-approx} depend on the phase difference of signals,
i.e. two multiplier blocks ($\otimes$) on the upper and lower branches
operate as phase detectors.
The obtained expressions \eqref{phi12-approx} with $m_1(t)\equiv1$
coincide with well-known (see, e.g., \cite{Viterbi-1966,Best-2007})
phase detector characteristic of the classic PLL with multiplier/mixer phase-detector
for sinusoidal signals.

By Assumptions~\ref{as-lpf-initstate} and \ref{as-lpf-data}
the low-pass filters outputs can be approximated as
\begin{equation}\label{g12-approx}
\begin{aligned}
  & I_2(t)
  \approx m_1(t)\cos(\theta_{e}(t)),
  & Q_2(t)
  \approx m_1(t)\sin(\theta_{e}(t)).
\end{aligned}
\end{equation}
Since $m_1^2(t) \equiv 1$, the input of the loop filter (LF)  is
\begin{equation}
\label{loop-filter-input-approx}
 u_d(t) = I_2(t)Q_2(t) \approx \varphi(\theta_{e}(t)) = \frac{m_1(t)^2}{2} \sin(2\theta_{e}(t)).
\end{equation}
Such an approximation
is called a \emph{phase detector characteristic of the Costas loop}.

Since an ideal low-pass filter is hardly realized,
its use in the mathematical analysis
requires additional justification.
Thus, the impact of the low-pass filters on the lock acquisition process must be studied rigorously.

The relation between the input $u_d(t)$
and the output $u_f(t)$ of the loop filter has the form 
\begin{equation}\label{loop-filter}
 \begin{aligned}
 & \dot x = A x + b u_d(t),
 \ u_f(t) = c^*x + hu_d(t),
 \end{aligned}
\end{equation}
where $A$ is a constant matrix,
the vector $x(t)$ is the loop filter state,
$b,c$ are constant vectors, h is a number.
The filter transfer function has the form:
\begin{equation}
  H(s) = -c^*(A-sI)^{-1}b+h.
\end{equation}
The control signal $u_f(t)$ is used to adjust the VCO frequency to
the frequency of the input carrier signal
\begin{equation} \label{vco first}
   \dot\theta_2(t) = \omega_2(t) = \omega_{\text{free}} + K_0u_f(t).
\end{equation}
Here $\omega_{free}$ is the free-running frequency of the VCO
and $K_0$ is the VCO gain.
The solution of \eqref{loop-filter} with initial data $x(0)$
(the loop filter output for the initial state $x(0)$) is as follows
\begin{equation}\label{loop-filter-int}
 \begin{array}{c}
 u_f(t,x(0)) = \alpha_0(t,x(0)) +
 \int\limits_0^t
 \gamma(t - \tau)\varphi(\tau)
 {\rm d}\tau
 + h u_d(t),
 \end{array}
\end{equation}
where $\gamma(t - \tau)=c^*e^{A(t-\tau)}b + h$
is the impulse response of the loop filter
and $\alpha_0(t,x(0))= c^*e^{At}x(0)$
is the zero input response of the loop filter,
i.e. when the input of the loop filter is zero.

\begin{assumption}[analog of Assumption~\ref{as-lpf-initstate}]\label{as-lf-initstate}
  Zero input response of loop filter $\alpha_0(t,x(0))$
  does not affect the synchronization of the loop
  (one of the reasons is that
  $\alpha_0(t,x(0))$ is an exponentially damped function for a stable matrix $A$).
\end{assumption}
\smallskip

Consider a constant frequency of the input carrier:
\begin{equation}\label{omega1-const}
   \dot\theta_1(t) = \omega_1(t) \equiv \omega_1,
\end{equation}
and introduce notation
\begin{equation}\label{initial-frequency-diff}
  \Delta\omega_{0} = \omega_1-\omega_{free}.
\end{equation}
Then Assumption~\ref{as-lf-initstate}
allows one to obtain the classic mathematical model of PLL-based circuit
\begin{figure}[H]
\centering
\includegraphics[scale=1.0]{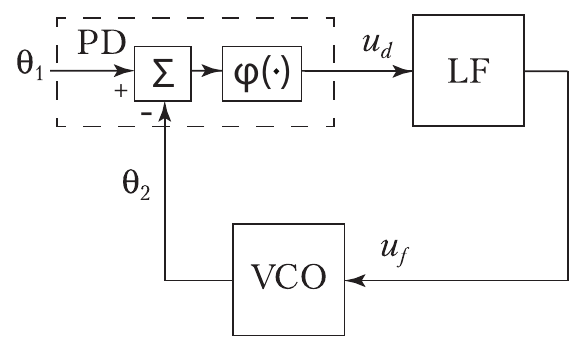}
\caption{Phase model of Costas loop
}\label{f2-1b}
\end{figure}
in signal's phase space
(see Figure~\ref{f2-1b}):
\begin{equation}\label{mathmodel-class-simple}
 \begin{aligned}
   & \dot\theta_{e} =
   \Delta\omega_{0}-K_0\int_0^t
   \gamma(t - \tau)\varphi(\theta_{e}(\tau)){\rm d}\tau
   - K_0 h \varphi(\theta_{e}(t)).
 \end{aligned}
\end{equation}

\begin{figure}[H]
\centering
\includegraphics[scale=1.0]{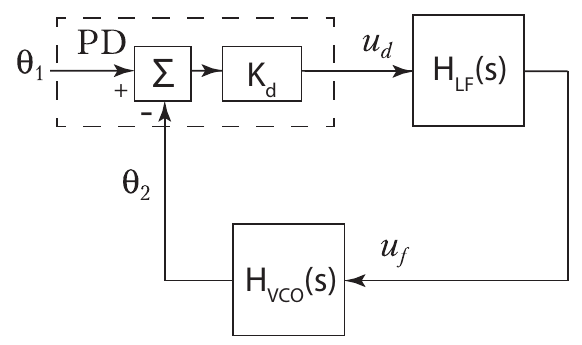}
\caption{Linear model of Costas loop
}\label{linear-pll}
\end{figure}

 For the locked state a linear PLL model can be derived, which is shown in Figure~\ref{linear-pll}. This model is useful for approximation of hold-in range.
In the locked state both reference and VCO frequencies are approximately the same, hence the input of the lowpass filter is a very low frequency signal. Therefore the lowpass filter can be ignored when setting up the linear model of the Costas loop. The linear model is made up of three blocks, the phase detector PD, the loop filter LF and the VCO. In digital Costas loops the VCO is replaced by a DCO (digital controlled oscillator). This will be discussed in later sections. For these building blocks the transfer functions are now defined as follows.

{\bf Phase detector (PD).} In the locked state, the phase error $\theta_e$ is very small so by \eqref{loop-filter-input-approx} we can write
\begin{equation}\label{3}
u_d(t)\approx m_1^2(t)\theta_e=K_d\theta_e
\end{equation}
with $K_d$ called phase detector gain.
\begin{equation}\label{4}
H_{PD}(s)=\frac{U_d(s)}{\Theta_e(s)}=K_d.
\end{equation}

Note that the uppercase symbols are Laplace transforms of the corresponding lower case signals.

{\bf Loop filter (LF).}
For the loop filter we choose a PI (proportional + integral) filter whose transfer function has the from
\begin{equation}\label{loop-filter-tf}
H_{LF}(s)=\frac{U_f(s)}{U_d(s)}=\frac{1+s\tau_2}{s\tau_1}.
\end{equation}
This filter type is the preferred one because it offers superior performance compared with lead-lag or lag filters.

{\bf VCO}.
The transfer function of the VCO is given by
\begin{equation}\label{6}
H_{VCO}(s)=\frac{\Theta_2(s)}{U_f(s)}=\frac{K_0}{s}
 \end{equation}
 where $K_0$ is called VCO gain.

\begin{figure}[H]
 \centering
 \includegraphics[width=0.7\textwidth]{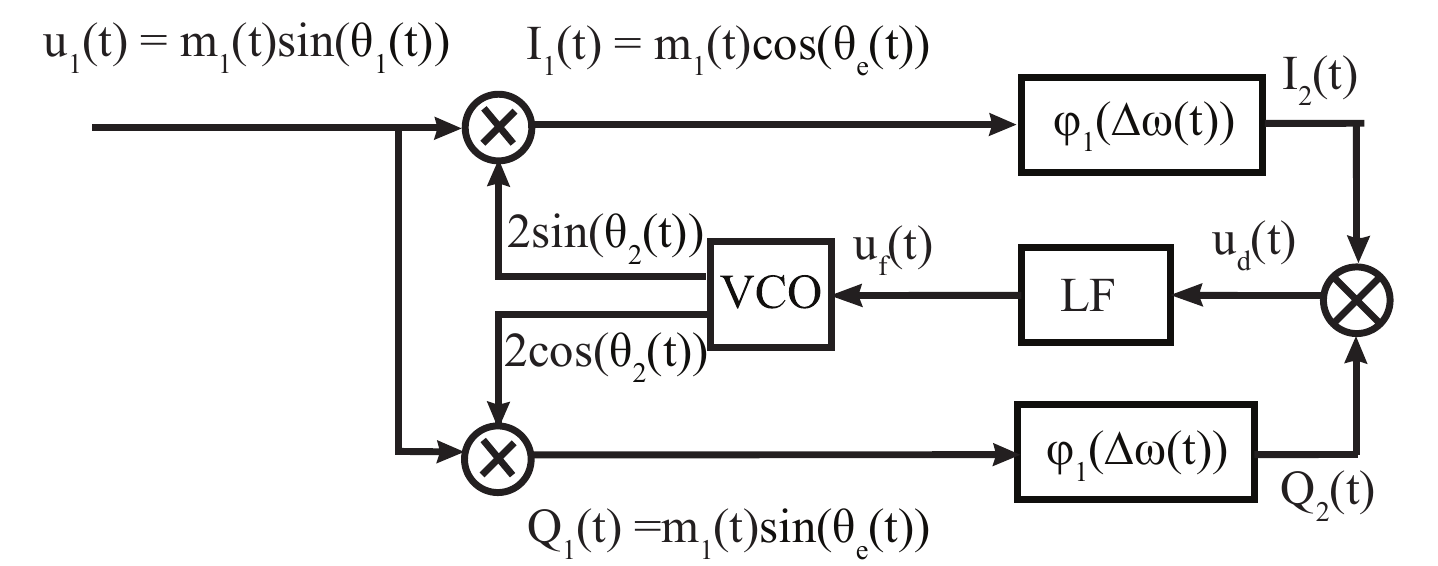}
 \caption{Model of Costas loop with delays}
 \label{delay-costas}
\end{figure}

Consider another non linear model of Costas loop in Figure~\ref{delay-costas} (delay model).
Here we use Assumtions 1--3 (initial states of filters are omitted, double-frequency terms are completely filtered by LPFs, and $m_1(t)$ doesn't affect synchronization) and
filters LPFs are replaced by the corresponding phase-delay blocks
$\varphi_1(\dot\theta_e(t)) = \varphi_1(\Delta\omega(t))$.
Outputs of low-pass filters are
\begin{equation}
\begin{aligned}
& I_2(t) = \cos(\theta_e(t) + \varphi_1(\dot\theta_e(t))),
\\
& Q_2(t) = \sin(\theta_e(t) + \varphi_1(\dot\theta_e(t))),
\end{aligned}
\end{equation}
where
\begin{equation}
\begin{aligned}
& \varphi_1(\omega) = \arg(H_{LPF}(j\omega)).
\end{aligned}
\end{equation}
Then after multiplication of $I_2(t)$ and $Q_2(t)$ we have
\begin{equation}
\label{ud}
\begin{aligned}
& u_d(t) = I_2(t)Q_2(t) = \frac{1}{2}\sin(2\theta_e(t) + 2\varphi_1(\dot\theta_e(t)))
\end{aligned}
\end{equation}
and the output $u_f(t)$ of the loop filter \eqref{loop-filter-tf}
satisfies the following equations
\begin{equation}
\begin{aligned}
& \dot x = \frac{1}{2}\sin(2\theta_e(t) + 2\varphi_1(\dot\theta_e(t))),\\
& u_f(t) = \frac{1}{\tau_1}x + \frac{\tau_2}{2\tau_1}\sin(2\theta_e(t) + 2\varphi_1(\dot\theta_e(t))).
\end{aligned}
\end{equation}
Equations of Costas loop in this case are
\begin{equation}
\label{phase-shift-costas-eq}
\begin{aligned}
& \dot x = \frac{1}{2}\sin(2\theta_e + 2\varphi_1(\dot\theta_e)),\\
& \dot\theta_e = \Delta\omega_0 -K_0\Big(
    \frac{1}{\tau_1}x
    + \frac{\tau_2}{2\tau_1}\sin(2\theta_e + 2\varphi_1(\dot\theta_e))\Big).
\end{aligned}
\end{equation}
For LPF transfer functions
\begin{equation}\label{14}
H_{LPF}(s) = \frac{1}{1 + s/\omega_3}
\end{equation}
phase shift is equal to $\varphi_1(\dot\theta_e) = - \arctan(\dot\theta_e/\omega_3)$.
Therefore \eqref{phase-shift-costas-eq} is equal to the following system
\begin{equation}
\label{best-model}
\begin{aligned}
& \dot x = \frac{1}{2}\sin\big(2\theta_e - 2\arctan(\dot\theta_e/\omega_3)\big),\\
& \dot\theta_e = \Delta\omega_0
    - \frac{K_0}{\tau_1}x
    - \frac{K_0\tau_2}{2\tau_1}\sin\big(2\theta_e - 2\arctan(\dot\theta_e/\omega_3)\big),
\end{aligned}
\end{equation}
where $$\arctan(\dot\theta_e/\omega_3) \in (-\frac{\pi}{2}, \frac{\pi}{2})$$.

\begin{figure}[H]
\centering
\includegraphics[scale=0.9]{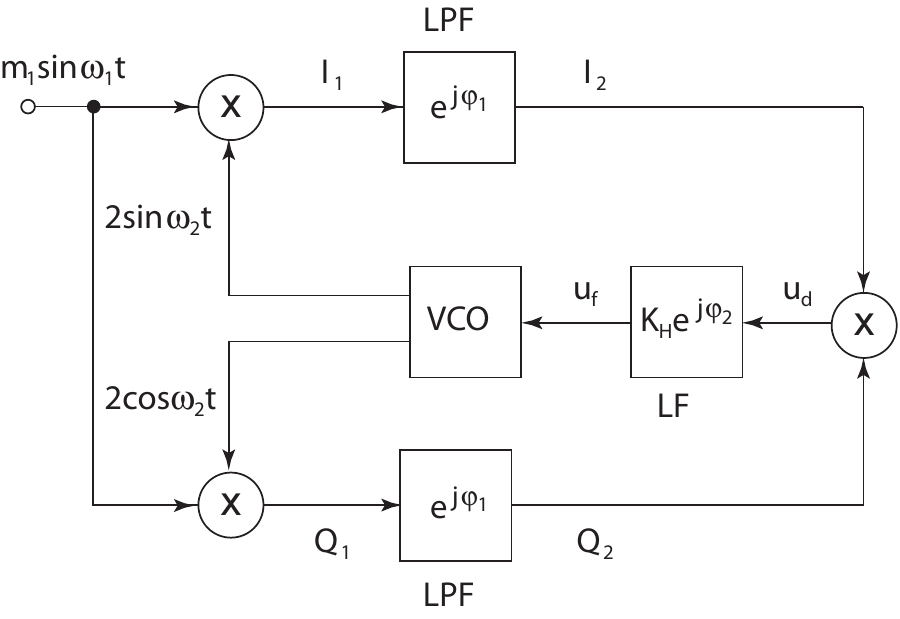}
\caption{Model of the Costas loop with delays in complex exponent form
}\label{f2-4a}
\end{figure}

Equation \eqref{best-model} is hard to analyze both numerically and analytically,
however this model is still useful.
In the following discussion it is used to approximate pull-in range and pull-in time.
For this purpose we need to simplify delay model shown in Figure~\ref{delay-costas}.
Consider block diagram in Figure~\ref{f2-4a}.
The lowpass filters (LPF) used in both $I$ an $Q$ branches are assumed to be first order filters having transfer function \eqref{loop-filter-tf}.
  As will be demonstrated later the corner frequency of these filters must be chosen such that the data signal I is recovered with sufficient accuracy, i.e. the corner frequency  $\omega_3$ must be larger than the symbol rate. Typically it is chosen twice the symbol rate, i.e. $f_3 = 2 f_S$  with $f_S =$ symbol rate and $f_3 = \omega_3/2\pi$. The output signal $I_1$ of the multiplier in the $I$ branch consists of two terms, one having the sum frequency  $\omega_1 + \omega_2$ and one having the difference frequency  $\omega_1  -  \omega_2$. Because the sum frequency term will be suppressed by the lowpass filter, only the difference term is considered. The same holds true for signal $Q_1$ in the $Q$ branch. It will show up that the range of difference frequencies is markedly below the corner frequency  $\omega_3$ of the lowpass filter. Hence the filter gain will be nearly 1 for the frequencies of interest. As will also be shown later the phase at frequency $\Delta\omega   =  \omega_1  -  \omega_2$ cannot be neglected. The lowpass filter is therefore represented as a delay block whose transfer function has the value $\exp(j \varphi_1)$, where $\varphi_1$ is the phase at frequency  $\Delta\omega$. The delayed signals $I_2$ and $Q_2$ are now multiplied by the product block at the right in the block diagram. Consequently the output signal $u_d(t)$ of this block will have a frequency of $2\Delta\omega$.
This signal is now applied to the input of the loop filter LF. Its transfer function has been defined in (\ref{loop-filter-tf}). The corner frequency of this filter is   $\omega_C= 1/\tau_2$.
Because the phase of the loop filter cannot be neglected, it is represented as a delay block characterized by
\begin{equation}\label{16}
H_{LF}(2\Delta\omega)=K_H\exp(j\varphi_2),
 \end{equation}
where  $\varphi_2$ is the phase of the loop filter at frequency $2\Delta\omega$.

\begin{figure}[H]
\centering
\includegraphics[scale=0.8]{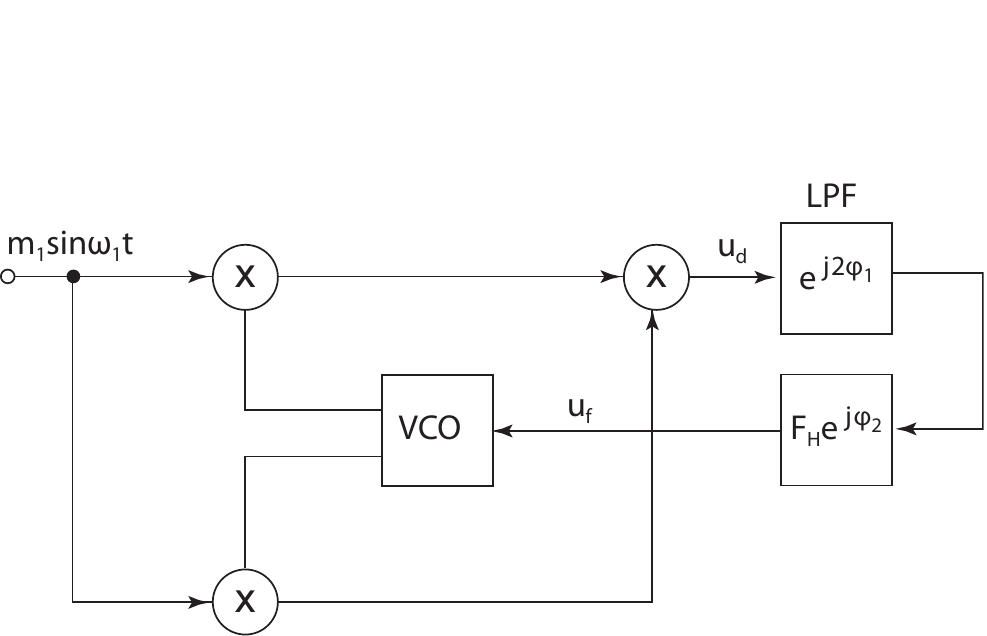}
\caption{Modified model of Costas loop, reversed order of blocks
}\label{f2-4b}
\end{figure}

The analysis of dynamic behavior becomes easier when the order of some blocks in Figure~\ref{f2-4a} is reversed (see Figure~\ref{f2-4b}), i.e. when we put the multiplying block before the lowpass filter.
Because the frequency of signal $u_d(t)$ in Figure~\ref{f2-4a} is twice the frequency of the signals $I_2$ and $Q_2$, the phase shift created by the lowpass filter at frequency $2\Delta\omega$    is now twice the phase shift at frequency   $\Delta\omega$.
The LPF is therefore represented here by a delay block having transfer function $\exp(2 j \varphi_1)$.

\begin{figure}[H]
\centering
\includegraphics[width=0.8\textwidth]{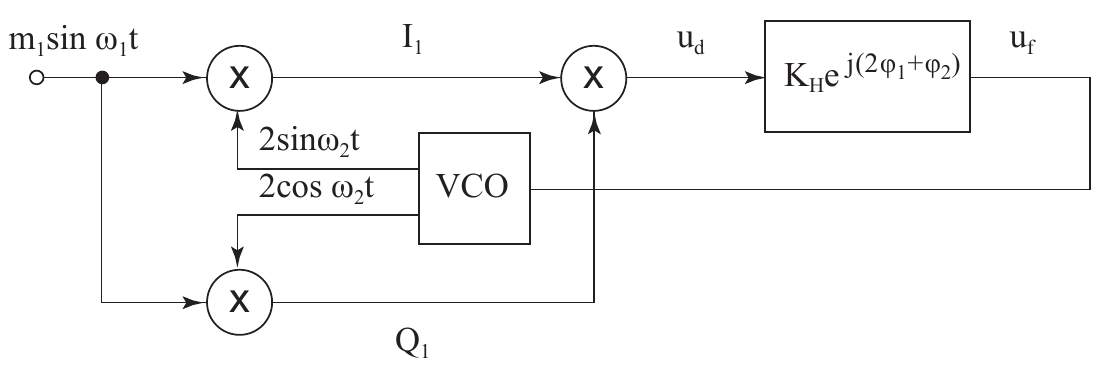}
\caption{ Modified model of Costas loop, concatenated blocks
}\label{f2-4c}
\end{figure}

We can simplify the block diagram even more by concatenating the lowpass filter and loop filter blocks. The resulting block delays the phase by $\varphi_{tot} = 2\varphi_1 + \varphi_2$. This is shown in Figure~\ref{f2-4c}. The output signal $u_f(t)$ of this delay block now modulates the frequency generated by the VCO.

\begin{figure}[H]
\centering
\includegraphics[width=0.4\textwidth]{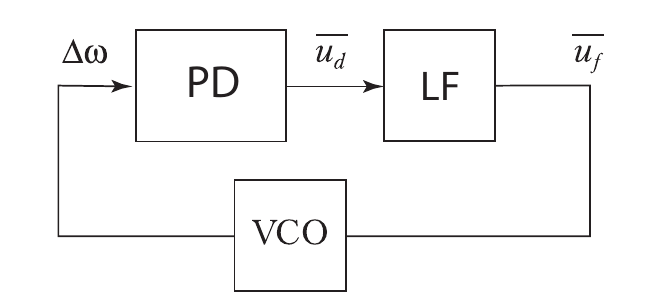}
\caption{Non linear model of Costas loop for computation of pull-in time
}\label{f2-7}
\end{figure}

To compute pull-in time we need to consider Costas loop model in Figure~\ref{f2-1b} with
averaged signals of phase detector output $u_d$ and  filter output $u_f$ (see Figure~\ref{f2-7}).
The model is built from three blocks. The first of these is labeled "phase-frequency detector". We have seen that in the locked state the output of the phase detector depends on the phase error $\theta_e$. In the unlocked state, however, the average phase detector output signal $\overline{u_d}$  is a function of frequency difference as will be shown in next section (Eq.~(\ref{20})), hence it is justified to call that block "phase-frequency detector". As we will recognize the pull-in process is a slow one, i.e. its frequency spectrum contains low frequencies only that are below the corner frequency  $\omega_C$ of the loop filter, cf. Eq.~(\ref{loop-filter-tf}). The loop filter can therefore be modeled as a simple integrator with transfer function
\begin{equation}\label{21a}
H_{LF}(s)\approx\frac{1}{s\tau_1}.
 \end{equation}
Therefore
\begin{equation}\label{21b}
  \overline{u_f(t)}=\frac{1}{\tau_1}\int\limits_0^t\overline{u_d(\tau)}d\tau,
\end{equation}
The frequency  $\omega_2$ of the VCO output signal is defined as
\begin{equation}\label{22}
  \omega_2=\omega_{free}+K_0\overline{u_f},
\end{equation}
where  $\omega_{free}$ is the free running frequency and $K_0$ is the VCO gain.
Now we define the instantaneous frequency difference $\Delta\omega$   as
\begin{equation}\label{24}
  \Delta\omega=\omega_1 - \omega_2.
\end{equation}
Substituting \eqref{initial-frequency-diff} and (\ref{24}) into (\ref{22}) finally yields
\begin{equation}\label{25}
\Delta\omega=\Delta\omega_0-K_0\overline{u_f}.
 \end{equation}

\subsubsection{QPSK Costas loop}\label{qpsk models}

\begin{figure}[H]
  \centering
  \includegraphics[scale=0.4]{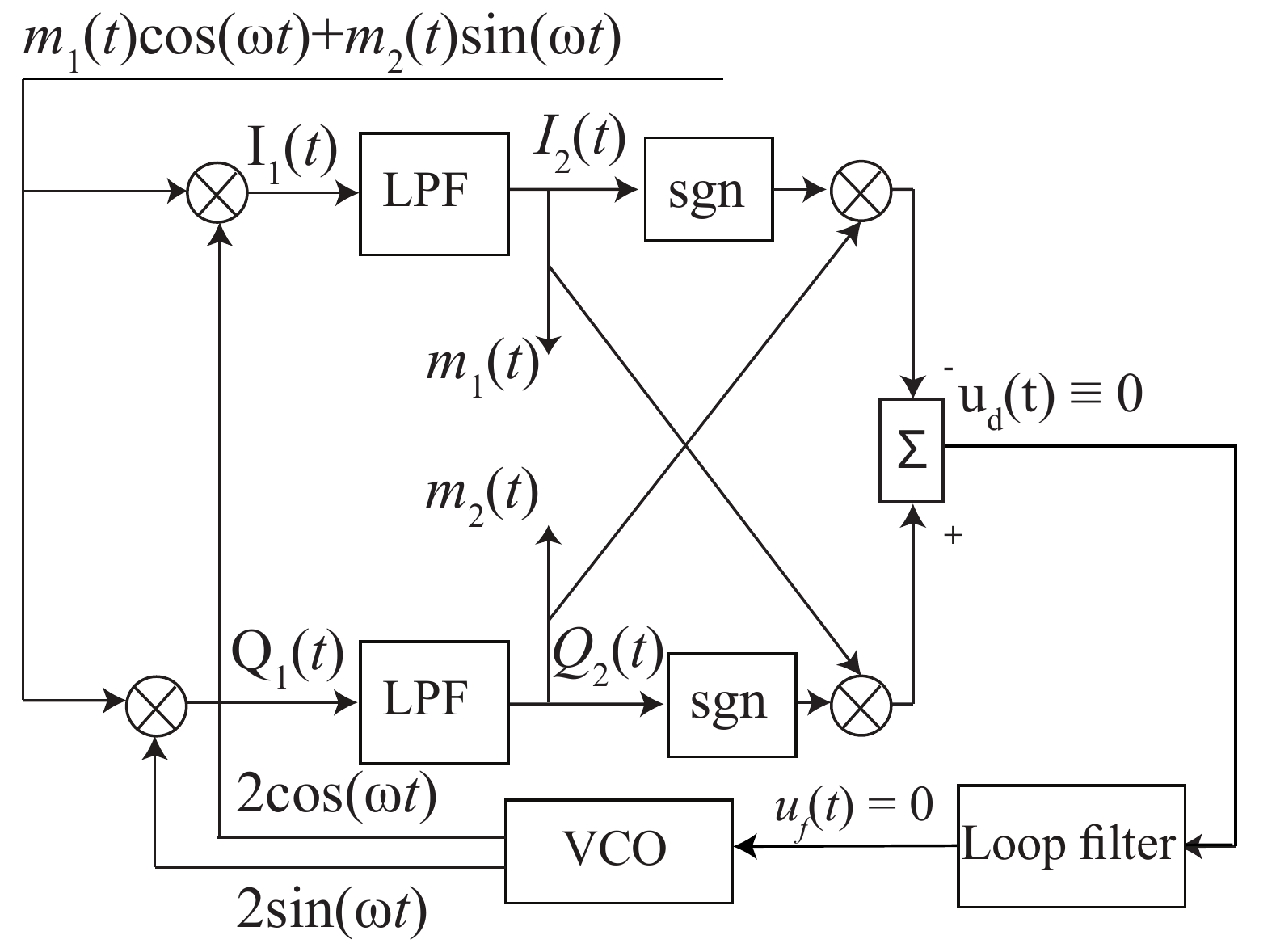}
  \caption{QPSK Costas loop after transient process.}
\label{costas_after_sync}
\end{figure}

Consider QPSK Costas loop operation (see Figure~\ref{costas_after_sync})
for the sinusoidal carrier and VCO
in lock state for the same initial frequencies $\omega_1 = \omega_2 = \omega$.
By \eqref{qpsk signal}, the input QPSK signal has the form
\begin{equation}
\notag
\begin{aligned}
m_1(t)\cos(\omega t) + m_2(t)\sin(\omega t),
\end{aligned}
\end{equation}
where $m_{1,2}(t) = \pm 1$ is the transmitted data,
$\sin(\omega t)$ and $\cos(\omega t)$ are sinusoidal carriers, $\theta_1(t) = \omega t$ --- phase of input signal. The outputs of the VCO are $2\cos(\omega t)$ and $2\sin(\omega t)$.

After multiplication of VCO signals
and the input signal by multiplier blocks ($\otimes$) on the upper $I$ branch one has
\begin{equation}
  \notag
  \begin{aligned}
      &
      I_1(t) =
      2\cos(\omega t)\Big(
        m_1(t)\cos(\omega t)
        + m_2(t)\sin(\omega t)\Big).
  \end{aligned}
\end{equation}
On the lower branch  the output signal of VCO is multiplied by the input signal:
\begin{equation}
  \notag
  \begin{aligned}
    &
      Q_1(t) =
      2\sin(\omega t)\Big(
        m_1(t)\cos(\omega t)
      + m_2(t)\sin(\omega t)\Big).
  \end{aligned}
\end{equation}

Here from an engineering point of view,
the high-frequency terms
$\cos(2\omega t)$ and $\sin(2\omega t)$ are removed by
ideal low-pass filters LPFs (see Assumption \ref{as-twice-frequency} in previews section).
In this case, the signals $I_2(t)$ and $Q_2(t)$ on the upper and lower branches
can be approximated as
\begin{equation}
  \begin{aligned}
    &
      I_2(t) \approx
       m_1(t)\cos(0) + m_2(t)\sin(0)
      = m_1(t),
    \\
    &
      Q_2(t) \approx
       - m_1(t)\sin(0) + m_2(t)\cos(0)
      = m_2(t).
  \end{aligned}
\end{equation}

\begin{figure}[H]
  \centering
  \includegraphics[scale=0.4]{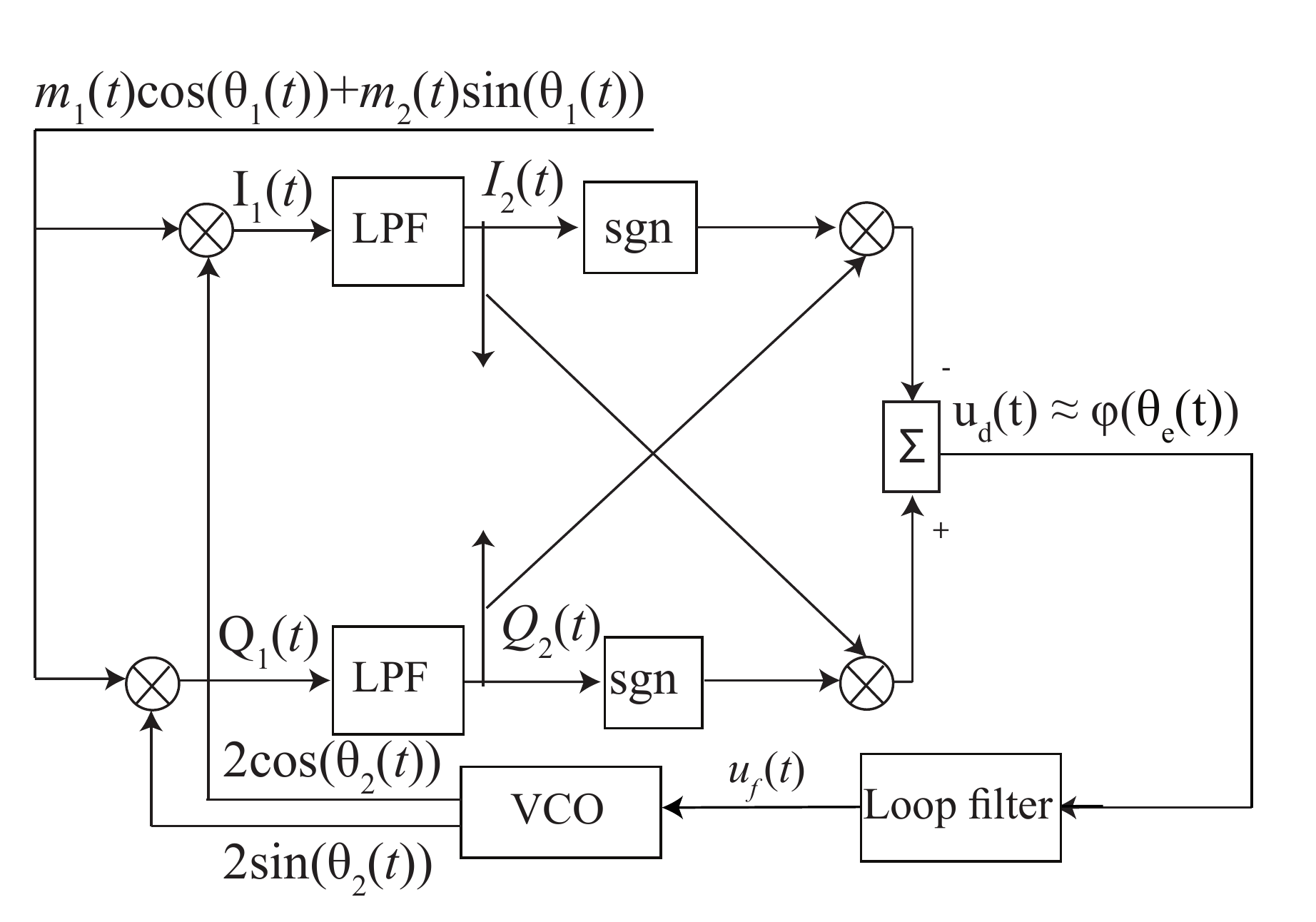}
\caption{QPSK Costas loop is out of lock, there is non zero phase difference.}
\label{qpsk_before_sync}
\end{figure}

Apart from considered case there are two possible cases:
1) the frequencies are different
or 2) the frequencies are the same but there is a constant phase difference.
Consider Costas loop before synchronization
(see Figure~\ref{qpsk_before_sync})
in the case when the phase of the input carrier $\theta_1(t)$ and
the phase of VCO $\theta_2(t)$ are different:
\begin{equation}\label{thetadelta}
  \theta_e(t) = \theta_1(t)-\theta_2(t) \not\equiv const.
\end{equation}

In this case, using Assumption 1,
the signals $I_2(t)$ and $Q_2(t)$ on the upper and lower branches
can be approximated as
\begin{equation}\label{g1g2-approx}
  \begin{aligned}
    &
      I_2(t) \approx
        m_1(t)\cos(\theta_e(t)) + m_2(t)\sin(\theta_e(t))
      ,
    \\
    &
      Q_2(t) \approx
        - m_1(t)\sin(\theta_e(t)) + m_2(t)\cos(\theta_e(t))
      .
  \end{aligned}
\end{equation}

\begin{figure}[H]
  \centering
  \includegraphics[scale=0.4]{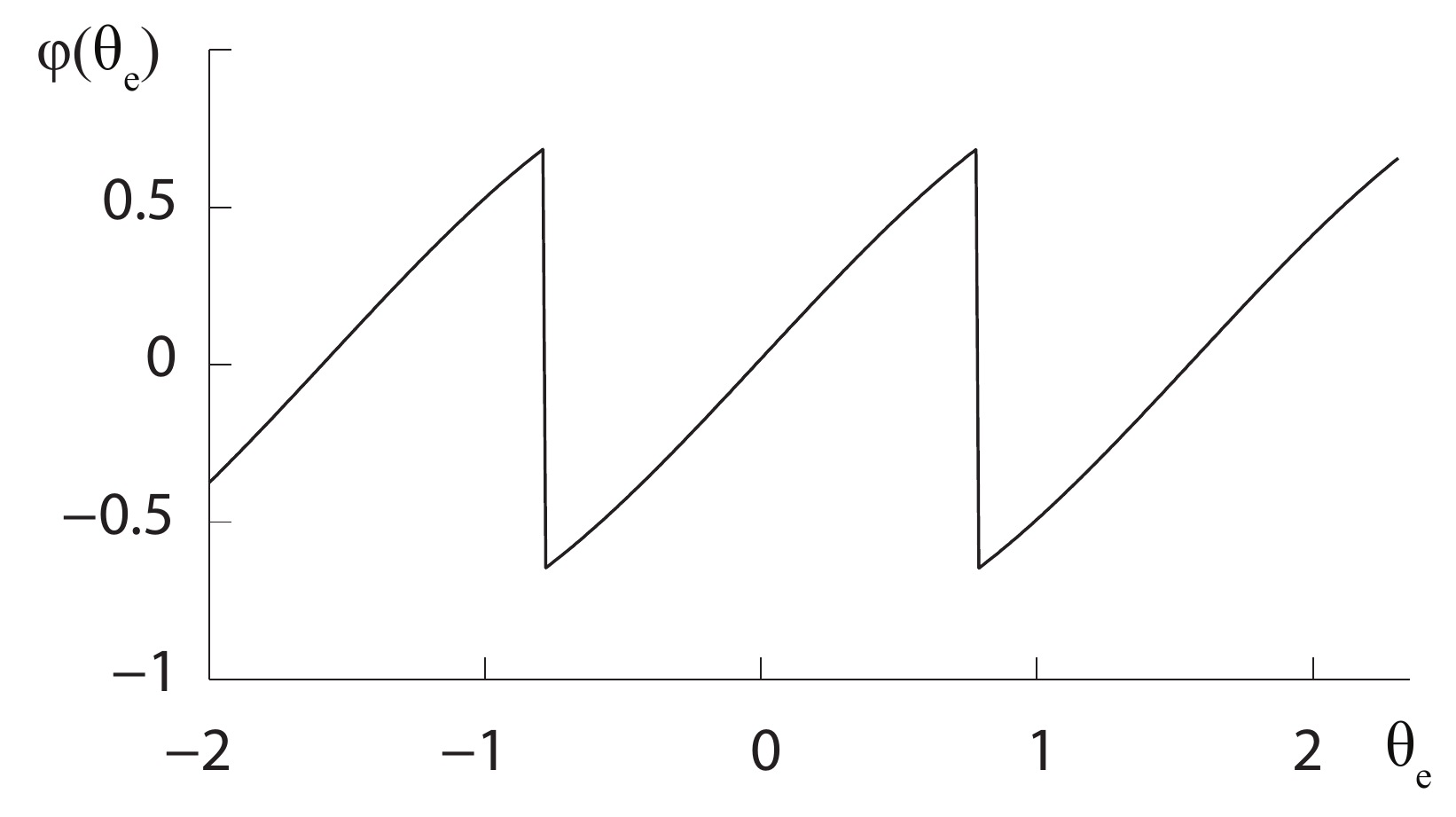}
\caption{$\varphi(\theta_e)$}
\label{qpsk-char}
\end{figure}

After the filtration, both signals, $I_1(t)$ and $Q_1(t)$,
pass through the limiters (sgn blocks).
Then the outputs of the limiters $\sign\big(I_2(t)\big)$ and $\sign\big(Q_2(t)\big)$
are multiplied with $Q_2(t)$ and $I_2(t)$, respectively.
By Assumption~2 and corresponding formula \eqref{g1g2-approx},
the difference of these signals
\begin{equation}
  \begin{aligned}
   &  u_d(t) = - Q_2(t)\sign\big(I_2(t)\big) + I_2(t)\sign\big(Q_2(t)\big)
  \end{aligned}
\end{equation}
can be approximated as
\begin{equation}\label{phi approx}
  \begin{aligned}
   &
    u_d(t)
    \approx
   \varphi(\theta_e(t)) =
   \left\{
      \begin{array}{ll}
        2m\sin(\theta_e(t)), &  -{\pi \over 4}< \theta_e(t) < {\pi \over 4}, \\
        -2m\cos(\theta_e(t)), &  {\pi \over 4}< \theta_e(t) < {3\pi \over 4}, \\
        -2m\sin(\theta_e(t)), &  {3\pi \over 4}< \theta_e(t) < {5\pi \over 4}, \\
        2m\cos(\theta_e(t)), &  {5\pi \over 4}< \theta_e(t) < -{\pi \over 4}, \\
      \end{array}
   \right.
  \end{aligned}
\end{equation}
with $m = |m_1|=|m_2|$.
Here $\varphi(\theta_e(t))$ is a piecewise-smooth function\footnote{It should be noted, that function $\varphi(\theta_e(t))$ depends on $m_{1,2}$ at the points $\theta_e = \pm {\pi \over 4}, \pm {3\pi \over 4}$.} shown in Figure~\ref{qpsk-char}.

The resulting signal $\varphi(t)$,
after the filtration by the loop filter, forms the
control signal $u_f(t)$  for the VCO.

To derive \emph{mathematical model in the signal space}
describing \emph{physical model} of QPSK Costas loop
one takes into account \eqref{loop-filter} and \eqref{vco first}:
\begin{equation} \label{full non linear}
    \begin{aligned}
        & \dot{x_1} = A_1 x_1 + 2b_1\cos(\omega_1 t - \theta_e)
          \big(m_1(t)\cos(\omega_1 t) + m_2(t)\sin(\omega_1 t)\big), \\
        & \dot{x_2} = A_2 x_2 + 2b_2\sin(\omega_1 t - \theta_e)
          \big(m_1(t)\cos(\omega_1 t) + m_2(t)\sin(\omega_1 t)\big), \\
        & \dot{x} = A x + b(\sign(c_2^*x_2)(c_1^*x_1)-\sign(c_1^*x_1)(c_2^*x_2)), \\
        & \dot\theta_e = \Delta\omega_0 - K_0(c^*x)
           - K_0h\big(\sign(c_2^*x_2)(c_1^*x_1)-\sign(c_1^*x_1)(c_2^*x_2)\big).
    \end{aligned}
\end{equation}
However equations \eqref{full non linear} are nonlinear and non autonomous with discontinuous right-hand side, which are extremely hard to investigate.
Therefore, the study of \eqref{full non linear} is outside of the scope of this work.

To derive linear model, we consider \eqref{phi approx} and the corresponding Figure~\ref{qpsk-char}.
The curve looks like a ``chopped'' sine wave. The Costas loop can get locked at four different values of  $\theta_e$\,, i.e. with  $\theta_e = 0,  \pi/2$,  $\pi$, or
$3 \pi/2$. To simplify the following analysis, we can define the phase error to be zero wherever the loop gets locked. Moreover, in the locked state the phase error is small, so we can write
\begin{equation}\label{43}
  u_d\approx 2m\theta_e=K_d\theta_e,
\end{equation}
i.e. the output signal of the adder block at the right of Figure~\ref{qpsk_before_sync} is considered to be the phase detector output signal $u_d$. The phase detector gain is then
\begin{equation}\label{44}
  K_d=2m.
\end{equation}
It is easily seen that the linear model for the locked state is identical with that of the Costas loop for BPSK, cf. Figure~\ref{linear-pll}. Because only small frequency differences are considered here, the lowpass filters can be discarded. The transfer functions of the loop filter and of the VCO are assumed to be the same as in the case of the Costas loop for BPSK, hence these are given by Eqs.~(\ref{loop-filter-tf}) and (\ref{6}).

\begin{figure}[H]
 \centering
 \includegraphics[width=0.5\textwidth]{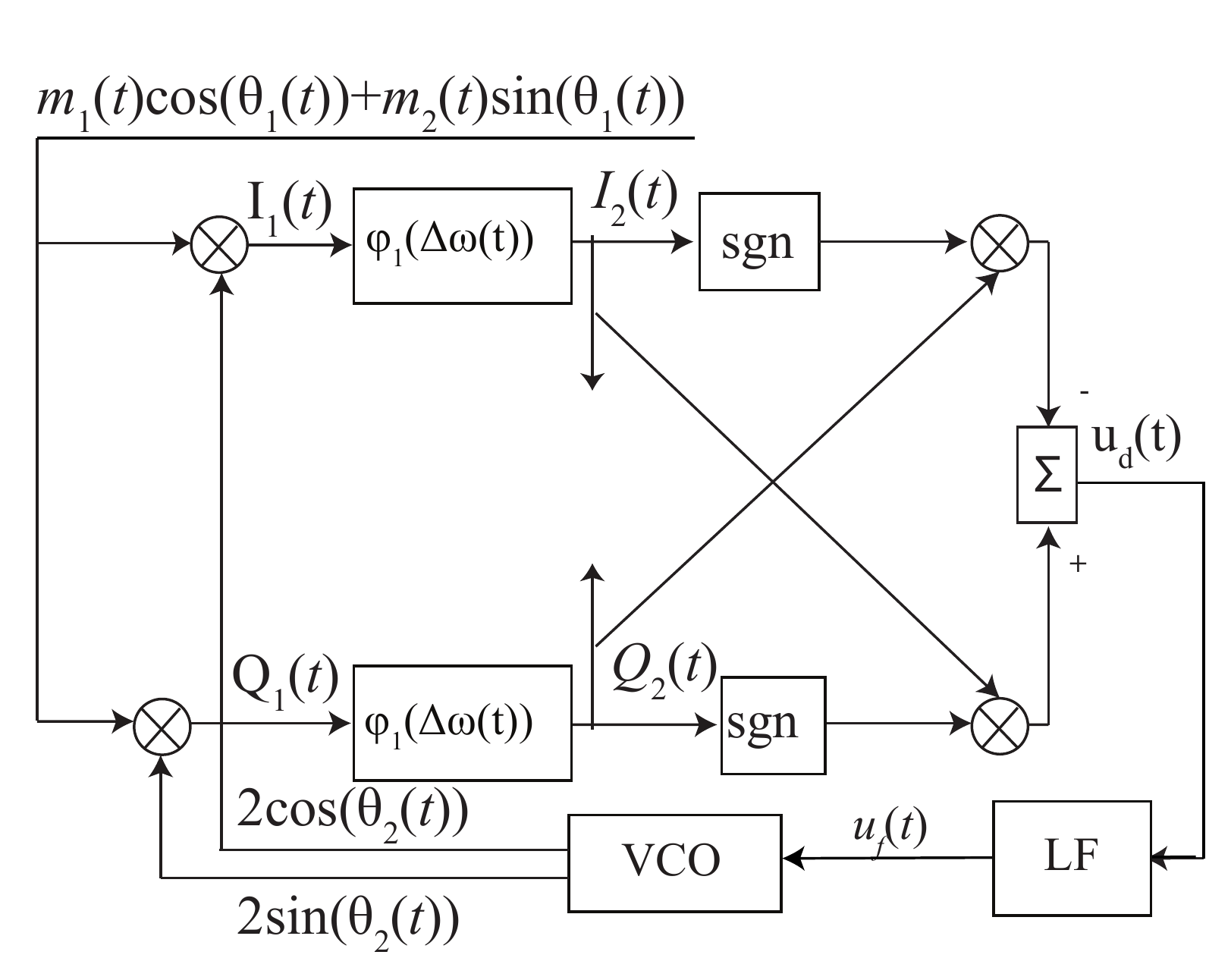}
 \caption{Model of QPSK Costas loop with delays}
 \label{delay-qpsk-costas}
\end{figure}

Similar to BPSK Costas loop, it is reasonable to consider delay model of QPSK Costas loop (see Figure~\ref{delay-qpsk-costas}).
Filters LPFs are replaced by the corresponding phase-delay blocks
$\varphi_1(\Delta\omega) = \arg(H_{LPF}(j\omega))$.
The outputs of low-pass filters are
\begin{equation}\label{g1g2-phase-approx}
  \begin{aligned}
    &
      I_2(t) \approx
       \cos(\theta_e(t)+\varphi_1(\Delta\omega(t))) + \sin(\theta_e(t)+\varphi_1(\Delta\omega(t)))
      ,
    \\
    &
      Q_2(t) \approx
        -\sin(\theta_e(t)+\varphi_1(\Delta\omega(t)) + \cos(\theta_e(t)+\varphi_1(\Delta\omega(t)))
      .
  \end{aligned}
\end{equation}
Then $u_d(t)$ can be approximated as
\begin{equation}
\begin{aligned}
& u_d(t)
    \approx
   \varphi(\theta_e(t) + \varphi_1(\Delta\omega(t))) =
   \\
   &
   \left\{
      \begin{array}{ll}
        2\sin(\theta_e(t)+ \varphi_1(\Delta\omega(t))), &  -{\pi \over 4}< \theta_e(t)+ \varphi_1(\Delta\omega(t)) < {\pi \over 4}, \\
        -2\cos(\theta_e(t)+ \varphi_1(\Delta\omega(t))), &  {\pi \over 4}< \theta_e(t) + \varphi_1(\Delta\omega(t))< {3\pi \over 4}, \\
        -2\sin(\theta_e(t)+ \varphi_1(\Delta\omega(t))), &  {3\pi \over 4}< \theta_e(t)+ \varphi_1(\Delta\omega(t)) < {5\pi \over 4}, \\
        2\cos(\theta_e(t)+ \varphi_1(\Delta\omega(t))), &  {5\pi \over 4}< \theta_e(t)+ \varphi_1(\Delta\omega(t)) < -{\pi \over 4}. \\
      \end{array}
   \right.
\end{aligned}
\end{equation}
Consider the loop filter transfer function \eqref{loop-filter-tf}.
Equations of delay model of QPSK Costas loop in this case are
\begin{equation}
\begin{aligned}
& \dot x = \varphi(\theta_e(t) + \varphi_1(\dot\theta_e)),\\
& \dot\theta_e = \Delta\omega_0 -K_0\Big(
    \frac{1}{\tau_1}x
    + \frac{\tau_2}{\tau_1}\varphi(\theta_e(t) + \varphi_1(\dot\theta_e))\Big).
\end{aligned}
\end{equation}

\begin{figure}[H]
\centering
\includegraphics[scale=0.6]{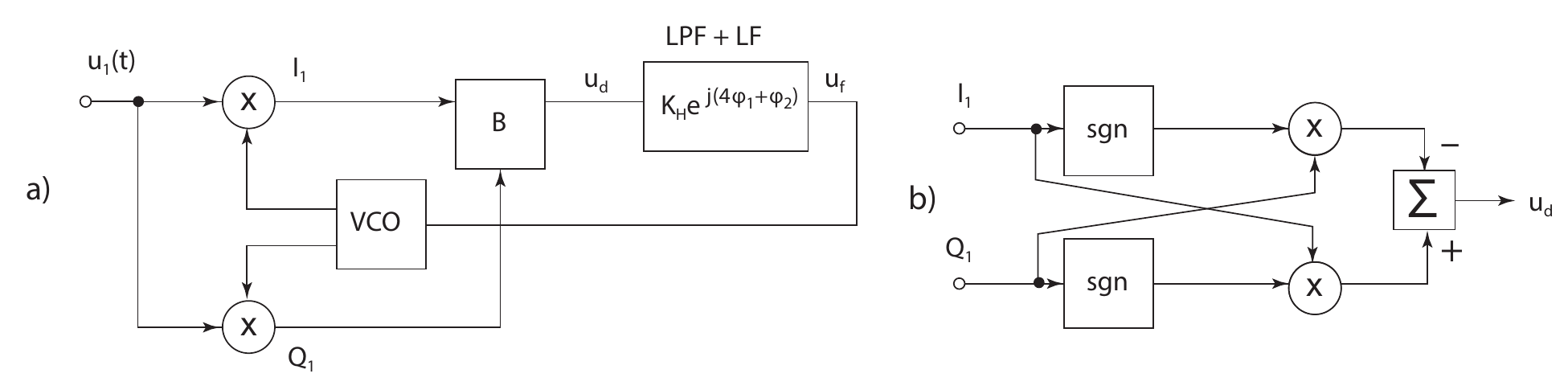}
\caption{Non linear model of the Costas loop for QPSK with delays in complex exponent form}
\label{f3-4}
\end{figure}

The non linear model of the Costas loop for QPSK is developed on the basis of the non linear model we derived for the Costas loop for BPSK, cf. Figure~\ref{f2-4c}. Here again the order of lowpass filters and the blocks shown at the right of Figure~\ref{costas_after_sync} is reversed. This results in the model shown in Figure~\ref{f3-4}a.
In the block labeled ``B'' the function blocks at the right of Figure~\ref{f3-4}a have been integrated, cf. Figure~\ref{f3-4}b. The output signal $u_d$ of block B is the ``chopped'' sine wave as shown in Figure~\ref{f3-2}. Its fundamental frequency is 4 times the frequency difference  $\omega_1 -  \omega_2$. The lowpass filters and the loop filter have been concatenated in the block labeled ``LPF + LF'' at the right of Figure~\ref{f3-4}a. Referring to Figure~\ref{costas_after_sync} signals $I_1$ and $Q_1$ are passed through lowpass filters. As in the case of the Costas loop for BPSK we assume here again that the difference frequency  $\Delta\omega$  is well below the corner frequency  $\omega_3$ of the lowpass filters, hence the gain of the lowpass filters is nearly 1 at  $\omega  = \Delta\omega$. Because the phase shift must not be neglected, we represent the lowpass filter by a delay, i.e. its frequency response at  $\omega  = \Delta\omega$   is
$$
H_{LPF}(\Delta\omega)=\exp(j\varphi_1),
$$
where  $\varphi_1$ is the phase of the lowpass filter.  Due to the arithmetic operations in block ``B'' (cf. Figure~\ref{f3-4}) the frequency of the $u_d$ is quadrupled, which implies that the phase shift at frequency $4\Delta\omega$    becomes $4 \varphi_1$. The frequency response of the loop filter at $\omega   = 4\Delta\omega$    is given by
$$
H_{LF}(4\Delta\omega)=\exp(j\varphi_2),
$$
where  $\varphi_2$ is the phase of the loop filter at frequency  $\omega  = 4\Delta\omega$.
Hence the cascade of lowpass filter and loop filter can be modeled by the transfer function $\exp(j[4\varphi_1 + \varphi_2])$ as shown in
Figure~\ref{f3-4}a.

\subsection{Mathematical models of Modified Costas loops}
\subsubsection{Modified Costas loop for BPSK}

\begin{figure}[H]
\centering
\includegraphics[scale=0.8]{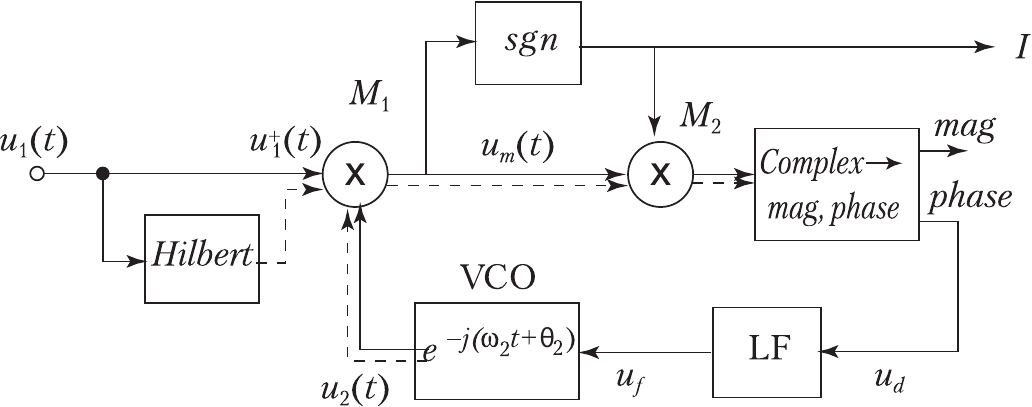}
\caption{Block diagram of modified Costas loop for BPSK
}\label{f4-1}
\end{figure}


The block diagram of the modified Costas loop for BPSK is shown in Figure~\ref{f4-1}. The input signal is given by
$$
u_1(t)=m_1(t)\cos(\omega_1t+\theta_1),
$$
where  $\theta_1$ is initial phase. The input signal is first converted into a pre-envelope signal, as explained in section \ref{ss1.1}. The output signal of the Hilbert transformer is
$$
\hat u_1(t)=H[m_1(t)\cos(\omega_1t+\theta_1)] = m_1(t)\sin(\omega_1t+\theta_1).
$$

Note that because the largest frequency of the spectrum of the data signal $m_1(t)$ is much lower than the carrier frequency  $\omega_1$, the Hilbert transform of the product $H[m_1(t)\cos(\omega_1t+\theta_1)]$  equals $m_1(t)H[\cos(\omega_1t+\theta_1)]$  [5]. The pre-envelope signal is obtained now from
\begin{equation}\label{56}
u_1^+(t)=u_1(t)+j\hat u(t)=m_1(t)\exp(j[\omega_1t+\theta_1]).
\end{equation}

The exponential in Eqn. (\ref{56}) is referred to as a ``complex carrier''. In Figure~\ref{f4-1} complex signals are shown as double lines. The solid line represents the real part, the dotted line represents the imaginary part. To demodulate the BPSK signal, the pre-envelope signal is now multiplied with the output signal of the VCO, which is here a complex carrier as well. The complex output signal of the VCO is defined as
\begin{equation}\label{complex vco}
u_2(t)=\exp(-j[\omega_2t+\theta_2]).
\end{equation}

In the locked state of the Costas loop both frequencies  $\omega_1$ and  $\omega_2$ are equal, and we also have  $\theta_1\approx\theta_2$. Hence the output signal of the multiplier $M_1$ is
\begin{equation}\label{58}
u_m(t)=m_1(t)\exp(j[(\omega_1-\omega_2)t+\theta_1-\theta_2])\approx m_1(t),
\end{equation}
i.e. the output of the multiplier is  the demodulated data signal $m_1(t)$. To derive the linear model of this Costas loop, it is assumed that  $\omega_1 =  \omega_2$ and  $\theta_1\ne \theta_2$. The output signal of multiplier $M_1$ then becomes
\begin{equation}\label{59}
u_m(t)=m_1(t)\exp(j[\theta_1-\theta_2]).
\end{equation}

\begin{figure}[H]
\centering
\includegraphics[scale=0.7]{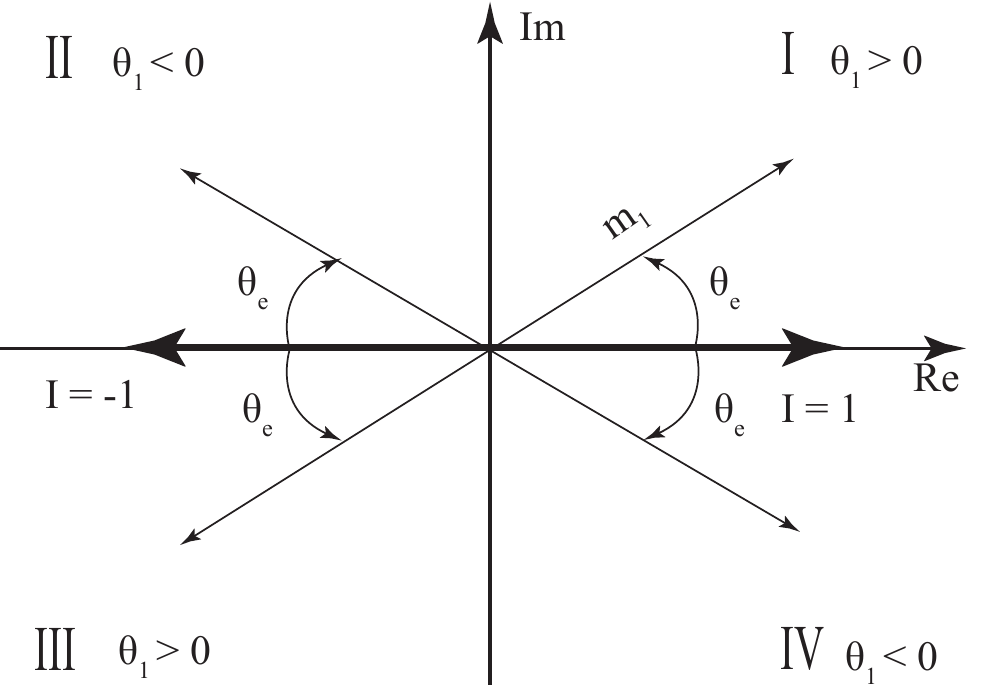}
\caption{Representation of phasor $u_m(t)$ in the complex plane
}\label{f4-2}
\end{figure}


This is a phasor having magnitude $|m_1(t)|$ and phase  $\theta_1 -  \theta_2$, as shown in Figure~\ref{f4-2}. Two quantities are determined from the phase of phasor $u_m(t)$, i.e. the demodulated data signal I and the phase error  $\theta_e$. The data signal is defined as
\begin{equation}\label{60}
I={\rm{sgn}}({\rm{Re}}[u_m(t)]),
\end{equation}
i.e. when the phasor lies in quadrants I or IV, the data signal is considered to be +1, and when the phasor is in quadrants II or III, the data signal is considered to be -1. This means that I can be either a phasor with phase 0 or a phasor with phase  $\pi$.

 These two phasors are plotted as thick lines in Figure~\ref{f4-2}.

The phase error  $\theta_e$ is now given by the difference of the phases of phasor $u_m(t)$ and phasor $I$, as shown in figure \ref{f4-2}, i.e.  $\theta_e$ is determined from
\begin{equation}\label{61}
\theta_e=phase(u_m(t) I)
\end{equation}

The product $u_m(t) I$  is computed by multiplier $M_2$ in Figure~\ref{f4-2}. The block labeled "Complex $\to$  mag, phase" is used to convert the complex signal delivered by $M_2$ into magnitude and phase. The magnitude is not used in this case, but only the phase. It follows from Eqn. (\ref{61}) that the phase output of this block is the phase error  $\theta_e$, hence the blocks $M_1,  M_2$, sgn, and Complex $\to$  mag, phase represent a phase detector with gain $K_d = 1$. The phase output of block Complex $\to$  mag, phase is therefore labeled $u_d$.
Figure~\ref{linear-pll} shows the complete linear model of the modified Costas loop for BPSK. The transfer functions of the loop filter and VCO have been defined in Eqs.~(\ref{loop-filter-tf}) and (\ref{6}). Note that with this type of Costas loop there is no additional lowpass filter, because the multiplication of the two complex carriers (cf. Eqn. (\ref{58})) does not create the unwanted double frequency component as found with the conventional Costas loops.

\subsubsection{Modified Costas loop for QPSK}\label{1.1.6}
\begin{figure}[H]
\centering
\includegraphics[scale=0.7]{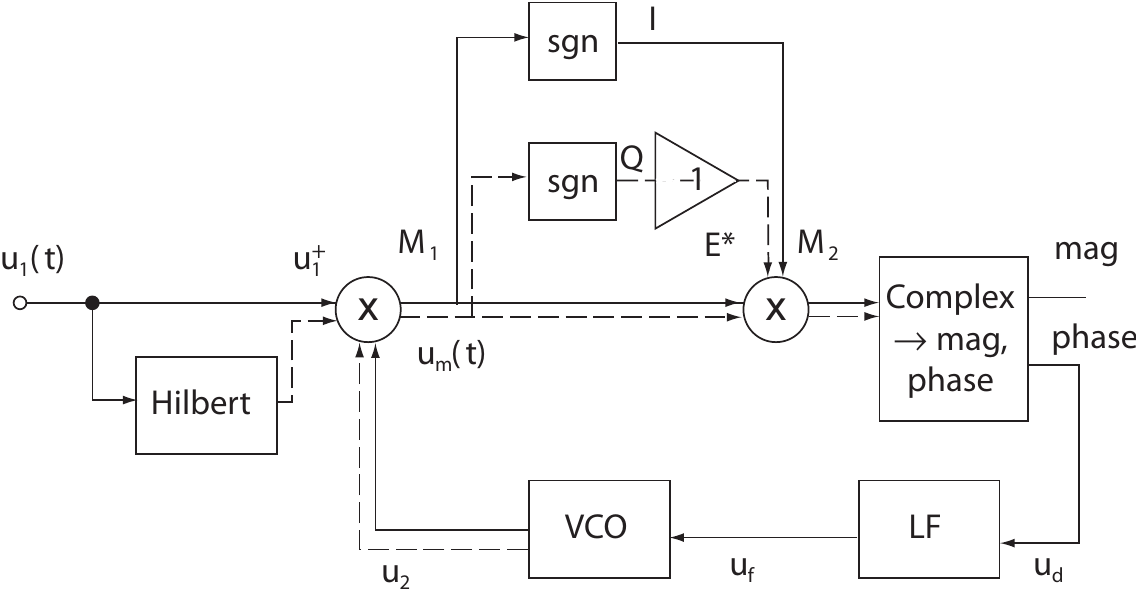}
\caption{Block diagram of modified Costas loop for QPSK
}\label{f5-1}
\end{figure}

Figure~\ref{f5-1} shows the block diagram of the modified Costas loop for QPSK.

The reference signal $u_1(t)$ is defined by
\begin{equation}\label{69}
  u_1(t)=m_1(t)\cos(\omega_1t+\theta_1)-m_2(t)\sin(\omega_1t+\theta_1),
\end{equation}
The Hilbert transformed signal is then given by
\begin{equation}\label{70}
\hat u_1(t)=m_1(t)\sin(\omega_1+\theta_1)+m_2(t)\cos(\omega_1+\theta_1)
\end{equation}
and the pre-envelope signal then becomes
\begin{equation}\label{71}\begin{aligned}
&u_1^+(t)=m_1(t)\cos(\omega_1t+\theta_1)-m_2(t)\sin(\omega_1t+\theta_1)+\\
&+jm_1(t)\sin(\omega_1t+\theta_1)+
jm_2(t)\cos(\omega_1t+\theta_1).\end{aligned}
\end{equation}

This can be rewritten as
\begin{equation}\label{72}
\begin{aligned}
&u_1^+(t)=(m_1(t)+jm_2(t))(\cos[\omega_1t+\theta_1]+j\sin[\omega_1t+\theta_1]=\\
&=(m_1(t)+jm_2(t))\exp(j[\omega_1t+\theta_1]).
\end{aligned}
\end{equation}
Herein the term $(m_1(t) + j m_2(t))$ is complex envelope, and the term $\exp(j\omega_1 t + \theta_1)$
is complex carrier. The VCO generates another complex carrier given by \eqref{complex vco}.
The multiplier $M_1$  creates signal $u_m(t)$ that is given by
\begin{equation}\label{74}
u_m(t)=(m_1(t)+jm_2(t))\exp(j[(\omega_1-\omega_2)t+(\theta_1-\theta_2)]).
\end{equation}

\begin{figure}[H]
\centering
\includegraphics[width=0.5\textwidth]{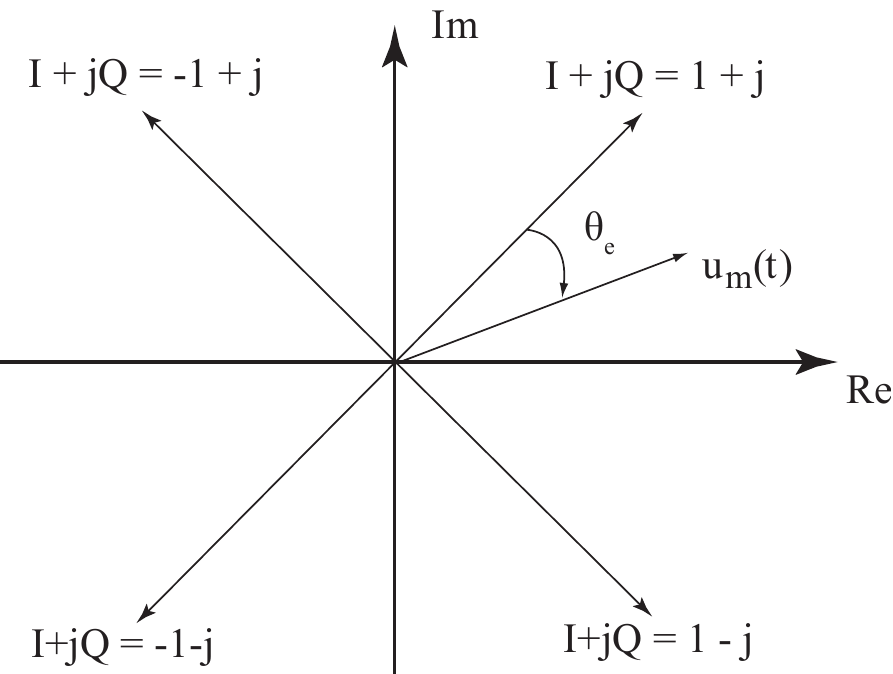}
\caption{Representation of phasor $u_m(t)$ in the complex plane
}\label{f5-2}
\end{figure}

When the loop has acquired lock,  $\omega_1 = \omega_2$, and  $\theta_1 \approx \theta_2$, so we have
\begin{equation}\label{75}
u_m(t)\approx (m_1(t)+jm_2(t))
\end{equation}
hence the output of $M_1$ is the complex envelope.
In the locked state, the complex envelope can take four positions, as shown in Figure~\ref{f5-2}.
When there is a phase error, $u_m(t)$  deviates from the ideal position, as demonstrated in the figure.
The phase error  $\theta_e$ then is the angle between $u_m(t)$ and the closest of the four possible positions.
When $u_m(t)$ is in quadrant I, e.g., phasor $1 + j$ is considered the estimate of the complex envelope. When $u_m(t)$ is in quadrant II, the estimate of the complex envelope is $-1 + j$ etc.
The estimates $I$ and $Q$ are taken from the output of sgn blocks, cf. Figure~\ref{f5-1}. The phase error is obtained from
\begin{equation}\label{76}
\theta_e=phase[u_m(t)(I-jQ)]
\end{equation}
where $I - jQ$ is the conjugate of the complex envelope. Multiplier $M_2$ delivers the product $u_m(t)   (I -jQ)$, and the block ``Complex $\to$  mag, phase'' is used to compute the phase of that complex quantity. Note that the magnitude is not required. The blocks $M_1$, sgn, Inverter, $M_2$, and Complex $\to$  mag, phase form a phase detector having gain $K_d = 1$. The phase output of block Complex $\to$  mag, phase is therefore labeled $u_d$.

Figure~\ref{linear-pll} shows the completed linear model of the modified Costas loop for QPSK, which is the same as for BPSK.
The transfer functions of the loop filter and VCO have been defined in Eqs.~(\ref{loop-filter-tf}) and (\ref{6}).

\subsection{Definitions of hold-in range, lock-in range, pull-in range.}

In the classic books on phase-locked loops \cite{Gardner-1966,Viterbi-1966,ShahgildyanL-1966} such concepts as hold-in pull-in lock-in and other frequency ranges for which PLL can achieve lock were introduced.
Usually in engineering literature non-rigorous definitions are given for these concepts.
In the following we introduce definitions, based on rigorous discussion in
\cite{KuznetsovLYY-2015-IFAC-Ranges,LeonovKYY-2015-TCAS}.

{\bf Definition of hold-in range.}
The largest interval $[0,\Delta\omega_{h})$ of frequency deviations $|\Delta\omega_0|$, such that
the loop re-achieves locked state after small perturbations of the filters' state, the phases and frequencies of VCO, and the input signals, is called a hold-in range.
This effect is also called steady-state stability.
In addition, for a frequency deviation within the hold-in range,
the loop in a locked state tracks
small changes in input frequency,
i.e. achieves a new locked state (\emph{tracking process}) \cite{KuznetsovLYY-2015-IFAC-Ranges,LeonovKYY-2015-TCAS}.

Assume that the loop power supply is initially switched off and then at $t = 0$ the power is switched on, and assume that the initial frequency difference is sufficiently large. The loop may not lock within one beat note, but the VCO frequency will be slowly tuned toward the reference frequency (acquisition process). This effect is also called a transient stability.
The pull-in range is used to name such frequency deviations that make the acquisition process possible.

{\bf Definition of pull-in range.}
The largest interval $[0,\Delta\omega_{P})$ of frequency deviations $|\Delta\omega_0|$, such that the loop achieves locked state for any initial states (filters and initial phase of VCO),
is called a pull-in range
\cite{KuznetsovLYY-2015-IFAC-Ranges,LeonovKYY-2015-TCAS}.
The largest frequency deviation $\Delta\omega_{P}$ is called a pull-in frequency\cite{KuznetsovLYY-2015-IFAC-Ranges,LeonovKYY-2015-TCAS}.

{\bf Definition of lock-in range.}
Lock-in range is a largest interval of frequency deviations
$|\Delta\omega_0| \in [0,\Delta\omega_L)$ inside pull-in range,
such that after an abrupt change of $\omega_1$
within a lock-in range the PLL re-acquires lock without cycle slipping, if it is not interrupted.
Here $\Delta\omega_L$ is called a lock-in frequency\cite{KuznetsovLYY-2015-IFAC-Ranges,LeonovKYY-2015-TCAS}\footnote{
The concept of the lock-in range was suggested by F. Gardner in 1966 \cite[p.40]{Gardner-1966} and it is widely used nowadays (see, e.g. \cite[p.34-35]{Best-1984},\cite[p.161]{Wolaver-1991},\cite[p.612]{HsiehH-1996},\cite[p.532]{Irwin-1997},\cite[p.25]{CraninckxS-1998-book}, \cite[p.49]{KiharaOE-2002},\cite[p.4]{Abramovitch-2002},\cite[p.24]{DeMuerS-2003-book},\cite[p.749]{Dyer-2004-book},\cite[p.56]{Shu-2005},\cite[p.112]{Goldman-2007-book},\cite[p.61]{Best-2007},\cite[p.138]{Egan-2007-book},\cite[p.576]{Baker-2011},\cite[p.258]{Kroupa-2012}). However later Gardner noticed that the lock-in range definition lacks rigor and requires clarification \cite[p.70]{Gardner-1979-book}, \cite[p.187-188]{Gardner-2005-book}. Recently a rigorous definition was suggested in \cite{KuznetsovLYY-2015-IFAC-Ranges,LeonovKYY-2015-TCAS}.}.

Finally, our definitions give
\(
  \Omega_{\text{lock-in}} \subset \Omega_{\text{pull-in}}
  \subset \Omega_{\text{hold-in}},
\)
\[
  [0,\Delta\omega_L) \subset [0,\Delta\omega_{P})
  \subset [0,\Delta\omega_H),
\]
which is in agreement with the classical consideration (see, e.g. \cite[p.34]{Best-1984},\cite[p.612]{HsiehH-1996},\cite[p.61]{Best-2007},\cite[p.138]{Egan-2007-book},\cite[p.258]{Kroupa-2012}).

\section{BPSK Costas loop}\label{s2}

\renewcommand{\thefigure}{2.\arabic{figure}}
\setcounter{figure}{0}

\renewcommand{\thesection}{\arabic{section}}
\setcounter{section}{2\,} 

\subsection{Lock-in range  $\Delta\omega_L$ and lock time $T_L$}\label{ss2.1}
Recall linear model of Costas loop in phase space (see Figure~\ref{linear-pll}).
By \eqref{3}, \eqref{4}, and \eqref{loop-filter-tf} we can derive the open loop transfer function of the Costas loop, which is defined by the ratio $\Theta_2 (s)/\Theta_1(s)$:
\begin{equation}\label{7}
G_{OL}(s)=K_d\frac{K_0}{s}\frac{1+s\tau_2}{s\tau_1}
 \end{equation}

\begin{figure}[H]
\centering
\includegraphics[scale=0.9]{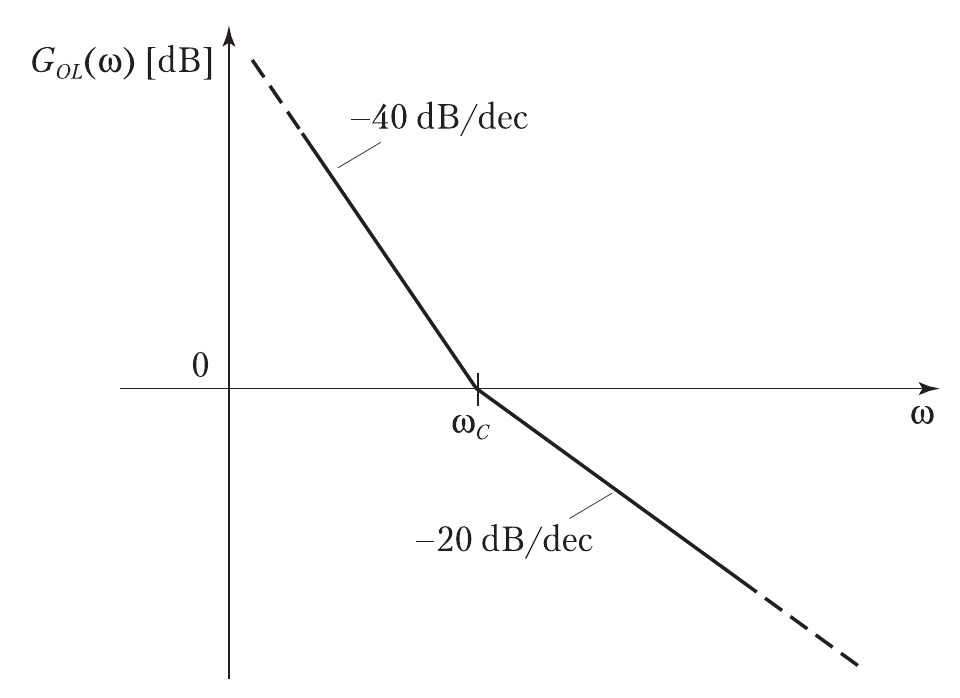}
\caption{Bode plot of magnitude of open loop gain $G_{OL}(\omega)$
}\label{f2-2}
\end{figure}


Figure~\ref{f2-2} shows a Bode plot of the magnitude of $G_{OL}$. The plot is characterized by the corner frequency  $\omega_C$, which is defined by $\omega_C = 1/\tau_2$, and gain parameters $K_d$ and $K_0$. At lower frequencies the magnitude rolls off with a slope of -- 40 dB/decade. At frequency $\omega_C$ the zero of the loop filter causes the magnitude to change its slope to -- 20 dB/decade. To get a stable system, the magnitude curve should cut the 0 dB line with a slope that is markedly less than -- 40 dB/decade. Setting the parameters such that the gain is just 0 dB at frequency  $\omega_C$ provides a phase margin of 45 degrees, which assures stability [2]. From the open loop transfer function we now can calculate the closed loop transfer function defined by
\begin{equation}\label{8}
G_{CL}(s)=\frac{\Theta_2(s)}{\Theta_1(s)}.
 \end{equation}

After some mathematical manipulations we get
\begin{equation}\label{9}
G_{CL}(s)=\frac{K_0K_d\frac{1+s\tau_2}{s\tau_1}}{s^2+s\frac{K_oK_d\tau_2}{\tau_1}+\frac{K_0K_d}{\tau_1}}.
 \end{equation}
It is customary to represent this transfer function in normalized form, i.e.
\begin{equation}\label{10}
G_{CS}(s)=\frac{2s\zeta\omega_n+\omega_n^2}{s^2+2s\zeta\omega_n+\omega_n^2}
 \end{equation}
with the substitutions
\begin{equation}
\label{omega zeta}
\omega_n=\sqrt{\frac{K_0K_d}{\tau_1}},\quad \zeta=\frac{\omega_n\tau_2}{2},
 \end{equation}
where  $\omega_n$ is called natural frequency and $\zeta$  is called damping factor. The linear model enables us to derive simple approximations for lock-in range  $\Delta\omega_L$ and lock time $T_L$.

\begin{figure}[H]
\centering
\includegraphics[scale=0.8]{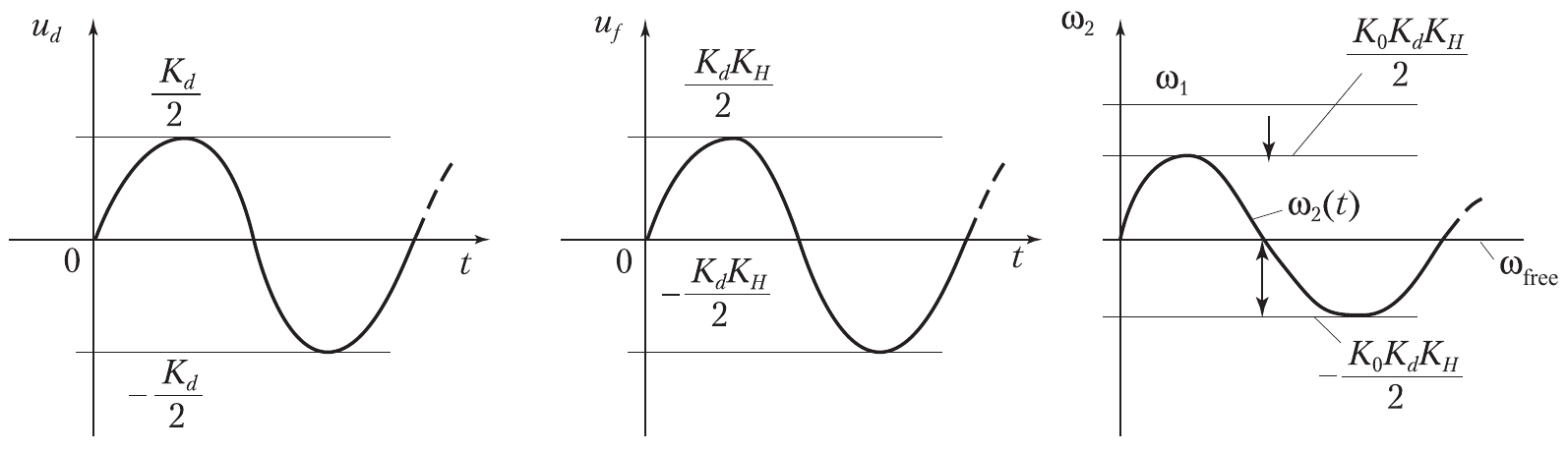}
\caption{Lock-in range of Costas loop
}\label{f2-3}
\end{figure}

For the following analysis we assume that the loop is initially out of lock. The frequency of the input signal (Figure~\ref{costas_before_sync}) is  $\omega_1$, and the frequency of the VCO is  $\omega_2$. The multiplier in the I branch therefore generates an output signal consisting of  a sum frequency term $\omega_1 +  \omega_2$ and a difference frequency term  $\omega_1 - \omega_2$. The sum frequency term is removed by the lowpass filter, and the frequency of the difference term is assumed to be much below the corner frequency $\omega_3$ of the lowpass filter, hence the action of this filter can be neglected for this case. Under this condition the phase detector output signal $u_d(t)$ will have the form (cf. Eqs. (\ref{ud}) and (\ref{3}))
\begin{equation}
\label{ud delta omega}
u_d(t)=\frac{K_d}{2}\sin(2\Delta\omega t)
\end{equation}

with  $\Delta\omega  =  \omega_1 -  \omega_2$. $u_d(t)$  is plotted in Figure~\ref{f2-3}, left trace. This signal passes through the loop filter. In most cases the corner frequency  $\omega_C=1\!/\tau_2$ is much lower than the lock-in range, hence we can approximate its transfer function by
\begin{equation}\label{KH}
H_{LF}(\omega)\approx\frac{\tau_2}{\tau_1}=K_H.
\end{equation}

Let us define the gain of this filter at higher frequencies by  constant $K_H$. Now the output signal of the loop filter is a sine wave having amplitude $K_d$ $K_H/2$ as shown by the middle trace in Figure~\ref{f2-3}.
 Consequently the frequency of the VCO will be modulated as shown in the right trace. The modulation amplitude is given by
$K_d K_0 K_H/2$.
In this figure the reference frequency and the initial frequency  $\omega_{free}$ of the VCO are plotted as horizontal lines.
 When  $\omega_1$ and  $\omega_{free}$ are such that the top of the sine wave just touches the  $\omega_1$  line, the loop acquires lock suddenly, i.e. the lock-in range  $\Delta\omega_L$ is nothing more than the modulation amplitude $K_d$ $K_0$ $K_H/2$. Making use of the substitutions (\ref{omega zeta}) we finally get
\begin{equation}\label{omega lock-in}
\Delta\omega_L=\zeta\omega_n
 \end{equation}

Now the lock process is a damped oscillation whose frequency is the natural frequency. Because the loop is assumed to lock within at most one cycle of that frequency, the lock time can be approximated by the period of the natural frequency, i.e. we have
\begin{equation}\label{13}
T_L\approx\frac{2\pi}{\omega_n}
 \end{equation}

\subsection{Pull-in range   $\Delta\omega_P$ and pull-in time $T_P$}\label{ss2.4}

We have seen that all signals found in this block diagram are sine functions, i.e. all of them seem to have zero average, hence do not show any $dc$ component. This would lead to the (erroneous) conclusion that a pull-in process would not be possible. In reality it will be recognized that some of the signals become asymmetrical, i.e. the duration of the positive half wave is different from the duration of the negative one. This creates a non zero $dc$ component, and under suitable conditions acquisition can be obtained. We are therefore going to analyze the characteristics of the signals in Figure~\ref{f2-4c}.

\begin{figure}[!ht]
\centering
\includegraphics[width=0.5\textwidth]{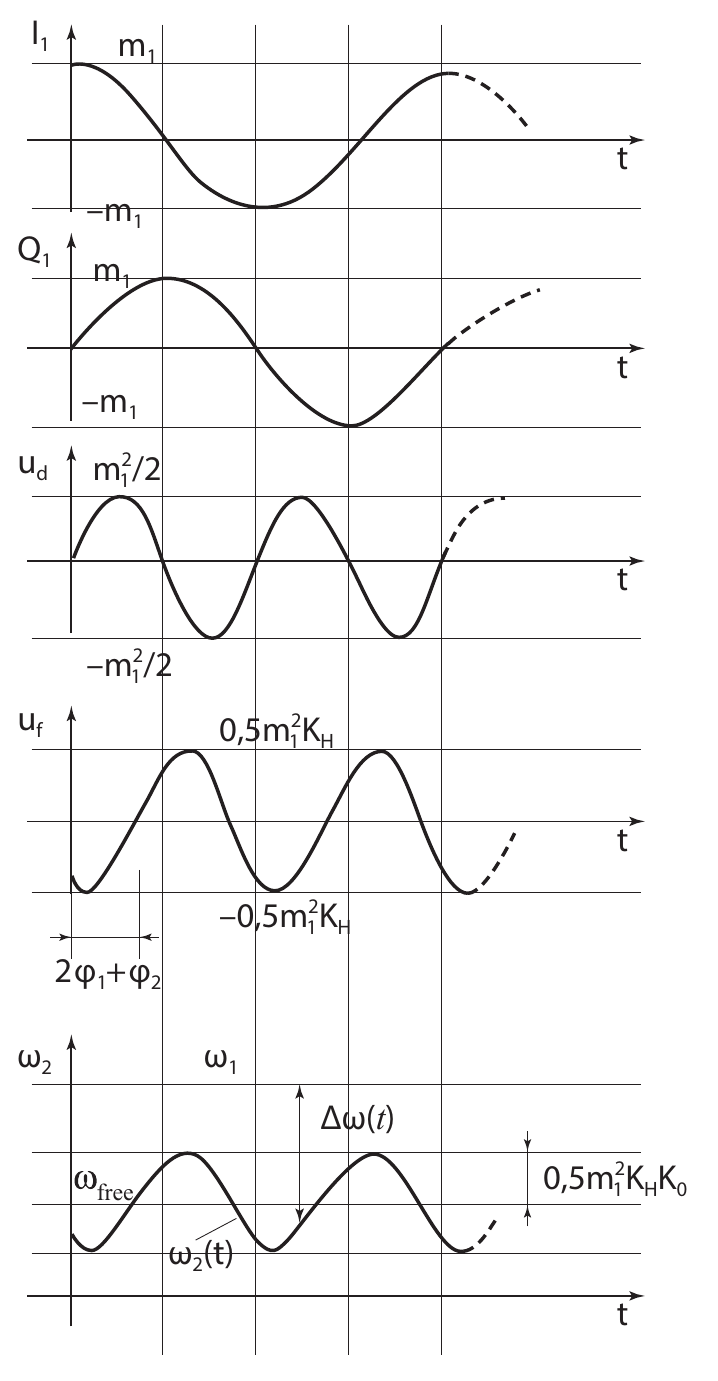}
\caption{Signals of the model in Figure~\ref{f2-4c}
}\label{f2-5}
\end{figure}


All considered signals are plotted in Figure~\ref{f2-5}. For signals $I_1$ and $Q_1$ we obtain
$$
I_1(t)=m_1(t)\cos(\Delta\omega t)
$$
$$
Q_1(t)=m_1(t)\sin(\Delta\omega t)
$$
The sum frequency terms are discarded because they are removed by the lowpass filter. The signal $u_d(t)$  is the product of $I_1$ and $Q_1$ and is given by \eqref{ud delta omega}.
For small arguments $2\Delta\omega t$ this can be written as
\begin{equation}
\notag
\begin{aligned}
& u_d(t) = m_1^2(t) \Delta\omega t = m_1^2(t)\theta_e(t),
\end{aligned}
\end{equation}
where $\theta_e = \Delta\omega t$. Because the phase detector gain is defined by
\begin{equation}
\notag
\begin{aligned}
& u_d(t) = K_d\theta_e(t),
\end{aligned}
\end{equation}
we have $K_d = m_1^2$.

\begin{figure}[H]
\centering
\includegraphics[width=0.5\textwidth]{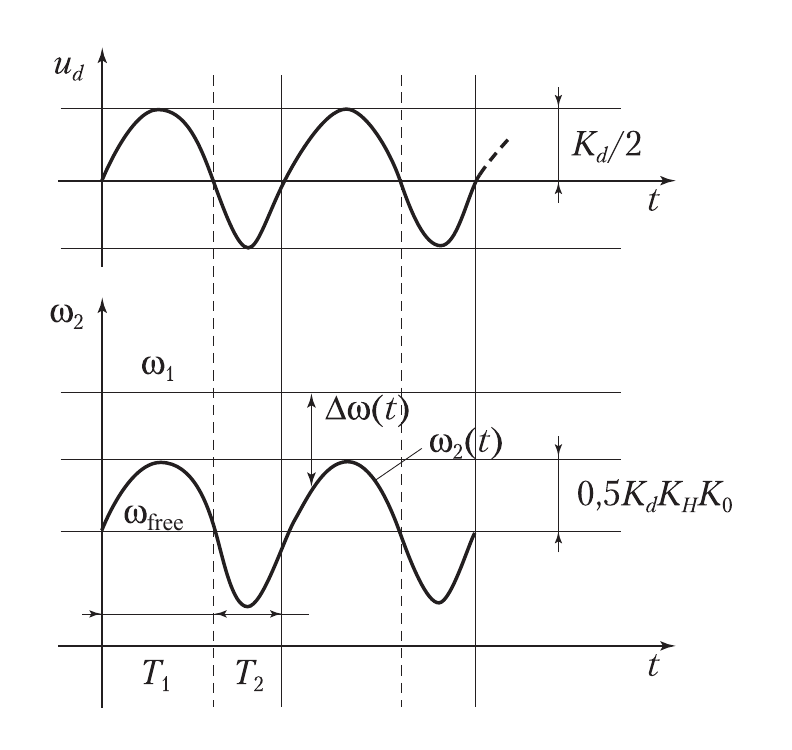}
\caption{Signals of the model in Figure~\ref{f2-4c} for $\varphi_{tot} = 0$
}\label{f2-6a}
\end{figure}

Next the loop filter output signal $u_f(t)$  is plotted. Its amplitude is $K_H$ $m_1^2/2$, and its phase is delayed by $\varphi_{tot} = 2\varphi_1 + \varphi_2$. This signal modulates the frequency of the VCO as shown in the bottom trace of Figure~\ref{f2-5}. The modulation amplitude is given by
$\frac{m_1^2 K_H K_0}{2}$. In order to get an estimate for the non zero $dc$ component of $u_d(t)$ we will have to analyze the asymmetry of the signal waveforms. It will be shown that $\overline{u_d}$  (the average of $u_d(t)$) is a function of frequency difference $\Delta\omega$    and phase  $\varphi_{tot}$. The analysis becomes easier when we first calculate $\overline{u_d}$ for some special values for $\varphi_{tot}$, i.e. for $\varphi_{tot} = 0; -\pi /2$; and $-\pi$. Let us start with  $\varphi_{tot} = 0$, cf. Figure~\ref{f2-6a}.

In Figure~\ref{f2-6a} the waveforms for $u_d(t)$ and  $\omega_2(t)$ are shown. The asymmetry of the signals is exaggerated in this plot. During the positive half cycle (duration $T_1$) the average value of VCO output frequency  $\omega_2(t)$ is increased, which means that the average difference frequency   $\Delta\omega(t)$ is lowered. Consequently the duration of the positive half wave becomes larger than half of a full cycle. During the negative half cycle (duration $T_2$), however, the average value of VCO output frequency  $\omega_2(t)$ is decreased, which means that the average difference frequency
$\Delta\omega (t)$  is increased. Consequently the duration of the negative half wave becomes less than half of a full cycle.
Next we are going to calculate the average frequency difference in both half cycles. The average frequency difference during half cycle $T_1$ is denoted  $\overline{\Delta\omega_{d+}}$, the average frequency difference during half cycle $T_2$ is denoted  $\overline{\Delta\omega_{d-}}$. We get
\begin{equation}\label{17a}
\overline{\Delta\omega_{d+}}=\Delta\omega-\frac{2}{\pi}\frac{K_0K_dK_H}{2},
 \end{equation}
\begin{equation}\label{17b}
\overline{\Delta\omega_{d-}}=\Delta\omega+\frac{2}{\pi}\frac{K_0K_dK_H}{2}.
 \end{equation}
For the durations $T_1$ and $T_2$ we obtain after some manipulations
\begin{equation}\label{18a}
T_1\approx\frac{\pi}{2\Delta\omega}\left(1+\frac{K_0K_dK_H}{\pi\Delta\omega}\right),
 \end{equation}
\begin{equation}\label{18b}
T_2\approx\frac{\pi}{2\Delta\omega}\left(1-\frac{K_0K_dK_H}{\pi\Delta\omega}\right).
 \end{equation}
Now the average value $\overline{u_d}$  can be calculated from
\begin{equation}\label{19}
\overline{u_d(t)}=\frac{K_0K_d^2K_H}{\pi^2\Delta\omega}.
 \end{equation}
 The average signal $\overline{u_d}$ is seen to be inversely proportional to the frequency difference  $\Delta\omega$. Because $\overline{u_d}$  is positive, the instantaneous frequency $\omega_2(t)$ is pulled in positive direction, i.e. versus  $\omega_1$, which means that a pull-in process will take place.

\begin{figure}[H]
\centering
\includegraphics[width=0.5\textwidth]{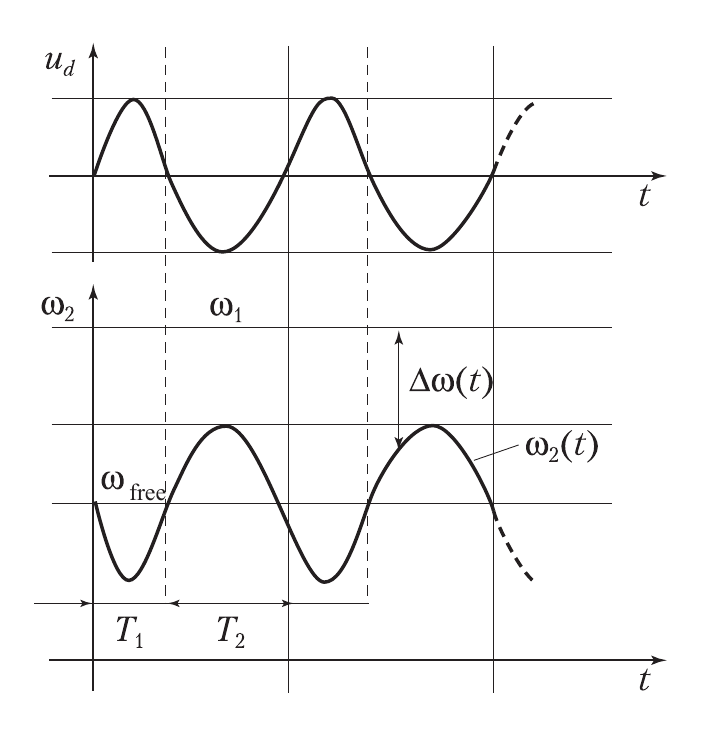}
\caption{Signals of the model in Figure~\ref{f2-4c} for $\varphi_{tot} = -\pi$
}\label{f2-6b}
\end{figure}


Next we are going to analyze the dependence of $\overline{u_d}$  on phase $\varphi_{tot}$. Let us consider now the case for $\varphi_{tot} = -\pi$, cf. Figure~\ref{f2-6b}. We observe that in interval $T_1$ the instantaneous frequency  $\omega_2(t)$ is pulled in negative direction, hence the average difference frequency $\overline{\Delta\omega_{d+}}$  becomes larger. Consequently interval $T_1$ becomes shorter. In interval $T_2$, however, the reverse is true. Here the instantaneous frequency $T_1$ the pulled in positive direction, hence the average $\overline{\Delta\omega_{d-}}$  is reduced, and interval $T_2$ becomes longer. The average $\overline{u_d}$  is now equal and opposite to the value of $\overline{u_d}$  for $\varphi_{tot} = 0$. Because it is negative under this condition, a pull-in process cannot take place, because the frequency of the VCO is "pulled away" in the wrong direction.

\begin{figure}[H]
\centering
\includegraphics[width=0.5\textwidth]{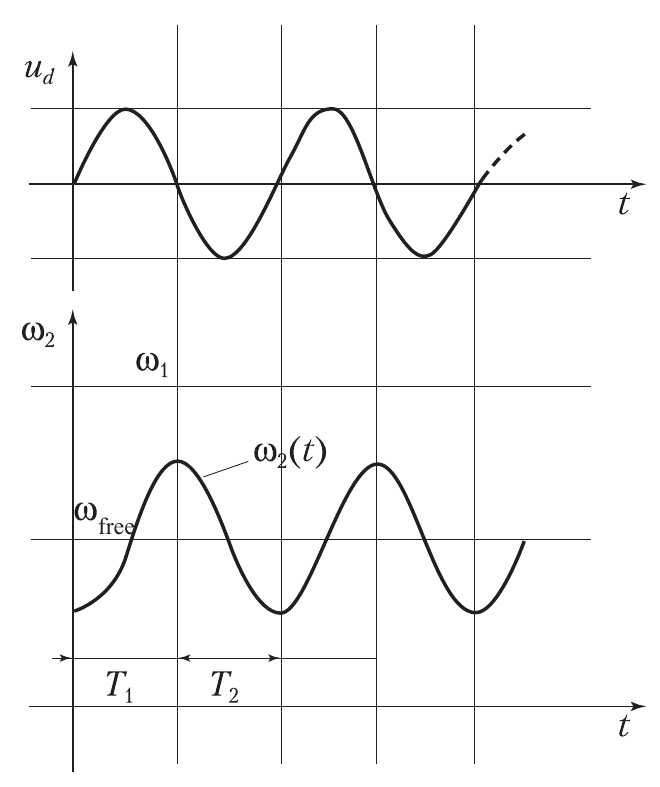}
\caption{Signals of the model in Figure~\ref{f2-4c} for $\varphi_{tot} = -\pi /2$
}\label{f2-6c}
\end{figure}

Last we consider the case  $\varphi_{tot} = -\pi /2$, cf. Figure~\ref{f2-6c}. In the first half of interval $T_1$ the instantaneous frequency  $\omega_2(t)$ is lowered, but in the second half it is increased. Consequently the average difference frequency   $\overline{\Delta\omega_{d+}}$ does not change its value during $T_1$. The same happens in interval $T_2$.  $\overline{\Delta\omega_{d-}}$ does not change either, and $\overline{u_d}$  remains 0.

It is easy to demonstrate that $\overline{u_d}$  varies with $\cos(\varphi_{ tot})$, hence we have
\begin{equation}\label{20}
\overline{u_d(t)}=\frac{K_0K_d^2K_H}{\pi^2\Delta\omega}\cos(\varphi_{tot}),\quad\varphi_{tot}=2\varphi_1+\varphi_2.
 \end{equation}
Eq.~(\ref{20}) tells us that the pull-in range is finite. The pull-in range can be found as the frequency difference for which phase $\varphi_{tot} = -\pi /2$. An equation for the pull-in range will be derived in  section \ref{ss2.4}. We also will have to find an equation for the pull-in time. The model shown in Figure~\ref{f2-7} will enable us to obtain a differential equation for the average frequency difference  $\overline{\Delta\omega}$  as a function of time.

Recall equations of filter output \eqref{21b}
\begin{equation}\notag
\overline{u_f(t)}=\frac{1}{\tau_1}\int\limits_0^t\overline{u_d(\tau)}d\tau
 \end{equation}
and frequency deviation \eqref{25}
\begin{equation}\notag
\Delta\omega=\Delta\omega_0-K_0\overline{u_f}.
 \end{equation}
 Eqs. (\ref{20}), (\ref{21b}), and (\ref{25}) enable us to compute the three variables  $\overline{u_d}$,   $\overline{u_f}$,  and $\Delta\omega$ as a function of time. This will be demonstrated in Section \ref{ss2.4}.

The pull-in range can be computed using Eq.~(\ref{20}). Lock can only be obtained when the total phase shift $\varphi_{tot}$ is not more negative than  $-\pi /2$. This leads to an equation of the form
\begin{equation}\label{26}
2\varphi_1(\Delta\omega_p)+\varphi_2(2\Delta\omega_p)=-\pi\!/2.
 \end{equation}
According to Eqs.~(\ref{loop-filter-tf}) and (\ref{14}) $\varphi_1$ and  $\varphi_2$ are given by
$$
\varphi_1(\omega)=-{\rm{arctg}}(\omega/\omega_3),
$$
$$
\varphi_2(\omega)=-\pi\!/2+{\rm{arctg}}(\omega/\omega_C)
$$
with $\omega_C = 1/ \tau_2$. Hence the pull-in range  $\Delta\omega_P$ can be computed from the transcendental equation
\begin{equation}\label{27}
2{\rm{arctg}}(\Delta\omega_P\!/\omega_3)={\rm{arctg}}(2\Delta\omega_P\!/\omega_C).
 \end{equation}

 To solve this equation for   $\Delta\omega_P$  we use the addition formula for the tangent function
 $$
 {\rm{tg}}(2\alpha)=\frac{2{\rm{tg}}\alpha}{1-{\rm{tg}}^2\alpha}
 $$
and can replace $2{\rm{arctg}}(\Delta\omega_P/\omega_3)$  by  ${\rm{arctg}}\frac{2\frac{\Delta\omega_P}{\omega_3}}{1-\frac{\Delta\omega_P^2}{\omega_3^2}}$. Eq.~(\ref{27}) can now be
rewritten as ${\rm{arctg}}\dfrac{2\frac{\Delta\omega_P}{\omega_3}}{1-\frac{\Delta\omega_P^2}{\omega_3^2}}=
{\rm{arctg}}2\frac{\Delta\omega_P}{\omega_C}$.

When the arctg expressions on both sides of the equation are equal, their arguments must also be identical, which leads to
$$
\frac{\frac{2\Delta\omega_P}{\omega_3}}{1-\frac{\Delta\omega_P^2}{\omega_3^2}}=
2\frac{\Delta\omega_P}{\omega_C}.
$$
Hence we get for the pull-in range
\begin{equation}\label{28}
\Delta\omega_P
=\omega_3\sqrt{\frac{\frac{\omega_3}{\omega_C}-1}
{\frac{\omega_3}{\omega_C}}}.
 \end{equation}
Last an equation for the pull-in time $T_P$ will be derived. Eqs. (\ref{20}), (\ref{21b}), and (\ref{25}) describe the behavior of the three building blocks in Figure~\ref{f2-7} and enable us to compute the three variables  $\overline{u_d}$,  $\overline{u_f}$, and   $\Delta\omega$. We only need to know the instantaneous  $\Delta\omega$   vs. time, hence we eliminate $\overline{u_d}$  and $\overline{u_f}$  from Eqs. (\ref{21b}) and (\ref{25}) and obtain the differential equation
\begin{equation}\label{29}
\frac{d}{dt}\Delta\omega\tau_1+\frac{1}{\Delta\omega}\frac{K_0^2K_d^2K_H}{\pi^2}\cos(\varphi_{tot})=0.
 \end{equation}
This differential equation is non linear, but the variables  $\Delta\omega$   and $t$ can be separated, which leads to an explicit solution. Putting all terms containing  $\Delta\omega$ to the left side and performing an integration we get

\begin{equation}\label{30}
\frac{\tau_1\pi^2}{K_0^2K_d^2K_H}
\int\limits_{\Delta\omega_0}^{\Delta\omega_L}
\dfrac{\Delta\omega}{\cos(\varphi_{tot})} d\Delta\omega=-\int\limits_0^{T_P}dt.
 \end{equation}
The limits of integration are $\Delta\omega_0$ and $\Delta\omega_L$ on the left side, because the pull-in process starts with an initial frequency offset $\Delta\omega=\Delta\omega_0$ and ends when $\Delta\omega$ reaches the value $\Delta\omega_L$, which is the lock-in range. Following that instant a lock-in process will start. The integration limits on the right side are $0$ and $T_P$, respectively, which means that the pull-in process has duration $T_P$, and after that interval  (fast) lock-in process starts.

Performing the integration on the left imposes some considerable problems, when we remember that $\cos(\varphi_{tot})$ is given by
\[
  \cos(\varphi_{tot}) = \cos(-2\arctg\dfrac{\Delta\omega}{\omega_3}
  -\dfrac{\pi}{2}+\arctg\dfrac{2\Delta\omega}{\omega_C}).
\]
Finding an explicit solution for the integral seems difficult if not impossible, but the $\cos$ term can be drastically simplified.
When we plot $\cos(\varphi_{tot})$ vs. $\Delta\omega$ we observe that within the range
 $\Delta\omega_L<\Delta\omega<\Delta\omega_0$  the term
 $\cos(\varphi_{tot})$ is an almost perfect straight line.
 Hence we can replace $\cos(\varphi_{tot})$ by
 \[
   \cos(\varphi_{tot}) \approx 1 - \dfrac{\Delta\omega}{\Delta\omega_P}.
 \]
Inserting that substitution into Eq. \eqref{30} yields a rational function
of  $\Delta\omega$  on the left side, which is easily integrated. After some mathematical procedures we obtain for the pull-in time $T_P$
\begin{equation}\label{31}
  T_P=\frac{\Delta\omega_P \pi^2\tau_1}{2K_0^2K_d^2K_H}
  \bigg[
    \Delta\omega_P\ln\dfrac{\Delta\omega_P-\Delta\omega_L}{\Delta\omega_P-\Delta\omega_0}
    -\Delta\omega_0+\Delta\omega_L
  \bigg].
 \end{equation}
Making use of Eqs. \eqref{omega zeta} and \eqref{KH} we have
\[
   K_H = \dfrac{\tau_2}{\tau_1},\quad
   \omega_n^2 = \dfrac{K_0K_d}{\tau_1}, \quad
   \zeta = \dfrac{\omega_n\tau_2}{2}.
\]
Using these substitutions Eq. \eqref{31} can be rewritten as
\begin{equation}\label{32}
  T_P=\dfrac{\Delta\omega_P \pi^2}{2\zeta\omega_n^3}
  \bigg[
    \Delta\omega_P\ln\dfrac{\Delta\omega_P-\Delta\omega_L}{\Delta\omega_P-\Delta\omega_0}
    -\Delta\omega_0+\Delta\omega_L
  \bigg].
 \end{equation}
This equation is valid for initial frequency  offsets in the range $\Delta\omega_L<\Delta\omega_0<\Delta\omega_P$.
For lower frequency offsets, a fast pull-in process will occur,
and Eq. \eqref{13} should be used.

%
%

\subsection{Numerical example 1: Designing an analog Costas loop for BPSK}\label{ss2.5}

An analog Costas loop for BPSK shall be designed in this section. It is assumed that a binary signal is modulated onto a carrier. The carrier frequency is set to 400 kHz, i.e. the Costas loop will operate at a center frequency $\omega_0 = 2\pi$    400'000 = 2'512'000\, rad $s^{-1}$.  The symbol rate is assumed to be $f_S = 100'000$ symbols/s. Now the parameters of the loop (such as time constants  $\tau_1$ and  $\tau_2$, corner frequencies $\omega_C$ and  $\omega_3$, and gain parameters such as $K_0, K_d$)  must be determined. (Note that these parameters have been defined in Eqs.~(\ref{4}), (\ref{loop-filter-tf}), (\ref{6}) and (\ref{13})).

The modulation amplitude is set $m_1 = 1$. According to Eq.~(\ref{3}) the phase detector gain is then $K_d = 1$. It has proven advantageous to determine the remaining parameters by using the open loop transfer function $G_{OL}(s)$ of the loop [2]. This is given by
\begin{equation}\label{33}
G_{OL}(s)=\frac{K_0K_d}{s}\frac{1+s\!/\omega_c}{s\tau_1}\frac{1}{1+s\!/\omega_3}
 \end{equation}

\begin{figure}[H]
\centering
\includegraphics[width=0.6\textwidth]{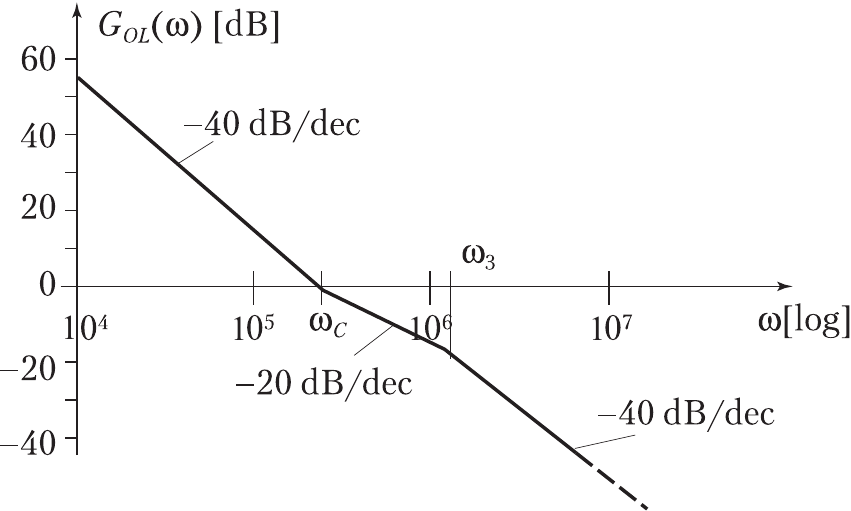}
\caption{ Bode plot of open loop transfer function of Costas loop
}\label{f2-9}
\end{figure}


The magnitude $|G_{OL}(\omega )|$ (Bode diagram) is plotted in Figure~\ref{f2-9}. The magnitude curve crosses the 0 dB line at the so called transit frequency  $\omega_T$. It is common practice to choose  $\omega_T$ to be about
$(0.05\omega_0 \ldots 0.1\omega_0)  $. Here we set  $\omega_T = 0.1\omega_0$, i.e.  $\omega_T = 251'200$ rad $s^{-1}$. Furthermore we set corner frequency  $\omega_C =\omega_T$. When doing so, the slope of the asymptotic magnitude curve changes from -- 40 dB/decade to -- 20 dB/decade at $\omega  = \omega_C$. Under this condition the phase of $G_{OL}(\omega)$ is -135$^{\circ}$ at  $\omega_C$. Consequently the phase margin of the loop becomes 45$^{\circ}$, which provides sufficient stability.
According to Eq.~(\ref{loop-filter-tf})  $\tau_2$ becomes $4 \mu s$. Next corner frequency  $\omega_3$ will be determined. The corner frequency of the lowpass filter must be chosen such that the demodulated data signal (i. e. the output of the lowpass filter in the I branch) is recovered with high fidelity. To fulfill this requirement,  $\omega_3$ should be chosen as large as possible. On the other hand, the lowpass filter should suppress the double frequency component (here at about 800 kHz) sufficiently, which means that  $\omega_3$ should be markedly less than $2 \omega_0$. It's a good compromise to set corner frequency to twice the symbol rate, i.e.  $\omega_3 = 2 \cdot 2\pi  \cdot 100'000 = 1'256'000\, {\rm{rad}} s^{-1}$. Last the remaining parameters  $\tau_1$ and $K_0$ must be chosen. They have to be specified such that the open loop gain becomes 1 at frequency $\omega  = \omega_C$.
According to Eq.~(\ref{33}) we can set
\begin{equation}\label{34}
G_{OL}(\omega_C)=1\approx\frac{K_oK_d}{\omega_C^2\tau_1}.
\end{equation}
 Because 2 parameters are still undetermined, one of those can be chosen arbitrarily, hence we set  $\tau_1 = 20 \mu s$. Finally from (\ref{34}) we get $K_0 = 1'262'000\, s^{-1}$.

The design of the Costas loop is completed now, and we can compute the most important loop parameters. For the natural frequency and damping factor we get from \eqref{omega zeta}
$$\begin{aligned}
&  n = 251'000 {\rm{rad}}/s \quad (f_n = 40 {\rm{kHz}}),\\
&\zeta  = 0.5.\end{aligned}
$$
From \eqref{omega lock-in} the lock-in range becomes
$$
  \Delta\omega_L = 125'000 {\rm{rad}} s\quad  (\Delta f_L = 20 {\rm{kHz}})
$$
and from (\ref{13}) the lock time becomes
$$
  T_L = 25 \mu s.
$$
Next we want to compute the pull-in range. Eq.  (\ref{28})  yields
$$
\Delta\omega_P  = 1'086'440\, {\rm{rad}} s^{-1}
\quad  (\Delta f_P  = 173\, {\rm{kHz}}).
$$

\subsection{Numerical example 2: Designing a digital Costas loop for BPSK}\label{ss2.6}

To convert the analog loop into a digital one, we first must define a suitable sampling  frequency $f_{samp}$ (or sampling interval $T = 1/f_{samp}$). To satisfy the Nyquist theorem, the sampling frequency must be higher than twice the highest frequency that exists in the loop. In our case the highest frequency is found at the output of the multipliers in the $I$ and $Q$ branches (cf. Figure~\ref{costas_before_sync}). The sum frequency term is about twice the center frequency, hence $f_{samp}$ must be greater than 4 times the center frequency. A suitable choice would be $f_{samp} = 8$  $f_0 = 3.2$ MHz.

Next the transfer functions of the building block have to be converted into discrete transfer functions, i.e. $H(s)\to   H(z)$. For best results it is preferable to use the bilinear $z$ transform. Given an analog transfer function $H(s)$, this can be converted into a discrete transfer function $H(z)$ by replacing $s$ by
\begin{equation}\label{35}
s = \frac{2}{T}\frac{1-z^{-1}}{1+z^{-1}}.
\end{equation}
Now the bilinear z transform has the property that the analog frequency range from $0\ldots \infty$  is compressed to the digital frequency range from $0\ldots f_{samp}/2$. To avoid undesired ``shrinking'' of the corner frequencies ($\omega_C$ and  $\omega_3$), these must be ``prewarped'' accordingly, i.e. we must set
\begin{equation}\label{36}
\omega_{C,p}=\frac{2}{T}{\rm{tg}}\frac{\omega_CT}{2},
\end{equation}
\begin{equation}\label{37}
\omega_{3,p}=\frac{2}{T}{\rm{tg}}\frac{\omega_3T}{2},
\end{equation}
where  $\omega_{C,p}$  and $\omega_{3,p}$ are the prewarped corner frequencies. Now we can apply the bilinear z transform to the transfer functions of the lowpass filters (cf. Eq.~(\ref{14})) and of the loop filter (cf. Eq.~(\ref{loop-filter-tf})) and get
\begin{equation}\label{38}
H_{LPF}(z)=\frac{\left[1+\frac{2}{\omega_{3,p}T}\right]+\left[1-\frac{2}{\omega_{3,p}T}\right]z^{-1}}{1+z^{-1}},
\end{equation}
\begin{equation}\label{39}
H_{LF}(z)=\frac{\left[1+\frac{2}{\omega_{C,p}T}\right]+\left[1-\frac{2}{\omega_{C,p}T}\right]z^{-1}}
{\frac{2\tau_1}{T}-\frac{2\tau_1}{T}z^{-1}}.
\end{equation}
Because the VCO is a simple integrator, we can apply the discrete $z$ transform of an integrator, i.e.
\begin{equation}\label{40}
H_{VCO}(z)=\frac{K_0T}{1-z^{-1}}.
\end{equation}
The digital Costas loop is ready now for implementation. A Simulink model will be presented in section \ref{ss2.7}.

\subsection{Simulating the digital Costas loop for BPSK}\label{ss2.7}
\begin{figure}[ht!]
\centering
\includegraphics[width=\textwidth]{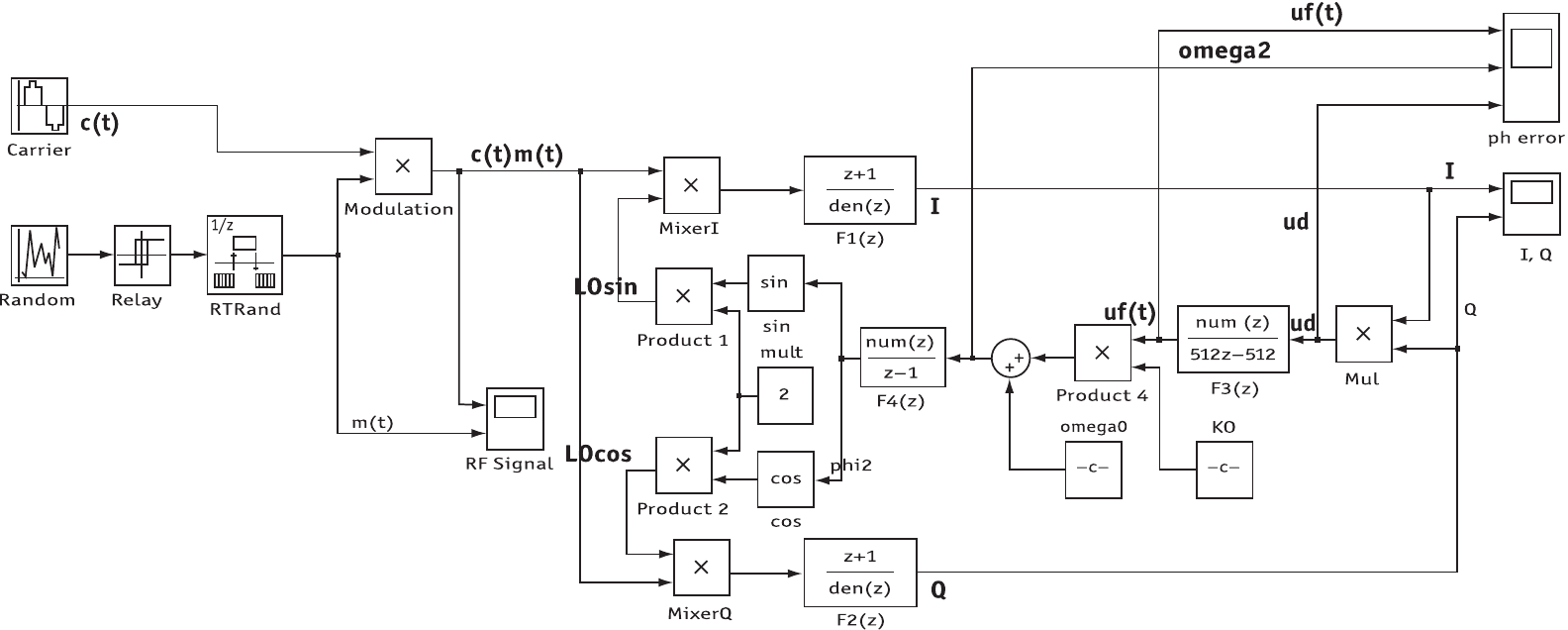}
\caption{Simulink model of the digital Costas loop for BPSK
}\label{f2-10}
\end{figure}

A Simulink model of a Costas loop for BPSK is shown in Figure~\ref{f2-10}.

A data signal is created by a random number generator at the left in the block diagram. The other blocks are self explanatory. The model is used now to check the validity of the approximations found for pull-in range and pull-in time.

Eq.~(\ref{28}) predicts a pull-in range
 $\Delta f_P = 173$ kHz. The simulation revealed a pull-in range of  $\Delta f_P = 133$ kHz., which shows that the theoretical result is a rather crude approximation. A series of other simulation delivered results for the pull-in time  $\Delta T_P$. The results are listed in Table 2-1.
\begin{table*}
\begin{center}
\begin{tabular}{|l|l|l|l|}
\hline
$\Delta f_0$ (Hz)   &$\Delta\omega_0$ (rad $s^{-1}$)  &$T_P$ (theory) ($\mu s$) & ($T_P$ (simulation) ($\mu s$)\\
\hline
50 kHz  &314'000& 33  &30\\
\hline
70 kHz &  439'000 &78 &85\\
\hline
100 kHz & 628'000 & 204&  200\\
\hline

\end{tabular}

\bigskip

\centering{Table 2-1. Comparison of predicted and simulated results for the pull-in range}
\end{center}
\end{table*}

We note that the predicted and simulated parameters are in good agreement.

\subsection{Remarks on simulation of BPSK Costas loop}
Note that a numerical simulation of various models of the same circuit
can lead to essentially different results if the corresponding mathematical assumptions, used for the models construction,
are not satisfied.
Also the errors caused by numerical integration (e.g. in MATLAB and SPICE)
can lead to unreliable results \cite{BestKKLYY-2015-ACC,KuznetsovKLNYY-2015-ISCAS,BianchiKLYY-2015,KuznetsovKLSYY-2014-ICUMT-BPSK}. The following examples demonstrate some limitations of numerical approach on simple models.

Next the following parameters are used in simulation:
low-pass filters transfer functions $H_{lpf}(s) = \frac{2}{s/\omega_3+1}$, $\omega_3 = 1.2566 \cdot 10^6$
and the corresponding parameters in system \eqref{loop-filter}  are
$A_{1,2} = -\omega_3$, $b_{1,2} = 1$, $c_{1,2} = \omega_3$;
loop filter transfer function $H_{lf}(s) = \frac{\tau_2 s + 1}{\tau_1 s}$,
$\tau_2 = 3.9789 \cdot 10^{-6}$, $\tau_1 = 2 \cdot 10^{-5}$,
and the corresponding parameters in system \eqref{loop-filter}  are
$A = 0$, $b = 1$, $c = \frac{1}{\tau_1}$, $h = \frac{\tau_2}{\tau_1}$;
carrier frequency $\omega_1=2 \cdot \pi \cdot 400000$;
VCO input gain $L=4.8 \cdot 10^6$;
and carrier initial phase $\theta_2(0)=\theta_1(0)=0$.

\begin{example}[double frequency and averaging]
In Figure~\ref{averaged_non_averaged}
it is shown that Assumption~\ref{as-twice-frequency} may not be valid:
mathematical model in signal's phase space (see Figure~\ref{f1-1} -- black color)
and physical model (see Figure~\ref{costas_before_sync} and system \eqref{loop-filter-int} -- red color)
after transient processes have different phases in the locked states.

Here VCO free-running frequency $\omega_{free}=2 \cdot \pi \cdot 400000-600000$;
initial states of filters are all zero: $x(0) = x_1(0)\equiv x_2(0) = 0$.

\begin{figure}[H]
  \centering\includegraphics[width=0.23\textwidth]{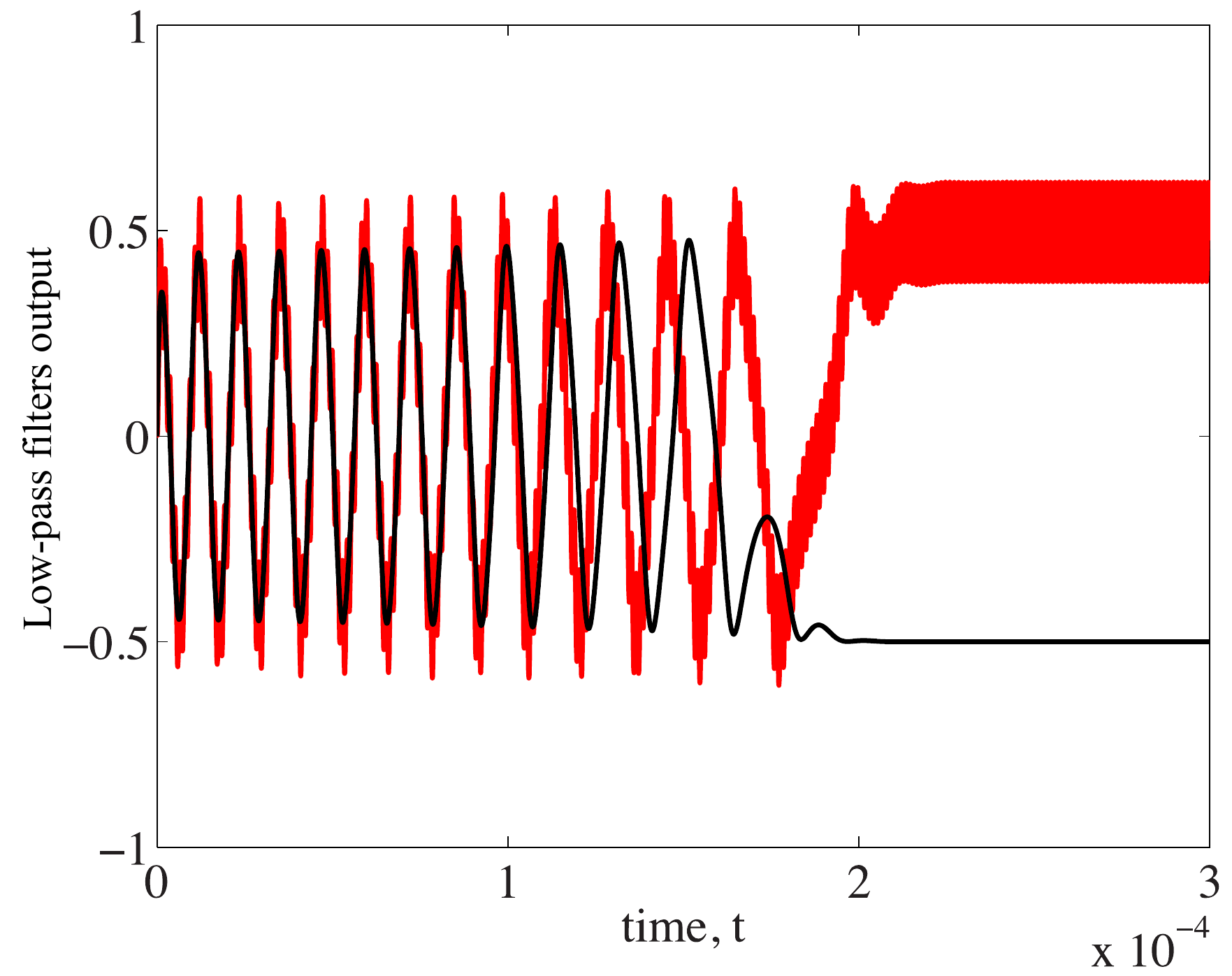}
  \centering\includegraphics[width=0.23\textwidth]{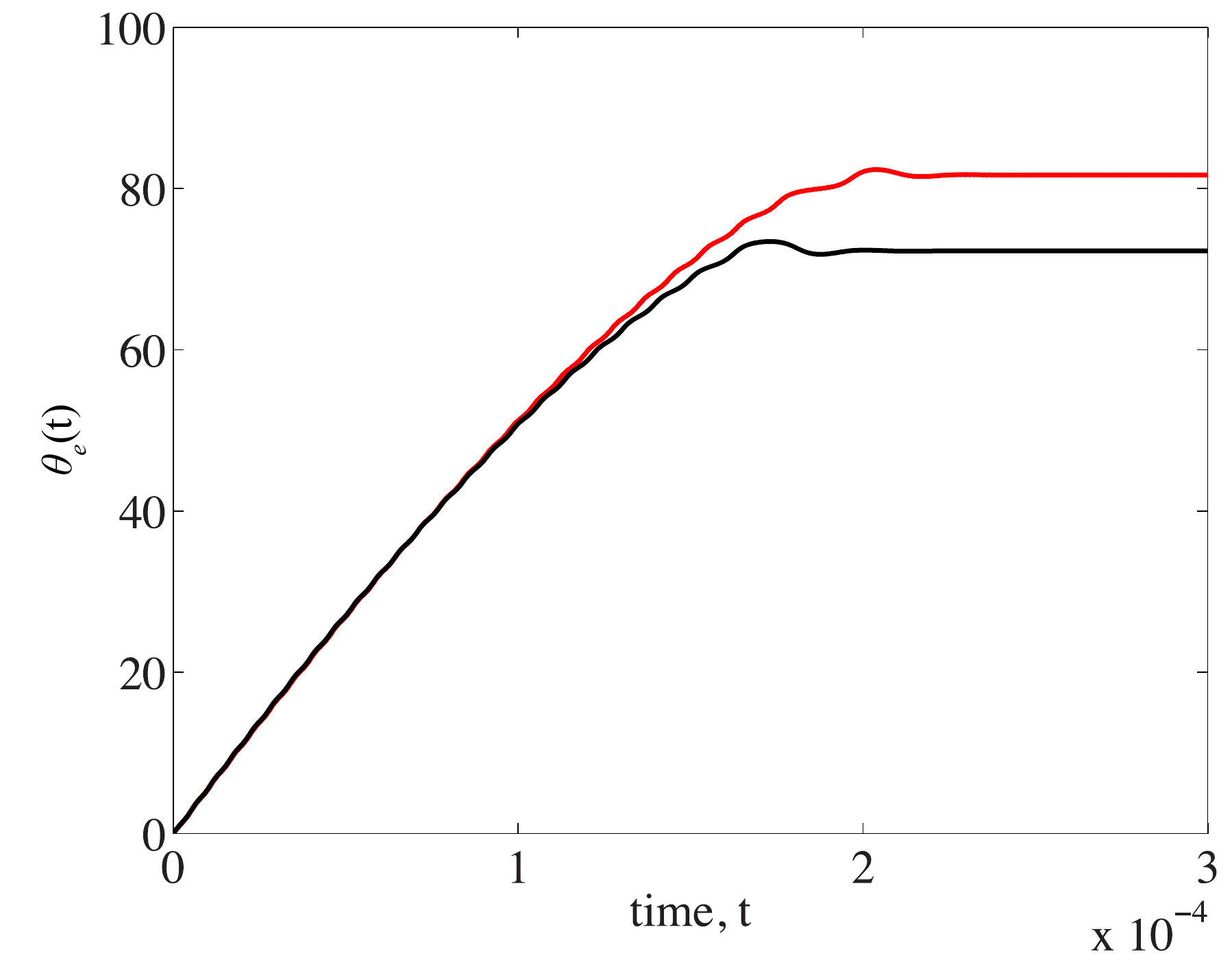}
  \caption{Low-pass filter outputs and phase difference for
  averaged model (black)
  and physical model (red) in Figure~\ref{costas_before_sync}.}
  \label{averaged_non_averaged}
\end{figure}
\end{example}

\begin{example}[numerical integration parameters]
In~Figure~\ref{semistable}
it is shown that standard simulation of the loop may not be valid: while
the classic mathematical model in signal's phase space
(Figure~\ref{f1-1}),
simulated in Simulink with predefined integration parameters: 'max step size' set to '1e-3',
is out of lock  (black),
the same model simulated in Simulink
with default integration parameters: 'max step size' set to 'auto', acquires lock (red).
Here Matlab chooses step from $5 \cdot 10^{-3}$ to $9\cdot10^{-2}$;
for the fixed step $2\cdot 10^{-2}$ the model acquires lock,
for the fixed step $1\cdot 10^{-2}$ the model doesn't acquire lock.

Here the initial loop filter state output is $x(0) = 0.0125$;
VCO free-running frequency $\omega_{free}=10000-89.45$;
VCO input gain $L=1000$;
initial phase shift $\theta_e(0)=-3.4035$.

\begin{figure}[H]
  \centering\includegraphics[width=0.23\textwidth]{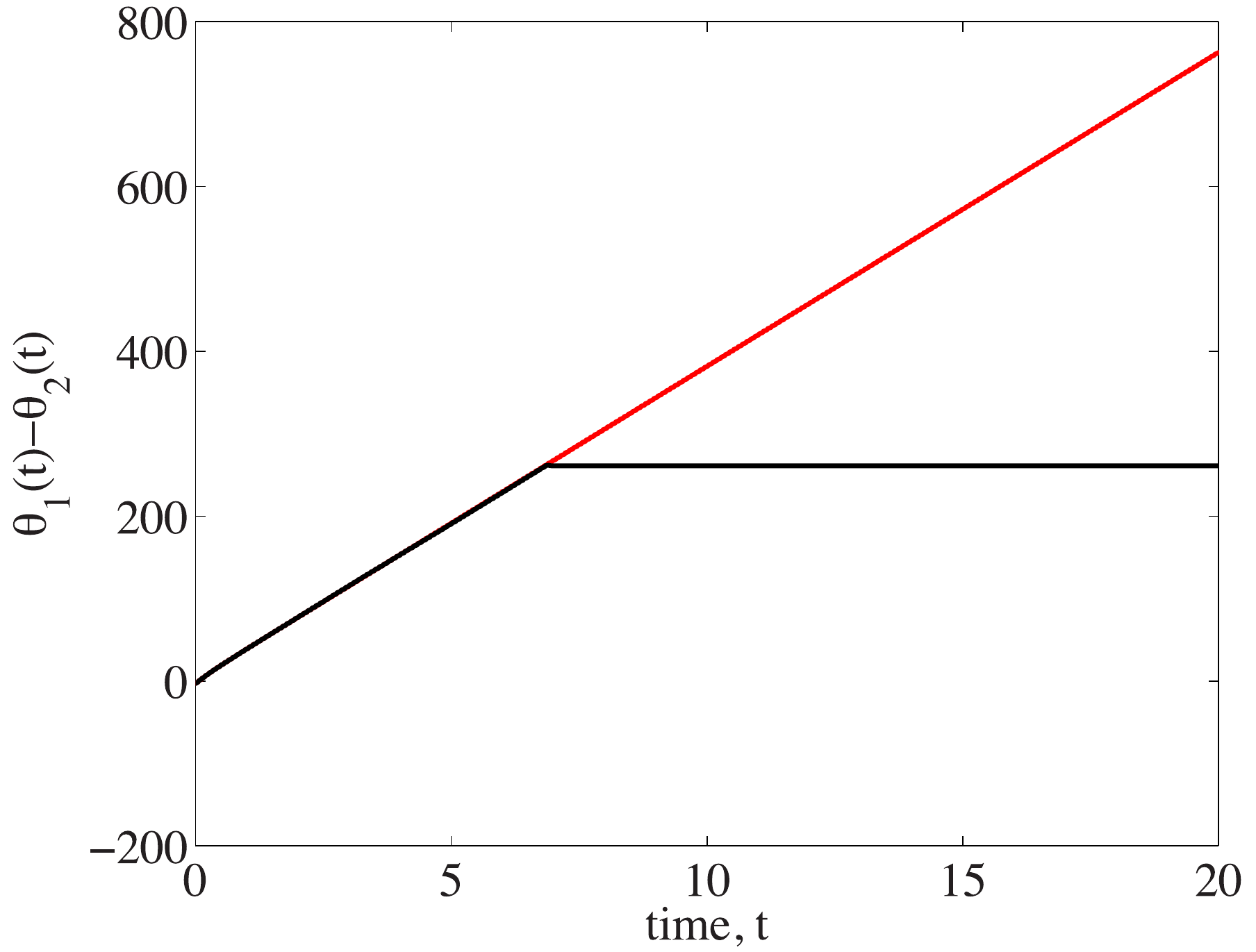}
  \centering\includegraphics[width=0.23\textwidth]{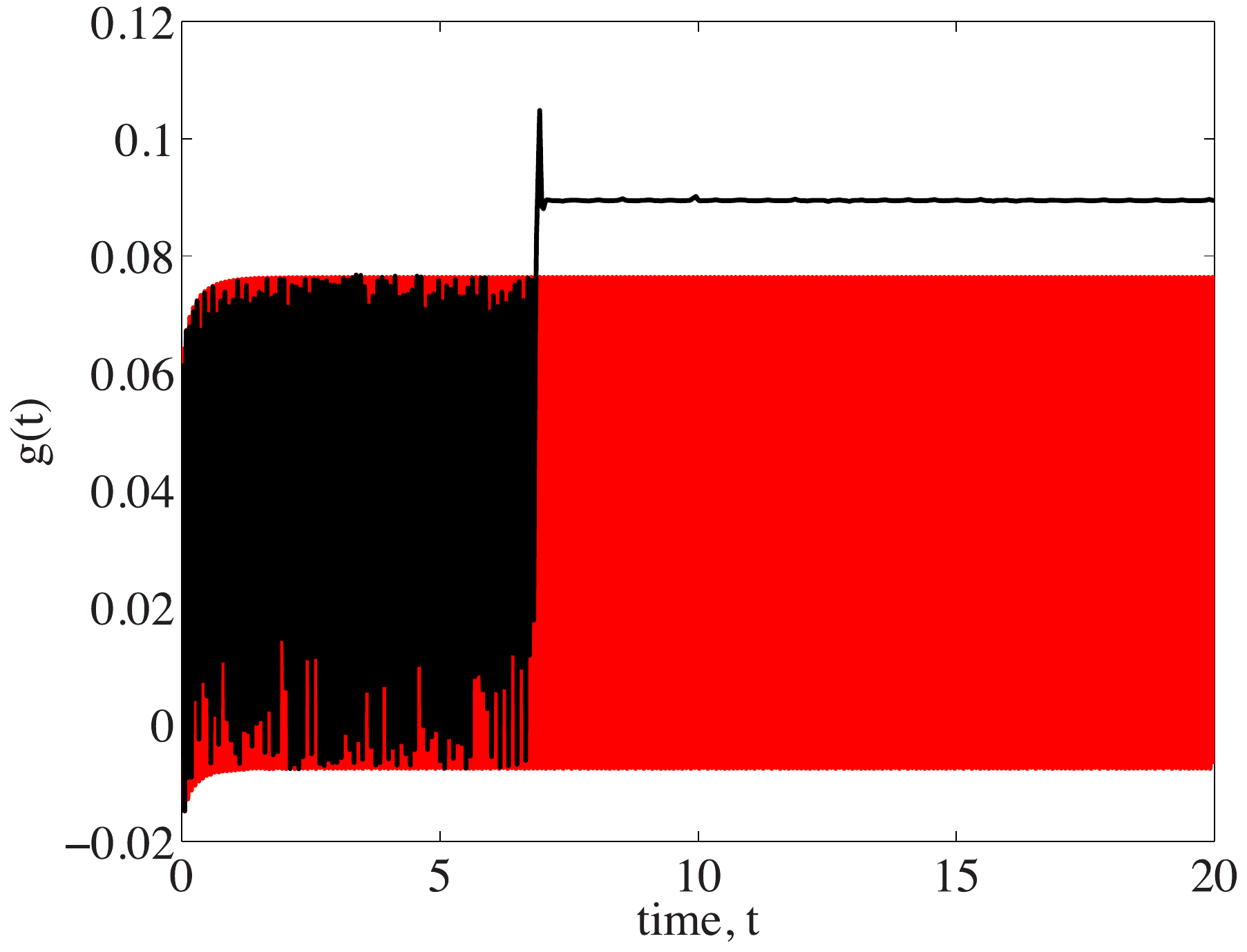}
  \caption{Filter outputs: default integration parameters in Simulink
  'max step size' set to 'auto' (black curve);
  Parameters configured manually 'max step size' set to '1e-3' (red curve).}
  \label{semistable}
\end{figure}
\end{example}

\begin{figure}[H]
  \centering\includegraphics[width=0.30\textwidth]{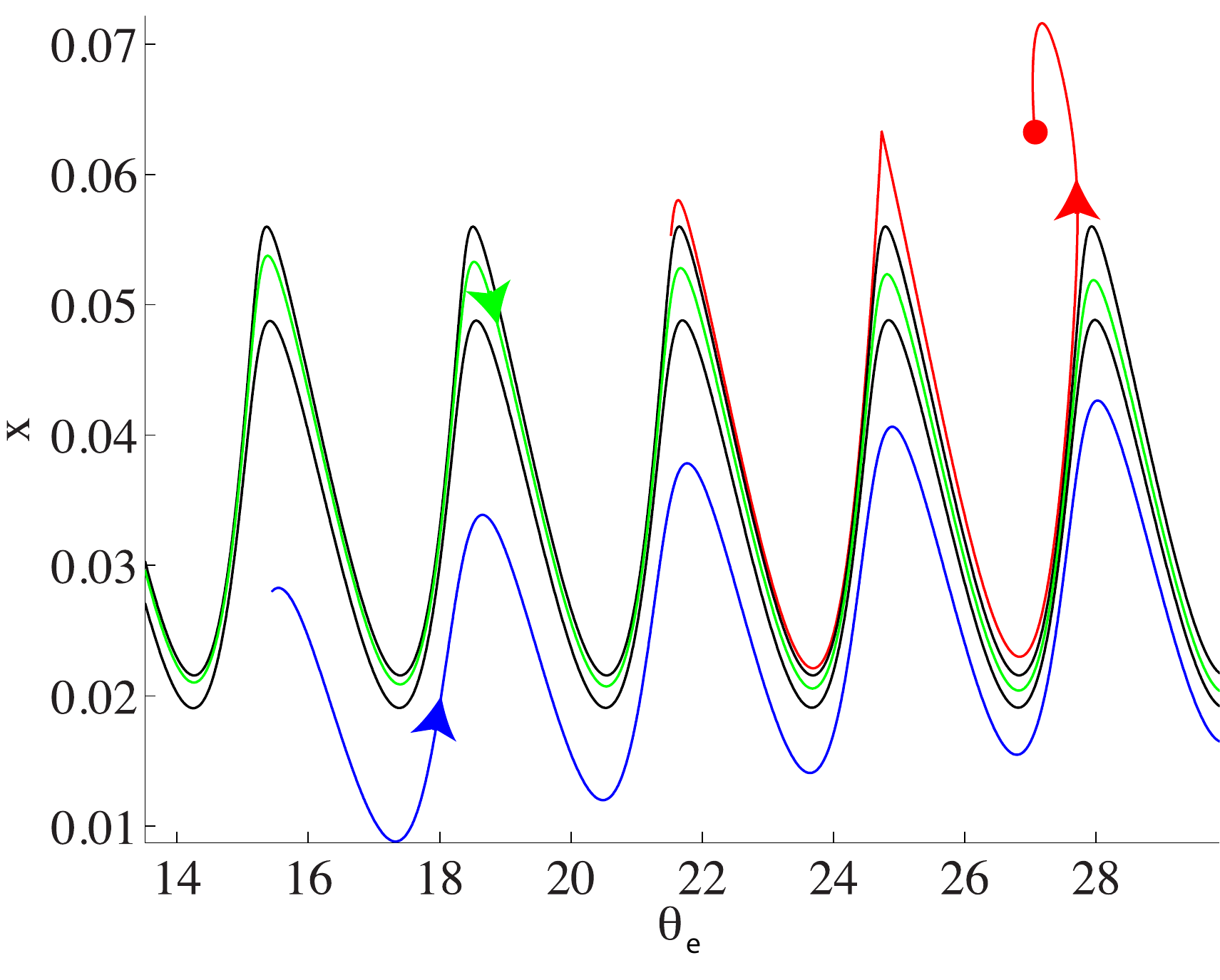}
  \caption{Phase portrait: coexistence of stable and unstable periodic solutions.}
  \label{semistable_phase}
\end{figure}

Consider now the corresponding phase portrait (see Figure~\ref{semistable_phase}).
Here the red trajectory tends to a stable equilibrium (red dot).
Lower and higher black trajectories are stable and unstable limit cycles, respectively.
The blue trajectory tends to a stable periodic trajectory (lower black periodic curve) and
in this case the model does not acquire lock.
All trajectories between black trajectories (see green trajectory) tend to the stable lower black trajectory.

If the gap between stable and unstable trajectories (black lines) is smaller than the discretization step,
the numerical procedure may slip through the stable trajectory
(blue trajectory may step over the black and green lines and begins to be attracted to the red dot).
In other words, the simulation may show that the Costas loop acquires lock
although in reality it does not.
The considered case corresponds to the coexisting attractors (one of which is a hidden oscillation)
and the bifurcation of birth of a semistable trajectory \cite{LeonovK-2013-IJBC}.

Note, that only trajectories (red) above the unstable limit cycle is attracted to the equilibrium.
Hence $\Delta\omega = 89.45$ does not belong to the pull-in range.

Corresponding limitations, caused by hidden oscillations, appear
in simulation of various phase-locked loop (PLL) based systems
\cite{LeonovK-2013-IJBC,KuznetsovLYY-2014-IFAC,KuznetsovKLNYY-2014-ICUMT-QPSK,KuznetsovKLSYY-2014-ICUMT-BPSK,KudryashovaKKLSYY-2014-ICINCO,KuznetsovKLNYY-2015-ISCAS,BestKKLYY-2015-ACC,BianchiKLYY-2015,LeonovKYY-2015-TCAS,Bianchi201645}.


\renewcommand{\thefigure}{3.\arabic{figure}}
\setcounter{figure}{0}
\renewcommand{\thesection}{\arabic{section}}
\setcounter{section}{2\,} \thispagestyle{empty}

\section{QPSK Costas loop}\label{s3}

\subsection{Lock-in range   $\Delta\omega_L$ and lock time $T_L$}\label{ss3.2}
\begin{figure}[H]
\centering
\includegraphics[scale=0.8]{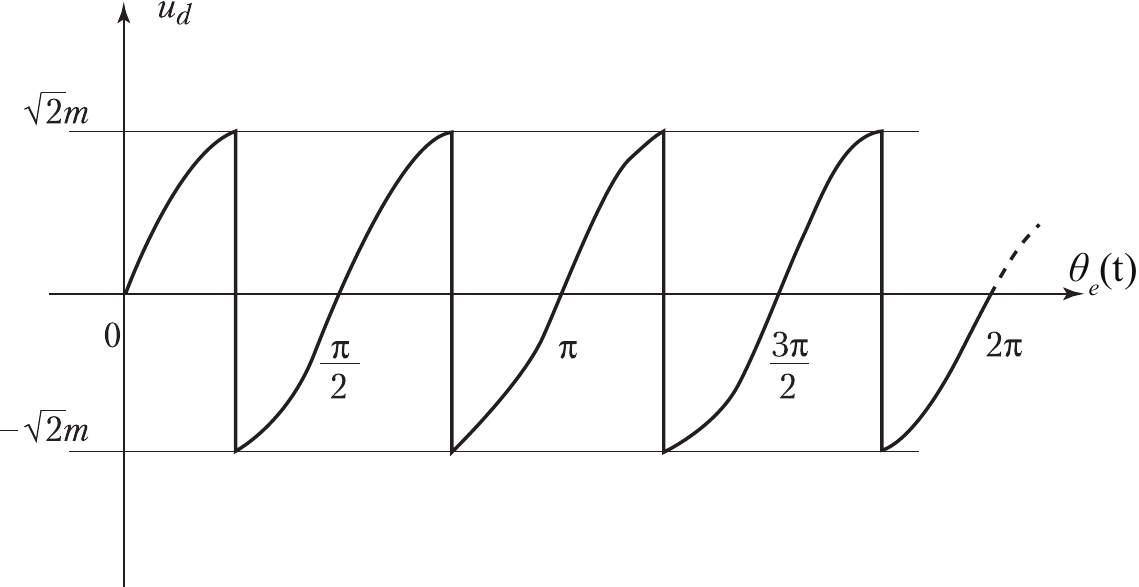}
\caption{ Phase detector output signal $u_d$ as a function of phase error  $\theta_e$
}\label{f3-2}
\end{figure}

\begin{figure}[!ht]
\centering
\includegraphics[width=0.9\linewidth]{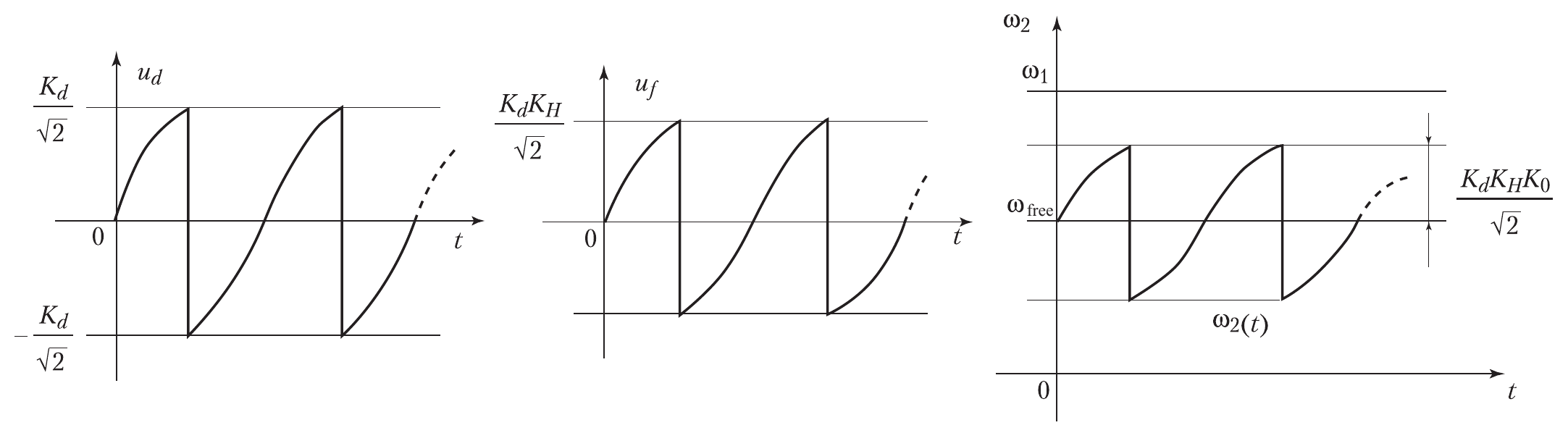}
\caption{Signals $u_d(t), u_f(t)$, and  $\omega_2(t)$ during the pull-in process
}\label{f3-3}
\end{figure}

The open loop transfer function is identical with that of the Costas loop for BPSK, cf. Eq.~(\ref{7}) and Figure~\ref{f2-2}. This holds true for the closed loop transfer function, too, cf. Eqs.~\eqref{9}, \eqref{10} , and \eqref{omega zeta}.
To determine the lock-in range, we assume that the loop is out of lock. Let the reference frequency be  $\omega_1$, and the initial VCO frequency  $\omega_{free}$. The difference frequency  $\omega_1 - \omega_2$ is called   $\Delta\omega$. When the loop has not acquired lock, the phase error  $\theta_e$  is a continuously rising function that increases towards infinity. The phase detector output signal $u_d$ is then a chopped sine wave as depicted in Figure~\ref{f3-2}. The fundamental frequency of this signal is four times the difference frequency, i.e. $4\Delta\omega$.
This signal is plotted once again in the left trace of Figure~\ref{f3-3}. The amplitude of this signal is $K_d/\sqrt 2$.
Because for the Costas loop for QPSK the phase detector gain is $K_d = 2m$,
this is equal to $\sqrt{2}m$.
The fundamental frequency of $u_d$ is assumed to be much higher than the corner frequency  $\omega_C$ of the loop filter, hence the transfer function of the loop filter can be approximated by
\begin{equation}\label{45}
H_{LF}(s)\approx \frac{\tau_2}{\tau_1}=K_H.
\end{equation}

Hence the output signal of the loop filter $u_f$  has an amplitude of $K_d K_H /\sqrt 2$, cf. middle trace of Figure~\ref{f3-3}. This signal modulates the output frequency of the VCO, and the modulation amplitude is given by $K_d K_H  K_0 /\sqrt 2$, cf. right trace in Figure~\ref{f3-3}. It is easily seen that the loop spontaneously locks when the peak of the  $\omega_2(t)$ waveform touches the  $\omega_1$ line, hence we have
\begin{equation}\label{46}
\Delta\omega_L=\frac{K_0K_dK_H}{\sqrt 2}.
\end{equation}

Making use of Eqs.~(\ref{omega zeta}) and (\ref{45}) this can be rewritten as
\begin{equation}\label{47}
\Delta\omega_L=\sqrt 2\zeta\omega_n.
\end{equation}

Because the transient response of the loop is a damped oscillation whose frequency is  $\omega_n$, the loop will lock in at most one cycle of  $\omega_n$, and we get for the lock time
\begin{equation}\label{48}
T_L\approx\frac{2\pi}{\omega_n}.
\end{equation}


\subsection{Pull-in range and pull-in time for QPSK}\label{ss4.4}
\begin{figure}[H]
\centering
\includegraphics[width=0.9\linewidth]{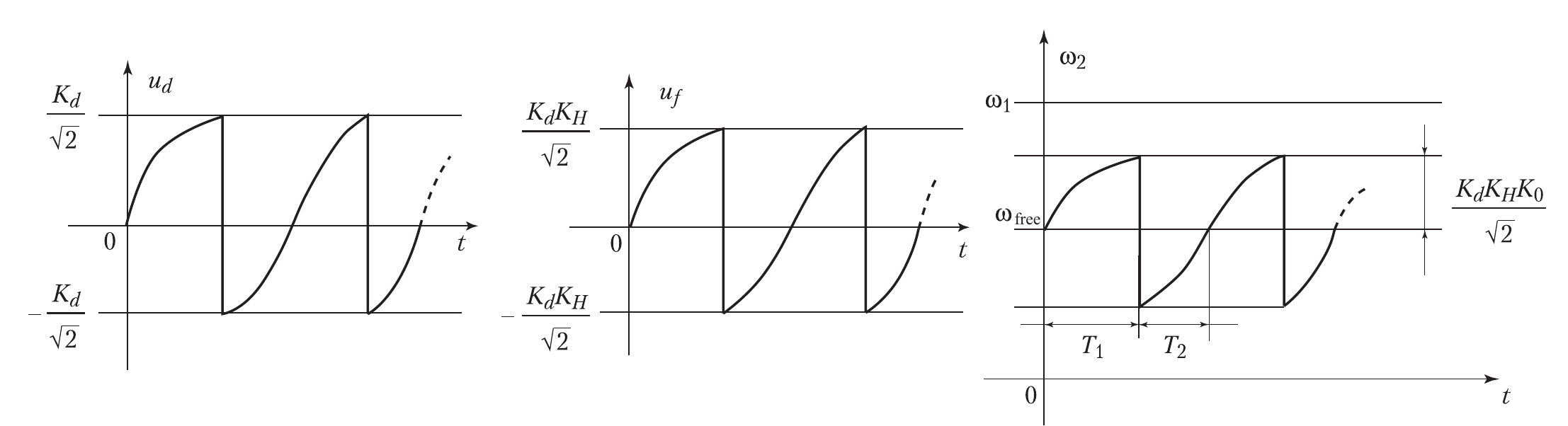}
\caption{Signals of the Costas loop for QPSK in the unlocked state
}\label{f3-5}
\end{figure}

Consider the simplified non linear model of QPSK Costas loop, cf section~\ref{qpsk models}.
 Let us define the total phase by $\varphi_{tot} = 4 \varphi_1 + \varphi_2$.
Next we are computing the average phase detector output signal $\overline{u_d}$  as a function of frequency difference and phase  tot. First we calculate $\overline{u_d}$  for the special case  $\varphi_{tot} = 0$. As shown in the right trace in Figure~\ref{f3-5} during interval $T_1$ the average frequency  $\omega_2$ is increased, hence the average difference $\Delta\omega$   becomes smaller. During next half cycle $T_2$ the reverse is true: the average difference $\Delta\omega$   becomes greater, hence for  $\varphi_{tot} = 0$  $T_1$ is longer than $T_2$. The modulating signal is therefore asymmetric, and because also $u_d(t)$ (left trace) is asymmetrical, its average $\overline{u_d}$  becomes non zero and positive. This asymmetry has been shown exaggerated in Figure~\ref{f3-5}.

Using the same mathematical procedure as for BPSK Costas loop, the average $u_d$ signal is given by
\begin{equation}\label{49}
\overline{u_d}=\frac{0.373^2K_d^2K_H}{\Delta\omega}\cos(4\varphi_1[\Delta\omega]+\varphi_2[4\Delta\omega]).
\end{equation}

As in case of the Costas loop for BPSK, here again Eq.~(\ref{49}) tells us that the pull-in range is finite. The pull-in range is the frequency difference for which phase  $\varphi_{tot} = -\pi\!/2$. An equation for the pull-in range will be derived here. We also will have to find an equation for the pull-in time. To derive the pull-in process, we will use the same non linear model as used for the Costas loop for BPSK, cf. Figure~\ref{f2-7}. The transfer functions for the loop filter and for the VCO have been given in Eqs.~(\ref{21b}) and (\ref{25}), respectively.

The pull-in range can be computed using Eq.~(\ref{49}). Lock can only be obtained when the total phase shift  tot is not more negative than  $-\pi\!/2$. This leads to an equation of the form
\begin{equation}\label{50}
4\varphi_1(\Delta\omega_p)+\varphi_2(4\Delta\omega_p)=-\pi\!/2.
\end{equation}

According to Eqs.~(\ref{loop-filter-tf}) and (\ref{14})  $\varphi_1$ and  $\varphi_2$ are given by
$$
\varphi_1(\omega)=-{\rm{arctg}}(\omega/\omega_3),
$$
$$
\varphi_2(\omega)=-\pi/2+{\rm{arctg}}(\omega/\omega_C)
$$
with  $\omega_C = 1/\tau_2$. Hence the pull-in range   $\Delta\omega_P$ can be computed from the transcendental equation
\begin{equation}\label{51}
4{\rm{arctg}}(\Delta\omega_P\!/\omega_3)={\rm{arctg}}(4\Delta\omega_p\!/\omega_C).
\end{equation}

Using the addition theorem of the tangent function
$$
{\rm{tg}}(4\alpha)=\frac{(1-{\rm{tg}}^2\alpha)4{\rm{tg}}\alpha}{1-6{\rm{tg}}^2\alpha+{\rm{tg}}^4\alpha}.
$$
the term $4{\rm{arctg}}(\Delta\omega_p/\omega_3)$ can be replaced by ${\rm{arctg}}\frac{\left[1-\left(\frac{\Delta\omega_p}{\omega_3}\right)^2\right]4\frac{\omega_p}{\omega_3}}
{1-6\left(\frac{\Delta\omega_p}{\omega_3}\right)^2+\left(\frac{\Delta\omega_p}{\omega_3}\right)^4}$.

Eq.~(\ref{51}) then reads
$$
{\rm{arctg}}\frac{\left[1-\left(\frac{\Delta\omega_p}{\omega_3}\right)^2\right]4\frac{\omega_p}{\omega_3}}
{1-6\left(\frac{\Delta\omega_p}{\omega_3}\right)^2+\left(\frac{\Delta\omega_p}{\omega_3}\right)^4}=
{\rm{arctg}}\frac{4\Delta\omega_p}{\omega_C}.
$$

When the $\arctg$ expressions on both sides are equal,
the arguments must be identical as well, hence we get
$$
\frac{\left[1-\left(\frac{\Delta\omega_p}{\omega_3}\right)^2\right]4\frac{\omega_p}{\omega_3}}
{1-6\left(\frac{\Delta\omega_p}{\omega_3}\right)^2+\left(\frac{\Delta\omega_p}{\omega_3}\right)^4}=
\frac{4\Delta\omega_p}{\omega_C}.
$$
Solving for  $\Delta\omega_P$ yields
\begin{equation}\label{52}
\Delta\omega_p=\omega_3\sqrt{\frac{6-\frac{\omega_C}{\omega_3}-\sqrt{\left[6-\frac{\omega_C}{\omega_3}\right]^2
-4\left(1-\frac{\omega_C}{\omega_3}\right)}}{2}}.
\end{equation}

Last an equation for the pull-in time $T_P$ will be derived. Based on the non linear model shown in Figure~\ref{f2-7} and in Eqs.~(\ref{21b}), (\ref{25}), and (\ref{49}) we can create a differential equation for the instantaneous difference frequency  $\Delta\omega$   as a function of time. For this type of Costas loop the differential equation has the form
\[
  \dfrac{d}{dt}
   \Delta\omega\tau_1 + \dfrac{\cos{\varphi_{tot}}}{\Delta\omega}
   0.373^2K^2_0K^2_dK_H =0
\]
with
\[
  \cos{\varphi_{tot}} = -4\arctg\dfrac{\Delta\omega}{\omega_3}
  -\dfrac{\pi}{2}
  +\arctg{\Delta\omega}{\omega_c}.
\]
Also here the $\cos$ term can be replaced by
\[
  \cos{\varphi_{tot}} \approx = 1 - \dfrac{\Delta\omega}{\Delta\omega_P}
\]
and, using similar procedures as in previews section,
we get for the pull-in time
\begin{equation}\label{53}
   T_P\approx\frac{\Delta\omega_P}{0.278\zeta\omega_n^3}
   \bigg[
     \Delta\omega_P\ln\dfrac{\Delta\omega_P-\Delta\omega_L}{\Delta\omega_P-\Delta\omega_0}
     -\Delta\omega_0+\Delta\omega_L
   \bigg],
\end{equation}
which again  is valid for initial frequency  offsets in the range
$\Delta\omega_L<\Delta\omega_0<\Delta\omega_P$.
For lower frequency offsets, a fast pull-in process will occur, and Eq. \eqref{48}
should be used.

\subsection{Numerical example: Designing a digital Costas loop for QPSK}\label{ss3.5}

A digital  Costas loop for QPSK shall be designed in this section. It is assumed that two  binary signals ($I$ and $Q$)  are modulated onto a quadrature carrier (cosine and sine carrier). The carrier frequency is set to 400 kHz, i.e. the Costas loop will operate at a center frequency  $\omega_0 = 2\pi$    400'000 = 2'512'000 rad $s^{-1}$.  The symbol rate is assumed to be $f_S$ = 100'000 symbols/s. Now the parameters of the loop (such as time constants  $\tau_1$ and  $\tau_2$, corner frequencies  $\omega_C$ and  $\omega_3$, and gain parameters such as $K_0, K_d$)  must be determined. (Note that these parameters have been defined in Eqs.~(\ref{4}), (\ref{loop-filter-tf}), (\ref{6}) and (\ref{13})).
It is possible to use the same parameters as for digital BPSK, i.e.
\begin{equation}
\begin{aligned}
& m_1 \equiv m_2 \equiv 1,\\
& K_d = 2, \\
& G_{OL}(s)=\frac{K_0K_d}{s}\frac{1+s\!/\omega_C}{s\tau_1}\frac{1}{1+s\!/\omega_3},\\
& \omega_T = 251'200,\\
& \tau_2 = 4 \mu s,\\
& \omega_3 = 2 * 2\pi * 100'000 = 1'256'000,\\
& \tau_1 = 20 \mu s,\\
& K_0 = 631'000 s^{-1},\\
&  \omega_n = 251'000\, {\rm{rad}}/s \quad (f_n = 40 {\rm{kHz}})\\
& \zeta  = 0.5,\\
& \Delta\omega_L = 177'483\,{\rm{rad}} s \quad (\Delta f_L = 20 {\rm{kHz}}).\\
\end{aligned}
\end{equation}

From (\ref{47}) the lock-in range becomes
$$
  \Delta\omega_L = 177'483\,{\rm{rad}} s \quad (\Delta f_L = 20 {\rm{kHz}})
$$
and from (\ref{48}) the lock time becomes
$$
  T_L = 25 \mu s.
 $$

Next we want to compute the pull-in range.
  Eq.  (\ref{52}) yields  $\Delta f_P = 73\, {\rm{kHz}})$.
In section \ref{ss3.6} we will simulate this Costas loop and compare the results of the simulation with the predicted ones.

In digital domain
$f_{samp} = 8$ and $f_0 = 3.2 {\rm{MHz}}$.
Transfer functions $H_{LPF}(z)$, $H_{VCO}$ and $H_{LF}(z)$ are defined in \eqref{38}, \eqref{39}, and \eqref{40}.
A Simulink model will be presented in section \ref{ss3.6}.

\subsection{Simulating the digital Costas loop for QPSK}\label{ss3.6}

\begin{figure}[H]
\centering
\includegraphics[width=0.95\textwidth]{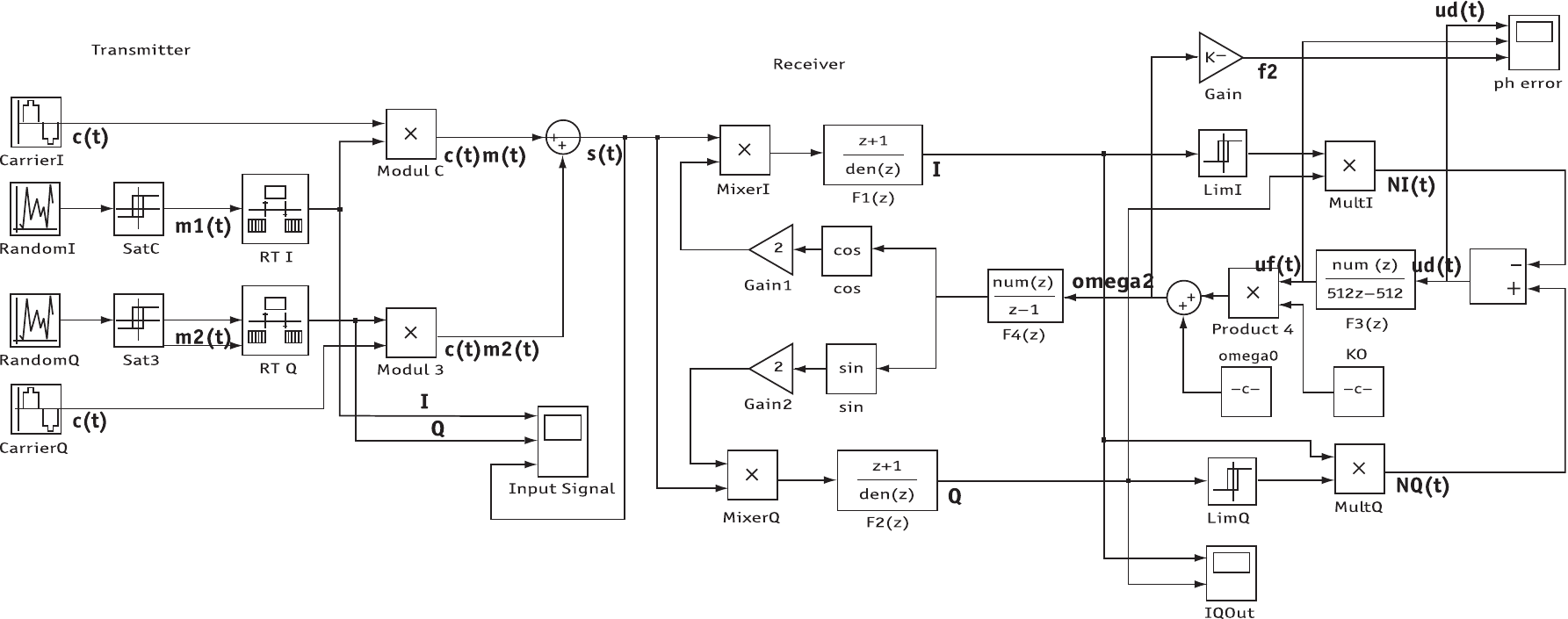}
\caption{Simulink model of the digital Costas loop for QPSK
}\label{f3-8}
\end{figure}

A Simulink model of a Costas loop for QPSK is shown in Figure~\ref{f3-8}.

Two data signals ($I$ and $Q$)  is created by random number generators at the left of the block diagram. The other blocks are self explanatory. The model is used now to check the validity of the approximations found for pull-in range and pull-in time.

Eq.~(\ref{52}) predicts a pull-in range   $\Delta f_P$ = 73~kHz.  The simulations revleads a value of 62~kHz.  A series of other simulations delivered results for the pull-in time  $\Delta T_P$. The results are listed in Table 3-1.
\begin{table*}
\begin{center}
  \begin{tabular}{|l|l|l|l|}
   \hline
   $\Delta f_0$ (Hz & $\Delta\omega_0$ (rad $s^{-1}$) & $T_P$ (theory) ($\mu s$) & ($T_P$ (simulation) ($\mu s$)\\
   \hline
   40 kHz &251'200& 14  &35\\
   \hline
   50 kHz & 314'000 &37 & 40\\
   \hline
   60 kHz & 376'800 & 86& 70\\
   \hline
 \end{tabular}
\bigskip

Table 3-1. Comparison of predicted and simulated results for the pull-in range
\end{center}
\end{table*}

At higher frequency offsets the results of the simulation are in good agreement with the predicted ones. The pull-in time for an initial frequency offset of 40~kHz is too low, however, but it should be noted that the lock time $T_L$ is about 25 $\mu s$, and the the total pull-in time cannot be less than the lock time.


\subsection{Remarks on simulation of QPSK Costas loop}
Similar problems to BPSK Costas loop simulation also exist for QPSK.
Different mathematical models can give qualitatively different results,
which shows the importance of analytical methods in studying QPSK Costas loops.

\renewcommand{\thefigure}{4.\arabic{figure}}
\setcounter{figure}{0}

\section{Modified Costas loop for BPSK}\label{s4}

\subsection{Lock-in range   $\Delta\omega_L$ and lock time $T_L$}\label{ss4.2}
From the model of Figure~\ref{linear-pll} with $K_d = 1$ the open loop transfer function is determined to be
\begin{equation}
\label{modified bpsk ol tf}
G_{OL}(s)=\frac{K_0}{s}\frac{1+s\tau_2}{s\tau_1}.
\end{equation}

Since open loop transfer function of Modified Costas loop is effectively the same as \eqref{7},
linear analysis is the same as for BPSK Costas loop.
Therefore
transfer function in normalized form is equal to
$$
G_{CS}(s)=\frac{2s\zeta\omega_n+\omega_n^2}{s^2+2c\zeta\omega_n+\omega_n^2},
$$
where
\begin{equation}\label{62}
\omega_n = \sqrt{\frac{K_0}{\tau_1}},\quad \zeta=\frac{\omega_n\tau_2}{2}.
\end{equation}
Here  $\omega_n$ is  natural frequency and $\zeta$  is damping factor.

\begin{figure}[H]
\centering
\includegraphics[width=0.9\linewidth]{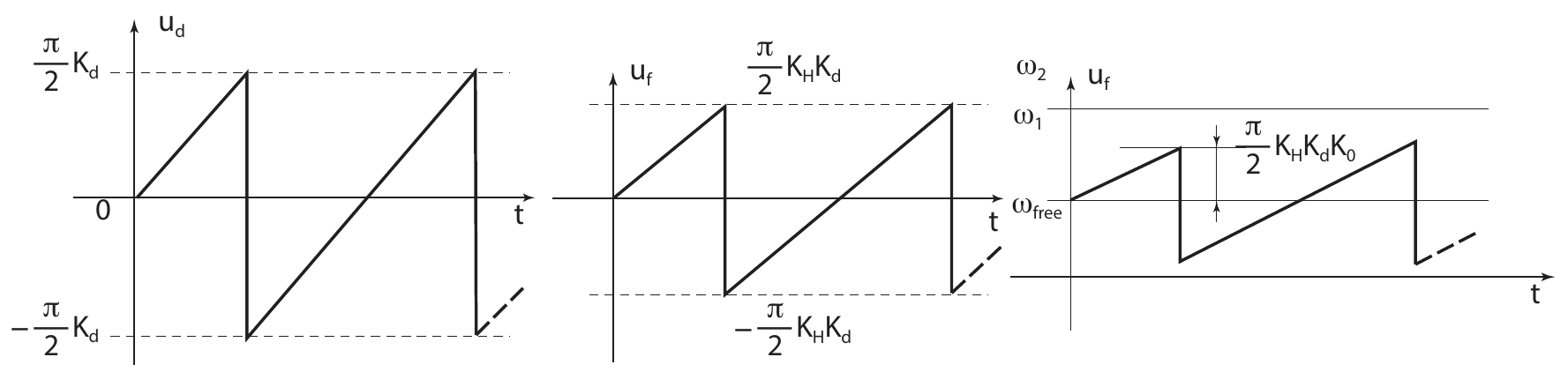}
\caption{Signals $u_d, u_f$, and  $\omega_2$ during the lock process
}\label{f4-5}
\end{figure}

For the following analysis we assume that the loop is initially out of lock. The frequency of the reference signal (Fig. \ref{f4-1}) is  $\omega_1$, and the frequency of the VCO is  $\omega_2$. The output signal of multiplier $M_1$ is then a phasor rotating with angular velocity  $\Delta\omega  =  \omega_1 -  \omega_2$. Consequently the phase output of block ``Complex $\to$  mag, phase'' is a sawtooth signal having amplitude ($\pi/2$) $K_d$ and fundamental frequency $2\Delta\omega$, as shown in the left trace of Figure~\ref{f4-5}. Because $2\Delta\omega$    is usually much higher than the corner frequency  $\omega_C$ of the loop filter, the transfer function of the loop filter at higher frequencies can be approximated again by
$$
H_{LF}(\omega)\approx\frac{\tau_2}{\tau_1}=K_H.
$$

The output signal $u_f$ of the loop filter is a sawtooth signal as well and has amplitude
($\pi/2$) $K_d K_H$, as shown in the middle trace of the figure \ref{f4-5}. This signal modulates the frequency  $\omega_2$ generated by the VCO. The modulation amplitude is given by
($\pi/2$)  $K_d$  $K_H$  $K_0$, cf. right trace. The Costas loop spontaneously acquires lock when the peak of the  $\omega_2$ waveform touches the  $\omega_1$ line, hence we have
$$
\Delta\omega_L=\frac{\pi}{2}K_dK_0K_H=\frac{\pi}{2}K_dK_0\frac{\tau_2}{\tau_1}.
$$

Making use of the substitutions Eqn. (\ref{62}) this can be rewritten as
\begin{equation}\label{63}
\Delta\omega_L=\pi\zeta\omega_n.
\end{equation}
Because the lock process is a damped oscillation having frequency  $\omega_n$  the lock time can be approximated by one cycle of this oscillation, i.e.
\begin{equation}\label{64}
T_L\approx\frac{2\pi}{\omega_n}.
\end{equation}

\subsection{Pull-in range and pull-in time of the modified Costas loop for BPSK}\label{ss4.4}

\begin{figure}[H]
\centering
\includegraphics[width=0.9\linewidth]{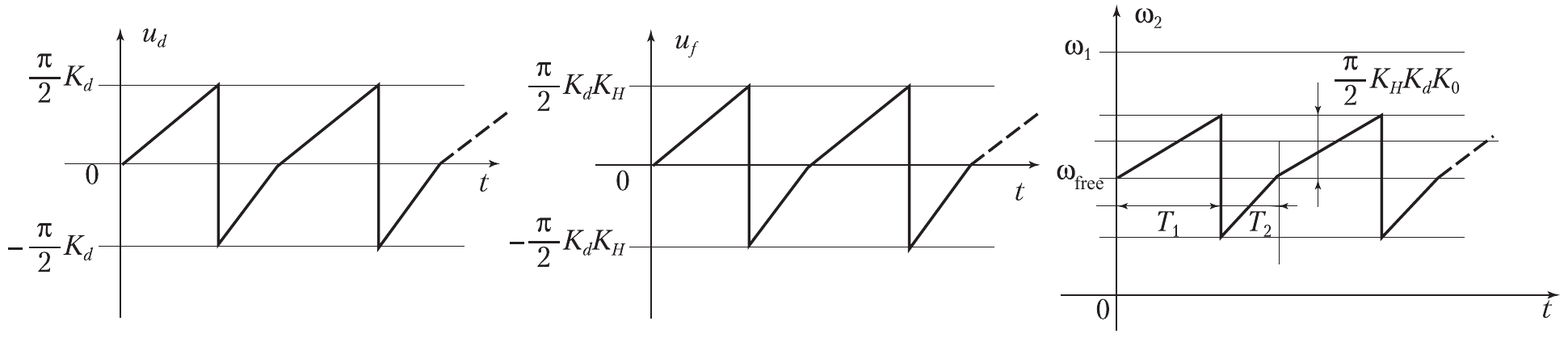}
\caption{Pull-in process of the modified Costas loop for BPSK
}\label{f4-6}
\end{figure}

Assume that the loop is not yet locked, and  $\Delta\omega  =  \omega_1 -  \omega_2$. As shown in section~\ref{ss4.2} (cf. also Figure~\ref{f4-5}) $u_d$ is a sawtooth signal having frequency $2\Delta\omega$, cf. left trace in Figure~\ref{f4-6}. As will be explained in short, this signal is asymmetrical, i.e. the duration of the positive wave $T_1$ is not identical with the duration $T_2$ of the negative. The middle trace shows the output signal of the loop filter, and the right trace shows the modulation of the VCO output frequency  $\omega_2$. From this waveform it is seen that during $T_1$ the average frequency difference $\Delta\omega$   becomes smaller, but during interval $T_2$ it becomes larger. Consequently the duration of $T_1$ is longer than the duration of $T_2$, and the average of signal $u_d$ is non zero and positive. Using the same mathematical procedure as in previews sections, the average $\overline{u_d}$ can be computed from
\begin{equation}\label{65}
\overline{u_d}=\frac{\pi^2K_dK_0K_H}{8\Delta\omega}.
\end{equation}

Because this type of Costas loop does not require an additional lowpass filter, the $u_d$ signal is not shifted in phase, and therefore there is no $\cos$ term in Eqn. (\ref{65}). This implies that there is no polarity reversal in the function $\overline{u_d}(\Delta\omega)$, hence the pull-in range becomes theoretically infinite. Of course, in a real circuit the pull-in range will be limited by the frequency range of the VCO is capable to generate. When the center frequency $f_0$ of the loop is 10 MHz, for example, and when the VCO can create frequencies in the range from $0\ldots 20$ MHz, then the maximum pull-in range  $\Delta f_P$ is 10 MHz, i. e.  $\Delta\omega_P = 6.28\cdot 10^6\,{\rm{rad}}/s$.

As seen in the last section, the pull-in range of this type of Costas loop can be arbitrarily large. Using the same model as for BPSK Costas loop (see Figure~\ref{f2-7}), we can derive an equation for the pull-in time:
\begin{equation}\label{68}
T_P\approx\frac{2}{\pi^2}\frac{\Delta\omega_0^2}{\zeta\omega_n^3}.
\end{equation}

\subsection{Designing a digital modified Costas loop for BPSK}\label{ss4.5}

The following design is based on the method we already used in section \ref{ss2.5}. It is assumed that a  binary signal $I$   is modulated onto a carrier. The carrier frequency is set to 400 kHz, i.e. the Costas loop will operate at a center frequency  $\omega_0 = 2\pi$    400'000 = 2'512'000\, rad $s^{-1}$.  The symbol rate is assumed to be $f_S = 100'000 \, {\rm{symbols}}/s$. Now the parameters of the loop (such as time constants  $\tau_1$ and  $\tau_2$, corner frequency  $\omega_C$, and gain parameters such as $K_0, K_d$)  must be determined. (Note that these parameters have been defined in Eqs.~(\ref{4}), (\ref{loop-filter-tf}), (\ref{6}), and (\ref{13})).

It has been shown in section 4.1 that for this type of Costas loop $K_d = 1$. The modulation amplitudes $m_1$ and $m_2$ are set to 1. It has proven advantageous to determine the remaining parameters by using the open loop transfer function $G_{OL}(s)$ of the loop,
 which is given here by \eqref{modified bpsk ol tf}.
The magnitude of $G_{OL}(\omega)$ has been shown in Figure~\ref{f2-2}.  As already explained in section \ref{ss2.5} the magnitude curve crosses the 0 dB line at the transit frequency  $\omega_T$. As in the case of the conventional Costas loop for BPSK/QPSK, we again set
\begin{equation}\label{costas parameters 1}
\begin{aligned}
& \omega_T = 0.1\omega_0, \\
& \omega_T = 251'200\, {\rm{rad}} s^{-1}, \\
& G_{OL}(\omega) = -135^{o}, \\
& \tau_2 = 4 \mu s, \\
& \tau_1 = 20 \mu s, \\
& K_0 = 1'262'000\, s^{-1}.
\end{aligned}
\end{equation}

For the natural frequency and damping factor we get from Eqn. (\ref{omega zeta})
\begin{equation}
\label{costas parameters 2}
\begin{aligned}
&\omega_n=251'000 {\rm{rad}}/s\quad (f_n=40 {\rm{kHz}})\\
&\zeta=0.5.\end{aligned}
\end{equation}

From (\ref{62}) lock-in range is
\begin{equation}
\label{costas parameters 3}
  \Delta\omega_L = 394'000 {\rm{rad}}\, s,
  \quad \Delta f_L = 62.7 {\rm{kHz}},
  \quad T_L = 25\mu  s.
\end{equation}

As done in section~\ref{ss2.6} a suitable sampling frequency $f_{samp}$ must be chosen for $z$-domain. As shown previously $f_{samp}$ must be greater than 4 times the center frequency of the Costas loop. Therefore $f_{samp} = 8$, $f_0 = 3.2$~MHz.
The transfer functions of the loop filter and VCO are the same as \eqref{39}
and \eqref{40}.

The digital Costas loop is ready now for implementation. A Simulink model will be presented in section \ref{ss4.6}.

\subsection{Simulating the modified digital Costas loop for BPSK}\label{ss4.6}

\begin{figure}[H]
\centering
\includegraphics[width=0.95\textwidth]{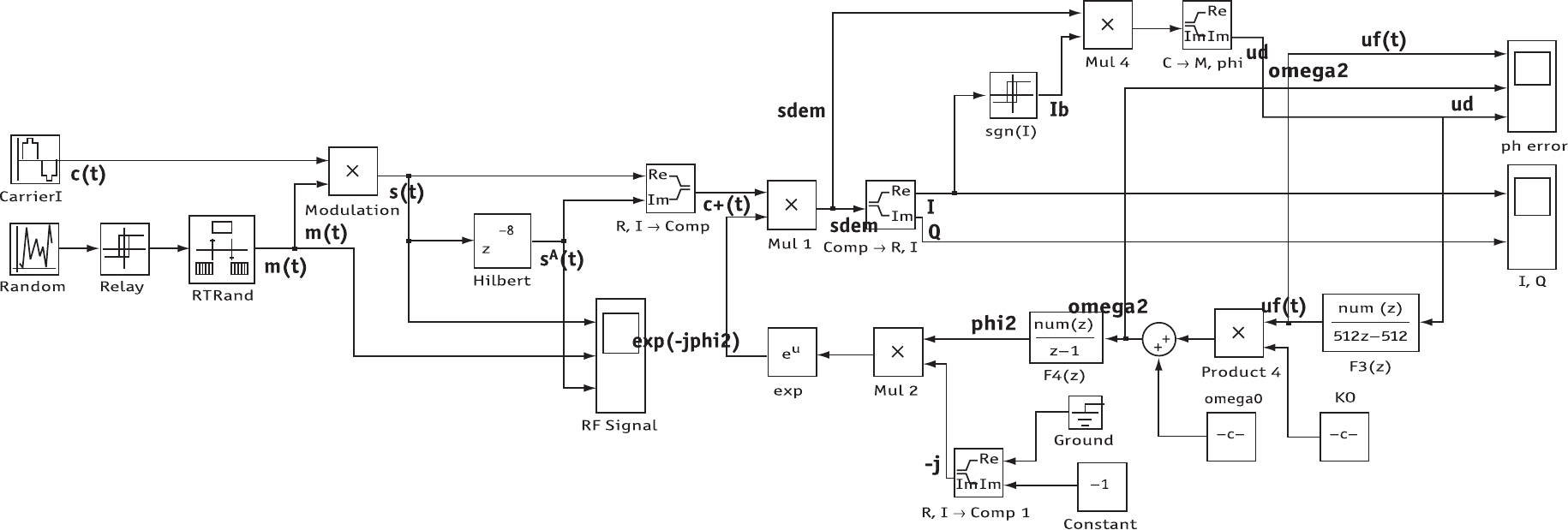}
\caption{Simulink model of the modified Costas loop for BPSK
}\label{f4-8}
\end{figure}

Figure~\ref{f4-8} shows the Simulink model of the Costas loop. Table 4-1 lists a number of results for the pull-in time $T_P$.
\begin{table*}
\begin{center}
\begin{tabular}{|l|l|l|l|}
\hline
$\Delta f_0$ (Hz)   &$\Delta\omega_0$ (rad $s^{-1}$)  &$T_P$ (theory) ($\mu s)$ & ($T_P$ (simulation) ($\mu s$)\\
\hline
50 kHz  &314'200& 2.5 &20\\
\hline
100 kHz & 628'000 &10 & 20\\
\hline
200 kHz & 1'256'000 & 40& 50\\
\hline
\end{tabular}

\bigskip
Table 4-1. Comparison of predicted and simulated results for the pull-in range
\end{center}
\end{table*}

The predictions for  $\Delta f_0 = 50$ kHz and 100 kHz are too low. As already mentioned in section \ref{ss3.6} the pull-in time cannot be lower than the lock time, and the latter is estimated  $\approx 25 \mu s$. The simulation results for these two difference frequencies are around $20 \mu s$, which roughly corresponds to the lock time. The simulation result for a frequency difference of 200 kHz comes close to the predicted value.

\subsection{Pull-in time for an alternative structure of the modified Costas loop for BPSK}\label{ss4.7}

\begin{figure}[H]
\centering
\includegraphics[width=0.35\textwidth]{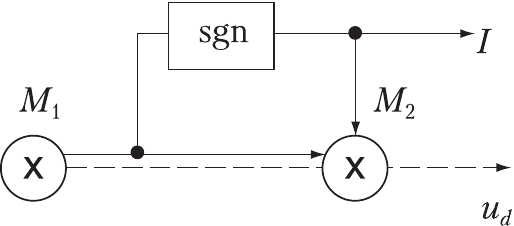}
\caption{Obtaining the phase error signal from multiplier $M_2$
}\label{f4-9}
\end{figure}

As demonstrated in Figure~\ref{f4-1} the phase error signal $u_d$ was obtained from the phase output of block "Complex $\to$  mag, phase". The phase of the complex input signal to this block can be obtained from the arc tg function. This imposes no problem when a processor is available. This is the case in most digital implementations of the Costas loop. As an alternative a phase error signal can also be obtained directly from the imaginary part of multiplier $M_2$; this is shown in Figure~\ref{f4-9}.

It is easily seen that here $u_d$ is given by
$$
u_d=m(t)\sin(\theta_e).
$$

The blocks shown in Figure~\ref{f4-9} therefore represent a phase detector having gain $K_d = m$. In cases when $m\ne  1$ this must be taken in account when specifying the open loop transfer function, cf. section~\ref{ss4.5}. For this design the pull-in time of the loop is given by
$$
T_p\approx\frac{\pi^2}{16}\frac{\Delta\omega_0^2}{\zeta\omega_n^3}.
$$

\subsection{A note on the design of Hilbert transformers}\label{ss4.8}

Hilbert transformers as used in the system of Figure~\ref{f4-1} are implemented in most cases by digital filters. In this application the maximum frequency in the spectrum of the modulating signal $m_1(t)$ is much lower than the carrier frequency $f_1$. Under this condition the Hilbert transformer can be replaced by a simple delay  block. All we have to do is to shift the input signal $u_1(t)$ by one quarter of a period of the carrier. When the sampling frequency $f_S$ is $n$ times the carrier frequency $f_1$, we would shift the input signal by $n/4$ samples. This implies that $n$ must be an integer multiple of 4.

\renewcommand{\thefigure}{5.\arabic{figure}}
\setcounter{figure}{0}

\section{Modified Costas loop for QPSK}\label{s5}
\subsection{Lock-in range   $\Delta\omega_L$ and lock time $T_L$}\label{ss5.2}
 From the model of Figure~\ref{linear-pll} the open loop transfer function is determined to be
\begin{equation}
\label{modified qpsk ol tf}
G_{OL}(s) = \frac{K_0}{s}\frac{1+s\tau_2}{s\tau_1}.
\end{equation}
 as explained in section~\ref{1.1.6}.
\begin{figure}[H]
\centering
\includegraphics[scale=0.8]{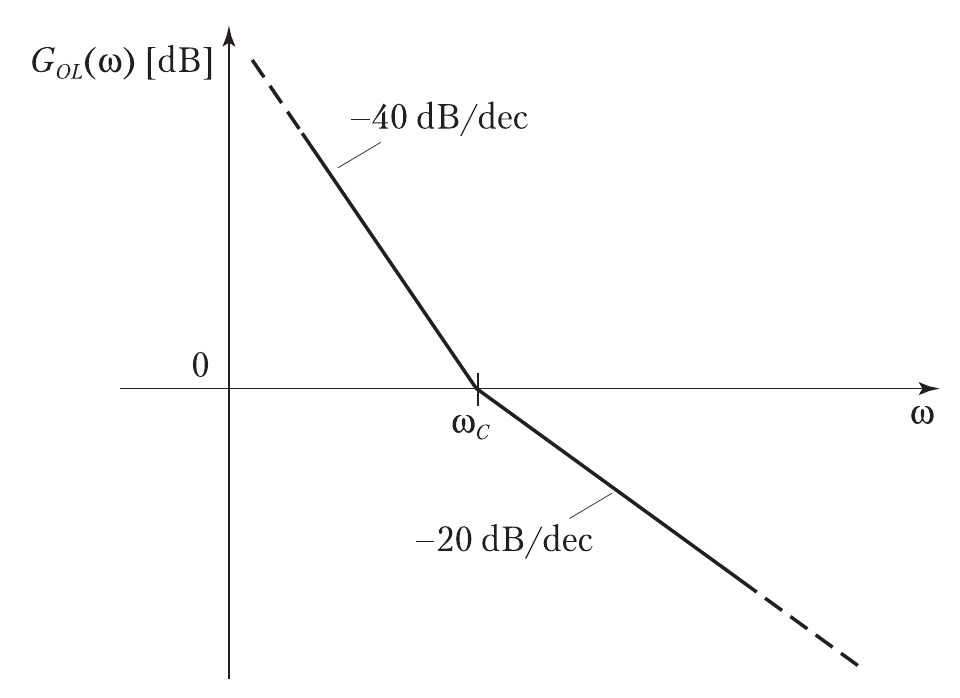}
\caption{Bode plot of magnitude of open loop gain $G_{OL}(\omega)$ for QPSK
}\label{f5-4}
\end{figure}


Figure~\ref{f5-4} shows a Bode plot of the magnitude of $G_{OL}$. The plot is characterized by the corner frequency  $\omega_C$, which is defined by  $\omega_C = 1/\tau_2$, and gain parameters $K_d$ and $K_0$. At lower frequencies the magnitude rolls off with a slope of -- 40 dB/decade. At frequency  $\omega_C$ the zero of the loop filter causes the magnitude to change its slope to -- 20 dB/decade. To get a stable system, the magnitude curve should cut the 0 dB line with a slope that is markedly less than -- 40 dB/decade. Setting the parameters such that the gain is just 0 dB at frequency  $\omega_C$ provides a phase margin of 45 degrees, which assures stability [2]. From the open loop transfer function we now can calculate the closed loop transfer function defined by
$$
G_{CL}(s)=\frac{\Theta_2(s)}{\Theta_1(s)}.
$$

After some mathematical manipulations we get
$$
G_{CL}(s)=\frac{K_0K_d\frac{1+s\tau_2}{s\tau_1}}{s^2+s\frac{K_0K_d\tau_2}{\tau_1}+\frac{K_0K_d}{\tau_1}}.
$$

It is customary to represent this transfer function in normalized form, i.e.
$$
G_{CS}(s)=\frac{2s\zeta\omega_n+\omega_n^2}{s^2+2s\zeta\omega_n+\omega_n^2}
$$
with the substitutions
\begin{equation}\label{77}
\omega_n=\sqrt\frac{K_0K_d}{\tau_1},\quad \zeta=\frac{\omega_n\tau_2}{2},
\end{equation}
where  $\omega_n$ is called natural frequency and $\zeta$   is called damping factor. The linear model enables us to derive simple expressions for lock-in range  $\Delta\omega_L$ and lock time $T_L$.

\begin{figure}[H]
\centering
\includegraphics[width=0.9\linewidth]{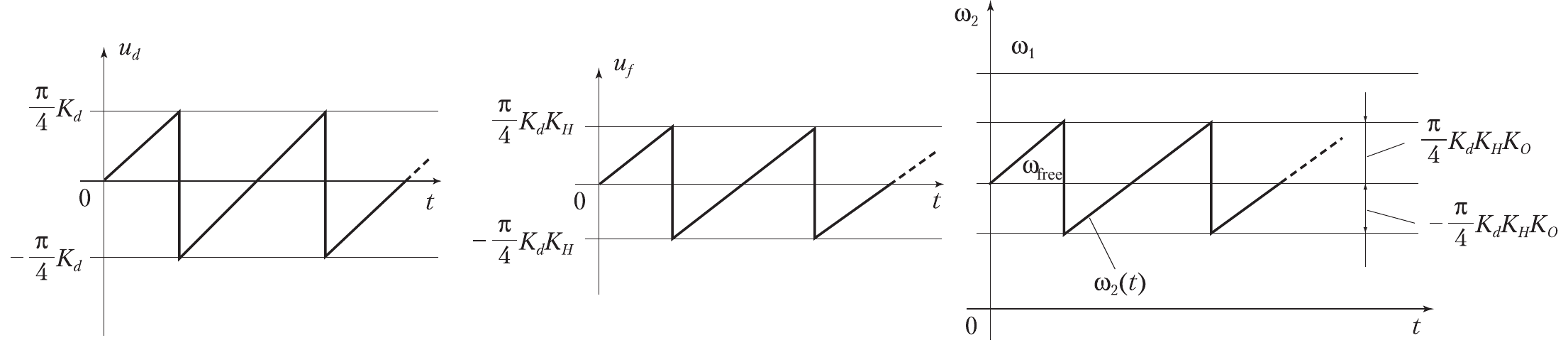}
\caption{Signals $u_d, u_f$, and  $\omega_2$ during the lock process
}\label{f5-5}
\end{figure}

For the following analysis we assume that the loop is initially out of lock. The frequency of the reference signal (Figure~\ref{f5-1}) is  $\omega_1$,and the frequency of the VCO is  $\omega_2$. The output signal of multiplier $M_1$ is then a phasor rotating with angular velocity  $\Delta\omega  =  \omega_1 -  \omega_2$. Consequently the phase output of block "Complex $\rightarrow$  mag, phase is a sawtooth signal having amplitude $(\pi/4)$ $K_d$ and fundamental frequency 4 $\Delta\omega$, as shown in the left trace of Figure~\ref{f5-5}. Because 4 $\Delta\omega$   is usually much higher than the corner frequency  $\omega_C$ of the loop filter, the transfer function of the loop filter at higher frequencies can be approximated again by
$$
H_{LF}(\omega)\approx\frac{\tau_2}{\tau_1}=K_H.
$$

The output signal $u_f$ of the loop filter is a sawtooth signal as well and has amplitude
$(\pi/4)$ $K_d K_H$, as shown in the middle trace of the figure. This signal modulates the frequency  $\omega_2$ generated by the VCO. The modulation amplitude is given by
$(\pi/4)$ $K_d K_H K_0$, cf. right trace. The Costas loop spontaneously acquires lock when the peak of the  $\omega_2$ waveform touches the  $\omega_1$ line, hence we have
\begin{equation}
\Delta\omega_L=\frac{\pi}{4}K_dK_0K_H=\frac{\pi}{4}K_dK_0\frac{\tau_2}{\tau_1}.
\end{equation}

Making use of the substitutions Eqn. (\ref{62}), this can be rewritten as
\begin{equation}\label{78}
\Delta\omega_L=\frac{\pi}{2}\zeta\omega_n.
\end{equation}

Because the lock process is a damped oscillation having frequency  $\omega_n$,  the lock time can be approximated by one cycle of this oscillation, i.e.
\begin{equation}\label{79}
T_L\approx\frac{2\pi}{\omega_n}.
\end{equation}

\subsection{Pull-in range and pull-in time of the modified Costas loop for QPSK}\label{ss5.4}

\begin{figure}[H]
\centering
\includegraphics[width=0.9\linewidth]{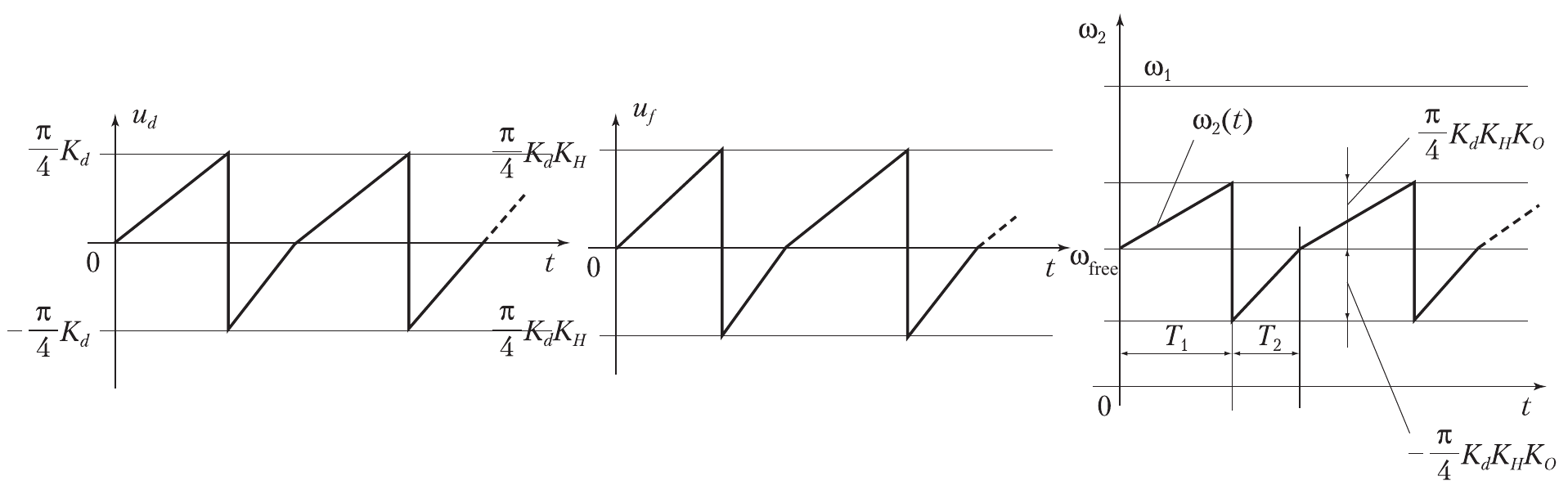}
\caption{Pull-in process of the modified Costas loop for QPSK
}\label{f5-6}
\end{figure}

Assume that the loop is not yet locked, and that the difference between reference frequency  $\omega_1$  and VCO output frequency  $\omega_2$  is $\Delta\omega   =  \omega_1 -  \omega_2$. As shown in section \ref{ss5.2} (cf. also Figure~\ref{f5-5}) $u_d$ is a sawtooth signal having frequency $4 \Delta\omega$, cf. left trace in Figure~\ref{f5-6}.

As will be explained in short, this signal is asymmetrical, i.e. the duration of the positive wave $T_1$ is not identical with the duration $T_2$ of the negative. The middle trace shows the output signal of the loop filter, and the right trace shows the modulation of the VCO output frequency  $\omega_2$. From this waveform it is seen that during $T_1$ the average frequency difference  $\Delta\omega$  becomes smaller, but during interval $T_2$ it becomes larger. Consequently the duration of $T_1$ is longer than the duration of $T_2$, and the average of signal $u_d$ is non zero and positive. Using the same mathematical procedure as in sections 2. 3 and 3.3 the average $\overline{u_d}$ can be computed from
\begin{equation}\label{80}
\overline{u_d}=\frac{\pi^2K_d^2K_0K_H}{64\Delta\omega}.
\end{equation}

Because this type of Costas loop does not require an additional lowpass filter, the $u_d$ signal is not shifted in phase, and therefore there is no $\cos$ term in Eqn. (\ref{80}). This implies that there is no polarity reversal in the function $\overline{u_d}(\Delta\omega)$, hence the pull-in range becomes theoretically infinite. Of course, in a real circuit the pull-in range will be limited by the frequency range of the VCO is capable to generate. When the center frequency $f_0$ of the loop is 10 MHz, for example, and when the VCO can create frequencies in the range from $0\ldots 20$ MHz, then the maximum pull-in range  $\Delta f_P$ is 10 MHz, i. e.  $\Delta\omega_P = 6.28\cdot 10^6\, {\rm{rad/s}}$.

As seen in the last section, the pull-in range of this type of Costas loop can be arbitrarily large. Using non linear model \eqref{f2-7} we can derive an equation for the pull-in range:
\begin{equation}\label{83}
\Delta\omega_p\approx\frac{16}{\pi^2}\frac{\Delta\omega_0^2}{\zeta\omega_n^3}.
\end{equation}

\subsection{Designing a digital modified Costas loop for QPSK}\label{ss5.5}

The following design is based on the method we already used in section \ref{ss4.5}. It is assumed that two  binary signals ($I$ and $Q$) are modulated onto a quadrature carrier (cosine and sine carrier). The carrier frequency is set to 400 kHz, i.e. the Costas loop will operate at a center frequency  $\omega_0 = 2\pi    400'000 = 2'512'000\, {\rm{rad}}\, s^{-1}$.  The symbol rate is assumed to be $f_S = 100'000\, {\rm{symbols}}/s$. Now the parameters of the loop (such as time constants  $\tau_1$ and  $\tau_2$, corner frequency  $\omega_C$, and gain parameters such as $K_0, K_d$)  must be determined. (Note that these parameters have been defined in Eqs.~(\ref{4}), (\ref{loop-filter-tf}), (\ref{6}), and (\ref{13})).

It has been shown in previews sections that for this type of Costas loop $K_d = 1$. The modulation amplitudes $m_1$ and $m_2$ are set to 1. It has proven advantageous to determine the remaining parameters by using the open loop transfer function $G_{OL}(s)$ of the loop, which is given here by \eqref{modified qpsk ol tf}.
The magnitude of $G_{OL}(\omega)$ has been shown in Figure~\ref{f5-4}.  As already explained in section \ref{ss2.5} the magnitude curve crosses the $0$ dB line at the transit frequency  $\omega_T$.
We again set parameters as in \eqref{costas parameters 1},\eqref{costas parameters 2} and \eqref{costas parameters 3}.
A Simulink model will be presented in section \ref{ss5.6}.

\subsection{Simulating the digital Costas loop for QPSK}\label{ss5.6}

\begin{figure}[H]
\centering
\includegraphics[width=0.95\textwidth]{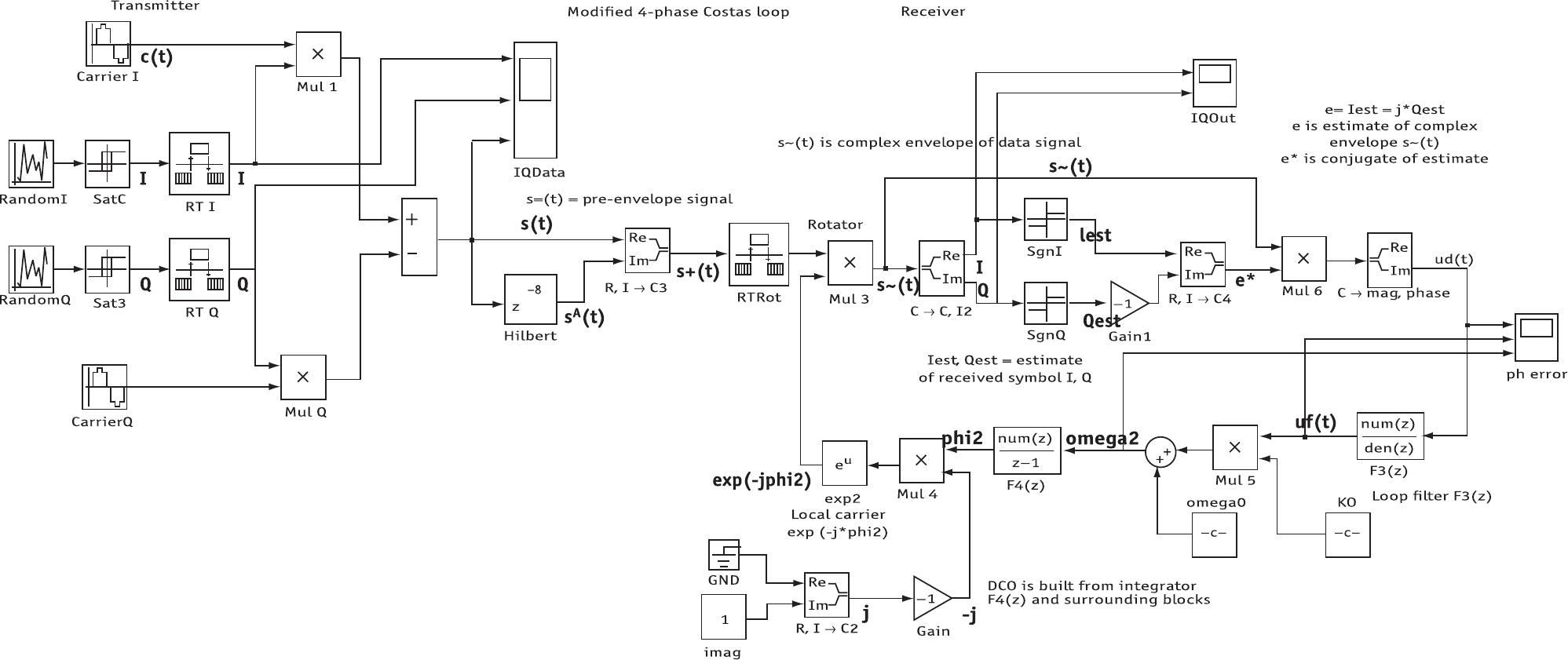}
\caption{Simulink model of the modified Costas loop for QPSK}
\label{f5-8}
\end{figure}

Figure~\ref{f5-8} shows the Simulink model of the Costas loop. Table 5-1 lists a number of results for the pull-in time $T_P$.
\begin{table*}
\begin{center}
\begin{tabular}{|l|l|l|l|}
\hline
$\Delta f_0$ (Hz)   &$\Delta\omega_0$ (rad $s^{-1}$)  &$T_P$ (theory) ($\mu s)$ & ($T_P$ (simulation) ($\mu s$)\\
\hline
50 kHz  &314'200& 20  &20\\
\hline
100 kHz & 628'000 &81 & 80\\
\hline
200 kHz & 1'256'000 & 327&  300\\
\hline
\end{tabular}

\bigskip

Table 5-1. Comparison of predicted and simulated results for the pull-in range

\end{center}
\end{table*}

The predictions come very close to the results obtained from the simulation.

\subsection{An alternative structure of the modified Costas loop for BPSK}\label{ss5.7}

\begin{figure}[H]
\centering
\includegraphics[width=0.4\textwidth]{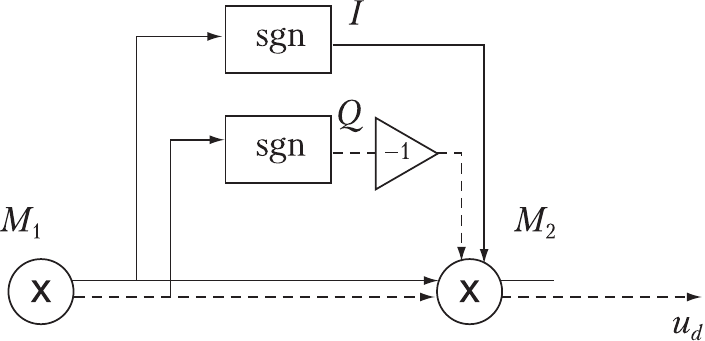}
\caption{Obtaining the phase error signal from multiplier $M_2$
}\label{f5-9}
\end{figure}

As demonstrated in Figure~\ref{f5-1} the phase error signal $u_d$ was obtained from the phase output of block "Complex $\to$  mag, phase". The phase of the complex input signal to this block can be obtained from the $\arctg$ function. This imposes no problem when a processor is available. This is the case in most digital implementations of the Costas loop. As an alternative a phase error signal can also be obtained directly from the imaginary part of multiplier $M_2$; this is shown in Figure~\ref{f5-9}.

It is easily seen that here $u_d$ is given by
$$
u_d=2m\sin(\theta_e).
$$

The blocks shown in Figure~\ref{f5-9} therefore represent a phase detector having gain $K_d = 2 m$. This must be taken in account when specifying the open loop transfer function, cf. section \ref{ss5.5}. For this design the pull-in time of the loop is given by
$$
T_p\approx\frac{1.78\Delta\omega_0^2}{\zeta\omega_n^3}.
$$

\section{Acknowledgements}
This work was supported by Russian Science Foundation (project 14-21-00041) and Saint-Petersburg State University

\section{Appendix}
\subsection{Hold-in range for lead-lag filter}
One needs to be cautious using model in Figure~\ref{f2-1b} even for calculating hold-in range for BPSK Costas.
Consider an example: Costas loop with lead-lag loop filter
\begin{equation}
  \begin{aligned}
    & F(s) =\frac{1+s\tau_2}{1+s\tau_1},\quad \tau_1 > \tau_2 > 0
  \end{aligned}
\end{equation}
and low-pass filters LPFs
\begin{equation}
  \begin{aligned}
    & H_{LPF}(s) = \frac{1}{1+\frac{s}{\omega_3}},\ \omega_3 > 0.
  \end{aligned}
\end{equation}

In locked state phase error $\theta_{e}$ satisfies
\begin{equation}
  \label{phase-error-equation}
  \begin{aligned}
    & \frac{\Delta\omega}{K_0K_d}=\frac{\sin(2\theta_{e})}{2},
  \end{aligned}
\end{equation}
therefore we get a bound for the hold-in range
\begin{equation}
\label{interv}
  \begin{aligned}
    & |\Delta\omega_0|
    <
    \frac{K_0K_d}{2}.
  \end{aligned}
\end{equation}
In order to find hold-in range we need to find poles of the closed-loop transfer function (roots of the characteristic polynomial) for the linearized model (small-signal model) of the system on Figure~\ref{costas_before_sync}.
Open-loop transfer function is
\begin{equation}
  \begin{aligned}
    & G_{OL} = \frac{K_0K_d}{s}\frac{1+s\tau_2}{1+s\tau_1}\frac{1}{1+\frac{s}{\omega_3}}\frac{\cos(2\theta_{eq})}{2}
  \end{aligned}
\end{equation}
\begin{equation}
  \label{pol-1}
  \begin{aligned}
    & \frac{1}{2}(1+\tau_2 s)K_0K_d\cos(2\theta_{eq}) + s(1+\frac{s}{\omega_3})(1+\tau_1 s).
  \end{aligned}
\end{equation}
Phase error $\theta_{eq}$ corresponds to hold-in range (see \eqref{phase-error-equation}) if
all roots of the polynomial \eqref{pol-1} have negative real parts (i.e. polynomial \eqref{pol-1} is stable).
Applying Routh-Hurwitz criterion to study stability of the polynomial, we get that for the following parameters
\begin{equation}
\begin{aligned}
  & \omega_3 \geq \frac{\tau_1 - \tau_2}{\tau_1\tau_2},
\end{aligned}
\end{equation}
polynomial \eqref{pol-1} is stable for all $|\Delta\omega_0|<\frac{K_0K_d}{8}$.
However, if
\begin{equation}
  \begin{aligned}
    &\omega_3 < \frac{\tau_1 - \tau_2}{\tau_1\tau_2}
  \end{aligned}
\end{equation}
the following condition is necessary for stability
\begin{equation}
  \begin{aligned}
    & \cos(2\theta_{eq})
    <
    \frac{2}{K_0K_d}
    \left(
    \frac{-1-\omega_3\tau_1}{-\tau_1+\tau_2+\omega_3\tau_1\tau_2}
    \right).
  \end{aligned}
\end{equation}
Then, taking into account static phase error equation \eqref{phase-error-equation}, we get different hold-in ranges for different values of $\omega_3$
\begin{equation}
\label{to-proof}
\left[
  \begin{aligned}
    & \frac{K_0K_d}{4}
    \sqrt{
    1-
    \left(
      \frac{2}{K_0K_d}
      \left(
        \frac{-1-\omega_3\tau_1}{-\tau_1+\tau_2+\omega_3\tau_1\tau_2}
      \right)
    \right)^2
    } < |\Delta\omega_0| < \frac{K_0K_d}{4},
    \\
    & \quad\quad\quad\quad\quad\quad\quad \text{if }
    \omega_3 < \frac{\tau_1 - \tau_2}{\tau_1\tau_2},\
     |\frac{2}{K_0K_d}
      \left(
        \frac{-1-\omega_3\tau_1}{-\tau_1+\tau_2+\omega_3\tau_1\tau_2}
      \right)|<1,\\
      & |\Delta\omega_0| < \frac{K_0K_d}{4}, \
      \text{if }
    \omega_3 < \frac{\tau_1 - \tau_2}{\tau_1\tau_2},\
     |\frac{2}{K_0K_d}
      \left(
        \frac{-1-\omega_3\tau_1}{-\tau_1+\tau_2+\omega_3\tau_1\tau_2}
      \right)|>1,\\
      & |\Delta\omega_0| < \frac{K_0K_d}{4},\ \text{if } \omega_3 \geq \frac{\tau_1 - \tau_2}{\tau_1\tau_2}.\\
  \end{aligned}
  \right.
\end{equation}

\end{document}